%% file: zeusF2bc.tex
\documentclass[zpreprint,zbstjhep,nomcite,pagesize,british]{zeus_Kpaper}

\usepackage{./zeusdet_siunitx}
\usepackage{booktabs}
\usepackage{xfrac}
\usepackage{enumerate}

\graphicspath{{./Figs/}}

\input{zeus_defs.tex}

\input{def.tex}

\newcommand*{\mbMSbarmeas}{\ensuremath{m_b(m_b) =
    \SI[parse-numbers=false]{4.07 \pm 0.14\,(\text{fit})
    ^{+0.01}_{-0.07}\,(\text{mod.})
    ^{+0.05}_{-0.00}\,(\text{param.})\,
    ^{+0.08}_{-0.05}\,(\text{theo.})}{\GeV}}\xspace}

\input{zeusF2bc-cit_nomcite.tex}
\begin{document}
%
%
\prepnum{DESY--14--083}
\prepdate{May 2014}

\addtolength{\zeustitlespace}{-0.5cm}
\zeustitle{%
  Measurement of beauty and charm production
  in deep inelastic scattering at HERA
  and measurement of the beauty-quark mass
}

\zeusauthor{ZEUS Collaboration}
\zeusdate{}

\maketitle

\begin{abstract}\noindent
  The production of beauty and charm quarks in $ep$
  interactions has been studied with the ZEUS detector at HERA for
  exchanged four-momentum squared $\num{5} < \Qsq < \SI{1000}{\GeV\squared}$ using an
  integrated luminosity of \SI{354}{\per\pb}.
  The beauty and charm content in events with at least one jet have
  been extracted using the invariant mass of charged tracks associated
  with secondary vertices and the decay-length significance of these
  vertices. Differential cross sections as a function of \Qsq,
  Bjorken $x$, jet transverse energy and pseudorapidity were measured
  and compared with next-to-leading-order QCD calculations. The beauty
  and charm contributions to the proton structure functions were
  extracted from the double-differential cross section as a function
  of $x$ and \Qsq.  The running beauty-quark mass, $m_{b}$ at the scale $m_{b}$, was determined
  from a QCD fit at next-to-leading order to HERA data for the first time and found to be
  \mbMSbarmeas.
\end{abstract}
\thispagestyle{empty}
\clearpage
\input{auth182a_out}
\clearpage
\mbox{}
\thispagestyle{empty}
\clearpage
\pagenumbering{arabic}
%
%
\input{zeusF2bc-txt}
\clearpage
{\raggedright
\input{zeusF2bc.bbl}

}
\input{zeusF2bc-tab}
\input{zeusF2bc-fig}
%
%
\end{document}

%% file: zeus_defs.tex
\chardef\usc=95
\chardef\til=126
\catcode`\@=11 
\DeclareRobustCommand\xdotspace{\futurelet\@let@token\@xdotspace}
\def\@xdotspace{%
  \ifx\@let@token.\else
  \ifx\@let@token\bgroup.\else
  \ifx\@let@token\egroup.\else
  \ifx\@let@token\/.\else
  \ifx\@let@token\ .\else
  \ifx\@let@token~.\else
  \ifx\@let@token!.\else
  \ifx\@let@token,.\else
  \ifx\@let@token:.\else
  \ifx\@let@token;.\else
  \ifx\@let@token?.\else
  \ifx\@let@token/.\else
  \ifx\@let@token'.\else
  \ifx\@let@token).\else
  \ifx\@let@token-.\else
  \ifx\@let@token\@xobeysp.\else
  \ifx\@let@token\space.\else
  \ifx\@let@token\@sptoken.\else
   .\space
   \fi\fi\fi\fi\fi\fi\fi\fi\fi\fi\fi\fi\fi\fi\fi\fi\fi\fi}
\catcode`\@=12 

\newcommand{\stru}[2]{%
   \relax\ifmmode\hbox{\vrule height#1 depth#2 width0pt}%
   \else\vrule height#1 depth#2 width0pt\fi}

\newcommand{\Ronum}[1]{\uppercase\expandafter{\romannumeral#1}}
\newcommand{\ronum}[1]{\expandafter{\romannumeral#1}}
\DeclareRobustCommand{\LaTeXZ}{%
  \LaTeX\kern-.05em4\kern-.1em
  {\raisebox{-0.2ex}{$\scriptstyle\text{ZEUS}$}}\xspace}

\newcommand{\slashfrac}[2]{%
  \raisebox{0.5ex}{\ensuremath #1}\kern-0.12em/\kern-0.08em
  \raisebox{-.8ex}{\ensuremath #2}}

\newcommand{\sqr}[3]{%
    {\vcenter{\hrule height.#3ex\hbox{\vrule width.#2ex height#1ex
     \kern#1ex\vrule width.#3ex}\hrule height.#2ex}}}

\catcode`\@=11 
\newcommand{\parenbar}{\mathpalette\p@renb@r}
\def\p@renb@r#1#2{\vbox{%
  \ifx#1\scriptscriptstyle \dimen@.7em\dimen@ii.2em\else
  \ifx#1\scriptstyle \dimen@.8em\dimen@ii.25em\else
  \dimen@1em\dimen@ii.4em\fi\fi \offinterlineskip
  \ialign{\hfill##\hfill\cr
    \vbox{\hrule width\dimen@ii}\cr
    \noalign{\vskip-.3ex}%
    \hbox to\dimen@{$\mathchar300\hfil\mathchar301$}\cr
    \noalign{\vskip-.3ex}%
    $#1#2$\cr}}}
\catcode`\@=12 

\ifzeusunit
  
\else
  
\fi

\newcommand{\MSbar}{\hbox{$\overline{\rm MS}$}\xspace}

\newcommand{\IP}{{\rm I$\kern-0.01667em$P}\xspace}

\mathchardef\qsm=63
\mathchardef\pls=43
\mathchardef\mns=512
\mathchardef\plm=518
\mathchardef\eql=61
\mathchardef\smallleft=300
\mathchardef\smallright=301
\mathchardef\les=316
\mathchardef\gre=318
\mathchardef\leq=532
\mathchardef\grq=533

\catcode`\@=11 
\newcounter{pict@width}
\newcounter{pict@height}
\newlength{\pict@scale}
\newcommand{\psfigadd}[4]{%
\setcounter{pict@width}{1*\ratio{#2+\pict@scale/2}{\pict@scale}}
\setcounter{pict@height}{1*\ratio{#3+\pict@scale/2}{\pict@scale}}
\setlength{\unitlength}{\pict@scale}
\hbox to #2{\hspace{-\fill}\begin{picture}(\thepict@width,\thepict@height)
\put(0,0){\psfig{figure=#1,width=#2,height=#3,clip=}}
\SetScale{0.283466457}
\SetWidth{1.763889}
{#4}
\end{picture}}
}
\newcounter{pict@widthfst}
\newcounter{pict@widthscd}
\newcounter{pict@widthtot}
\newcommand{\psfigaddtwo}[7]{%
\setcounter{pict@widthfst}{1*\ratio{#2+\pict@scale/2}{\pict@scale}}
\setcounter{pict@widthscd}{1*\ratio{#2+#4+\pict@scale/2}{\pict@scale}}
\setcounter{pict@widthtot}{1*\ratio{#2+#4+#6+\pict@scale/2}{\pict@scale}}
\setcounter{pict@height}{1*\ratio{#3+\pict@scale/2}{\pict@scale}}
\setlength{\unitlength}{\pict@scale}
\hbox{\hspace{-\fill}\begin{picture}(\thepict@widthtot,\thepict@height)
\put(0,0){\psfig{figure=#1,width=#2,height=#3,clip=}}
\put(\thepict@widthscd,0){\psfig{figure=#5,width=#6,height=#3,clip=}}
\SetScale{0.283466457}
\SetWidth{1.763889}
{#7}
\end{picture}}
}
\newcommand{\psfigror}[4]{%
\setcounter{pict@width}{1*\ratio{#2+\pict@scale/2}{\pict@scale}}
\setcounter{pict@height}{1*\ratio{#3+\pict@scale/2}{\pict@scale}}
\setlength{\unitlength}{\pict@scale}
\hbox{\begin{picture}(\thepict@width,\thepict@height)
\put(0,\thepict@height){\psfig{figure=#1,width=#3,height=#2,clip=,angle=270}}
\SetScale{0.283466457}
\SetWidth{1.763889}
{#4}
\end{picture}}
}
\newcommand{\psfigrol}[4]{%
\setcounter{pict@width}{1*\ratio{#2+\pict@scale/2}{\pict@scale}}
\setcounter{pict@height}{1*\ratio{#3+\pict@scale/2}{\pict@scale}}
\setlength{\unitlength}{\pict@scale}
\hbox{\begin{picture}(\thepict@width,\thepict@height)
\put(0,0){\psfig{figure=#1,width=#3,height=#2,clip=,angle=90}}
\SetScale{0.283466457}
\SetWidth{1.763889}
{#4}
\end{picture}}
}
\catcode`\@=12 
\newlength\listtextwidth

\catcode`\@=11 
\newlength{\@tabfninsert}
\newlength{\@tabfnwidth}
\newcommand{\tabfootnote}[2]{%
  \setlength{\@tabfninsert}{0.8em}
  \setlength{\@tabfnwidth}{\textwidth}
  \addtolength{\@tabfnwidth}{-\@tabfninsert}
  \addtolength{\@tabfnwidth}{-0.4em}
  \noindent\makebox[\@tabfninsert][r]{\footnotesize$^{#1}$\hfil}\hfill%
  \parbox[t]{\@tabfnwidth}{\footnotesize #2\hfill}}
\catcode`\@=12 

%% file: def.tex
\newcommand*{\sigredb}{\ensuremath{\sigma_{\text{r}}^{b\bar{b}}}\xspace}
\newcommand*{\sigredc}{\ensuremath{\sigma_{\text{r}}^{c\bar{c}}}\xspace}
\newcommand*{\sigredq}{\ensuremath{\sigma_{\text{r}}^{q\bar{q}}}\xspace}
\newcommand*{\Ftwob}{\ensuremath{F_{2}^{b\bar{b}}}\xspace}
\newcommand*{\Ftwoc}{\ensuremath{F_{2}^{c\bar{c}}}\xspace}
\newcommand*{\Ftwoq}{\ensuremath{F_{2}^{q\bar{q}}}\xspace}
\newcommand*{\Flq}{\ensuremath{F_{L}^{q\bar{q}}}\xspace}
\newcommand*{\Qsq}{\ensuremath{Q^{2}}\xspace}
\newcommand*{\yJB}{\ensuremath{y_{\text{JB}}}\xspace}
\newcommand*{\pT}{\ensuremath{p_{T}}\xspace}
\newcommand*{\pTjet}{\ensuremath{p_{T}^{\text{jet}}}\xspace}
\newcommand*{\ETjet}{\ensuremath{E_{T}^{\text{jet}}}\xspace}
\newcommand*{\etajet}{\ensuremath{\eta^{\text{jet}}}\xspace}
\newcommand*{\mvtx}{\ensuremath{m_{\text{vtx}}}\xspace}

\newcommand*{\alphaS}{\ensuremath{\alpha_{s}}\xspace}
\newcommand*{\alphaem}{\ensuremath{\alpha_{\text{em}}}\xspace}
\newcommand*{\dif}{\text{d}}

\newcommand*{\diffQsqb}{\ensuremath{\dif\sigma_{b}^{\text{jet}}/\dif\Qsq}\xspace}
\newcommand*{\diffxb}{\ensuremath{\dif\sigma_{b}^{\text{jet}}/\dif x}\xspace}
\newcommand*{\diffEtb}{\ensuremath{\dif\sigma_{b}^{\text{jet}}/\dif\ETjet}\xspace}
\newcommand*{\diffetab}{\ensuremath{\dif\sigma_{b}^{\text{jet}}/\dif\etajet}\xspace}
\newcommand*{\diffQsqc}{\ensuremath{\dif\sigma_{c}^{\text{jet}}/\dif\Qsq}\xspace}
\newcommand*{\diffxc}{\ensuremath{\dif\sigma_{c}^{\text{jet}}/\dif x}\xspace}
\newcommand*{\diffEtc}{\ensuremath{\dif\sigma_{c}^{\text{jet}}/\dif\ETjet}\xspace}
\newcommand*{\diffetac}{\ensuremath{\dif\sigma_{c}^{\text{jet}}/\dif\etajet}\xspace}

\newcommand*{\ddiffQsqxb}{\ensuremath{\dif^{2}\sigma_{b}^{\text{jet}}/\dif x\,\dif\Qsq}\xspace}
\newcommand*{\ddiffQsqxc}{\ensuremath{\dif^{2}\sigma_{c}^{\text{jet}}/\dif x\,\dif\Qsq}\xspace}

\newcommand*{\cqrad}{\ensuremath{C_{q}^{\text{rad}}}\xspace}
\newcommand*{\chad}{\ensuremath{C^{\text{had}}}\xspace}
\newcommand*{\deltahad}{\ensuremath{\Delta^{\text{had}}}\xspace}
\newcommand*{\crad}{\ensuremath{C^{\text{rad}}}\xspace}

\newcommand*{\ndf}{\ensuremath{\text{ndf}}\xspace}
\newcommand*{\stat}{\ensuremath{\text{stat.}}}
\newcommand*{\syst}{\ensuremath{\text{syst.}}}

\newcommand*{\extr}{\ensuremath{\text{extr.}}}
\renewcommand*{\MSbar}{\ensuremath{\overline{\text{MS}}}\xspace}
\newcommand*{\PYTHIA}{\textsc{Pythia}\xspace}
\newcommand*{\RAPGAP}{\textsc{Rapgap}\xspace}
\newcommand*{\DJANGOH}{\textsc{Djangoh}\xspace}
\newcommand*{\ARIADNE}{\textsc{Ariadne}\xspace}
\newcommand*{\HERACLES}{\textsc{Heracles}\xspace}
\newcommand*{\JETSET}{\textsc{Jetset}\xspace}

\newcommand*{\HVQDIS}{\textsc{Hvqdis}\xspace}
\newcommand*{\HERAFITTER}{HERAFitter\xspace}

\newcommand*{\numpmerr}[3]{\ensuremath{^{\num[round-precision=#3]{#1}}_{\num[round-precision=#3]{#2}}}}

\newcommand*{\numpmerrp}[3]{\ensuremath{^{\num[round-mode=places, round-precision=#3]{#1}}_{\num[round-mode=places, round-precision=#3]{#2}}}}

%% file: zeusF2bc-cit_nomcite.tex
\def\citeCTD{{\cite{%
nim:a279:290,npps:b32:181,nim:a338:254%
}}\xspace}
\def\citeMVD{{\cite{%
nim:a581:656%
}}\xspace}
\def\citeSTT{{\cite{%
nim:a535:191%
}}\xspace}
\def\citeCAL{{\cite{%
nim:a309:77,nim:a309:101,nim:a321:356,nim:a336:23%
}}\xspace}
\def\citeHES{{\cite{%
nim:a277:176%
}}\xspace}
\def\citeSixmT{{\cite{%
thesis:gosau:2007%
}}\xspace}
\def\citeBAC{{\cite{%
nim:a300:480%
}}\xspace}
\def\citePCAL{{\cite{%
desy-92-066,zfp:c63:391,acpp:b32:2025%
}}\xspace}
\def\citeSPECTRO{{\cite{%
nim:a565:572%
}}\xspace}

%% file: auth182a_out.tex
%
%
%
%

%
%
\def\3{\ss}
\pagenumbering{Roman}
                                                   %
\begin{center}
{                      \Large  The ZEUS Collaboration              }
\end{center}

{\small\raggedright


H.~Abramowicz$^{27, v}$, 
I.~Abt$^{21}$, 
L.~Adamczyk$^{8}$, 
M.~Adamus$^{34}$, 
R.~Aggarwal$^{4, b}$, 
S.~Antonelli$^{2}$, 
O.~Arslan$^{3}$, 
V.~Aushev$^{16, 17, p}$, 
Y.~Aushev$^{17, p, q}$, 
O.~Bachynska$^{10}$
% ...

\clearpage


{\setlength{\parskip}{0.4em}
\makebox[3ex]{$^{1}$}
\begin{minipage}[t]{14cm}
{\it INFN Bologna, Bologna, Italy}~$^{A}$
\end{minipage}

\makebox[3ex]{$^{2}$}
\begin{minipage}[t]{14cm}
{\it University and INFN Bologna, Bologna, Italy}~$^{A}$
\end{minipage}

\makebox[3ex]{$^{3}$}
\begin{minipage}[t]{14cm}
{\it Physikalisches Institut der Universit\"at Bonn,
Bonn, Germany}~$^{B}$
\end{minipage}

\makebox[3ex]{$^{4}$}
\begin{minipage}[t]{14cm}
{\it Panjab University, Department of Physics, Chandigarh, India}
\end{minipage}

\makebox[3ex]{$^{5}$}
\begin{minipage}[t]{14cm}
{\it Calabria University,
Physics Department and INFN, Cosenza, Italy}~$^{A}$
\end{minipage}

\makebox[3ex]{$^{6}$}
\begin{minipage}[t]{14cm}
{\it National Centre for Particle Physics, Universiti Malaya, 50603 Kuala Lumpur, Malaysia}~$^{C}$
\end{minipage}

\makebox[3ex]{$^{7}$}
\begin{minipage}[t]{14cm}
{\it The Henryk Niewodniczanski Institute of Nuclear Physics, Polish Academy of \\
Sciences, Krakow, Poland}~$^{D}$
\end{minipage}

\makebox[3ex]{$^{8}$}
\begin{minipage}[t]{14cm}
{\it AGH-University of Science and Technology, Faculty of Physics and Applied\\
Computer Science, Krakow, Poland}~$^{D}$
\end{minipage}

\makebox[3ex]{$^{9}$}
\begin{minipage}[t]{14cm}
{\it Department of Physics, Jagellonian University, Cracow, Poland}
\end{minipage}

\makebox[3ex]{$^{10}$}
\begin{minipage}[t]{14cm}
{\it Deutsches Elektronen-Synchrotron DESY, Hamburg, Germany}
\end{minipage}

\makebox[3ex]{$^{11}$}
\begin{minipage}[t]{14cm}
{\it Deutsches Elektronen-Synchrotron DESY, Zeuthen, Germany}
\end{minipage}

\makebox[3ex]{$^{12}$}
\begin{minipage}[t]{14cm}
{\it School of Physics and Astronomy, University of Glasgow,\\
Glasgow, United Kingdom}~$^{E}$
\end{minipage}

\makebox[3ex]{$^{13}$}
\begin{minipage}[t]{14cm}
{\it Hamburg University, Institute of Experimental Physics, Hamburg,
Germany}~$^{F}$
\end{minipage}

\makebox[3ex]{$^{14}$}
\begin{minipage}[t]{14cm}
{\it Institute of Particle and Nuclear Studies, KEK,
Tsukuba, Japan}~$^{G}$
\end{minipage}

\makebox[3ex]{$^{15}$}
\begin{minipage}[t]{14cm}
{\it Institute of Physics and Technology of Ministry of Education and Science of\\
Kazakhstan, Almaty, Kazakhstan}
\end{minipage}

\makebox[3ex]{$^{16}$}
\begin{minipage}[t]{14cm}
{\it Institute for Nuclear Research, National Academy of Sciences, Kyiv, Ukraine}
\end{minipage}

\makebox[3ex]{$^{17}$}
\begin{minipage}[t]{14cm}
{\it Department of Nuclear Physics, National Taras Shevchenko University of Kyiv, Kyiv, Ukraine}
\end{minipage}

\makebox[3ex]{$^{18}$}
\begin{minipage}[t]{14cm}
{\it Department of Physics, McGill University,
Montr\'eal, Qu\'ebec, Canada H3A 2T8}~$^{H}$
\end{minipage}

\makebox[3ex]{$^{19}$}
\begin{minipage}[t]{14cm}
{\it Meiji Gakuin University, Faculty of General Education,
Yokohama, Japan}~$^{G}$
\end{minipage}

\makebox[3ex]{$^{20}$}
\begin{minipage}[t]{14cm}
{\it Lomonosov Moscow State University, Skobeltsyn Institute of Nuclear Physics,\\
Moscow, Russia}~$^{I}$
\end{minipage}

\makebox[3ex]{$^{21}$}
\begin{minipage}[t]{14cm}
{\it Max-Planck-Institut f\"ur Physik, M\"unchen, Germany}
\end{minipage}

\makebox[3ex]{$^{22}$}
\begin{minipage}[t]{14cm}
{\it Department of Physics, University of Oxford,
Oxford, United Kingdom}~$^{E}$
\end{minipage}

\makebox[3ex]{$^{23}$}
\begin{minipage}[t]{14cm}
{\it INFN Padova, Padova, Italy}~$^{A}$
\end{minipage}

\makebox[3ex]{$^{24}$}
\begin{minipage}[t]{14cm}
{\it Dipartimento di Fisica dell' Universit\`a and INFN,
Padova, Italy}~$^{A}$
\end{minipage}

\makebox[3ex]{$^{25}$}
\begin{minipage}[t]{14cm}
{\it Polytechnic University, Tokyo, Japan}~$^{G}$
\end{minipage}

\makebox[3ex]{$^{26}$}
\begin{minipage}[t]{14cm}
{\it Dipartimento di Fisica, Universit\`a `La Sapienza' and INFN,
Rome, Italy}~$^{A}$
\end{minipage}

\makebox[3ex]{$^{27}$}
\begin{minipage}[t]{14cm}
{\it Raymond and Beverly Sackler Faculty of Exact Sciences, School of Physics, \\
Tel Aviv University, Tel Aviv, Israel}~$^{J}$
\end{minipage}

\makebox[3ex]{$^{28}$}
\begin{minipage}[t]{14cm}
{\it Department of Physics, Tokyo Institute of Technology,
Tokyo, Japan}~$^{G}$
\end{minipage}

\makebox[3ex]{$^{29}$}
\begin{minipage}[t]{14cm}
{\it Universit\`a di Torino and INFN, Torino, Italy}~$^{A}$
\end{minipage}

\makebox[3ex]{$^{30}$}
\begin{minipage}[t]{14cm}
{\it Universit\`a del Piemonte Orientale, Novara, and INFN, Torino,
Italy}~$^{A}$
\end{minipage}

\makebox[3ex]{$^{31}$}
\begin{minipage}[t]{14cm}
{\it Department of Physics, University of Toronto, Toronto, Ontario,\\
Canada M5S 1A7}~$^{H}$
\end{minipage}

\makebox[3ex]{$^{32}$}
\begin{minipage}[t]{14cm}
{\it Physics and Astronomy Department, University College London,
London, United Kingdom}~$^{E}$
\end{minipage}

\makebox[3ex]{$^{33}$}
\begin{minipage}[t]{14cm}
{\it Faculty of Physics, University of Warsaw, Warsaw, Poland}
\end{minipage}

\makebox[3ex]{$^{34}$}
\begin{minipage}[t]{14cm}
{\it National Centre for Nuclear Research, Warsaw, Poland}
\end{minipage}

\makebox[3ex]{$^{35}$}
\begin{minipage}[t]{14cm}
{\it Department of Particle Physics and Astrophysics, Weizmann Institute, Rehovot,\\
 Israel}
\end{minipage}

\makebox[3ex]{$^{36}$}
\begin{minipage}[t]{14cm}
{\it Department of Physics, York University, Ontario, Canada M3J 1P3}~$^{H}$
\end{minipage}
}
\vspace{3em}


{\setlength{\parskip}{0.4em}\raggedright
\makebox[3ex]{$^{ A}$}
\begin{minipage}[t]{14cm}
 supported by the Italian National Institute for Nuclear Physics (INFN)
\end{minipage}

\makebox[3ex]{$^{ B}$}
\begin{minipage}[t]{14cm}
 supported by the German Federal Ministry for Education and Research (BMBF), under
 contract No. 05 H09PDF
\end{minipage}

\makebox[3ex]{$^{ C}$}
\begin{minipage}[t]{14cm}
 supported by HIR grant UM.C/625/1/HIR/149 and UMRG grants RU006-2013, RP012A-13AFR and RP012B-13AFR from
 Universiti Malaya, and ERGS grant ER004-2012A from the Ministry of Education, Malaysia
\end{minipage}

\makebox[3ex]{$^{ D}$}
\begin{minipage}[t]{14cm}
 supported by the National Science Centre under contract No. DEC-2012/06/M/ST2/00428
\end{minipage}

\makebox[3ex]{$^{ E}$}
\begin{minipage}[t]{14cm}
 supported by the Science and Technology Facilities Council, UK
\end{minipage}

\makebox[3ex]{$^{ F}$}
\begin{minipage}[t]{14cm}
 supported by the German Federal Ministry for Education and Research (BMBF), under
 contract No. 05h09GUF, and the SFB 676 of the Deutsche Forschungsgemeinschaft (DFG)
\end{minipage}

\makebox[3ex]{$^{ G}$}
\begin{minipage}[t]{14cm}
 supported by the Japanese Ministry of Education, Culture, Sports, Science and Technology
 (MEXT) and its grants for Scientific Research
\end{minipage}

\makebox[3ex]{$^{ H}$}
\begin{minipage}[t]{14cm}
 supported by the Natural Sciences and Engineering Research Council of Canada (NSERC)
\end{minipage}

\makebox[3ex]{$^{ I}$}
\begin{minipage}[t]{14cm}
 supported by RF Presidential grant N 3042.2014.2 for the Leading Scientific Schools and by
 the Russian Ministry of Education and Science through its grant for Scientific Research on
 High Energy Physics
\end{minipage}

\makebox[3ex]{$^{ J}$}
\begin{minipage}[t]{14cm}
 supported by the Israel Science Foundation
\end{minipage}
}
\pagebreak[4]


{\setlength{\parskip}{0.4em}
\makebox[3ex]{$^{ a}$}
\begin{minipage}[t]{14cm}
now at University of the Punjab, Lahore, Pakistan
\end{minipage}

\makebox[3ex]{$^{ b}$}
\begin{minipage}[t]{14cm}
also funded by Max Planck Institute for Physics, Munich, Germany
\end{minipage}

\makebox[3ex]{$^{ c}$}
\begin{minipage}[t]{14cm}
also funded by Max Planck Institute for Physics, Munich, Germany, now at Sri Guru Granth Sahib World University, Fatehgarh Sahib
\end{minipage}

\makebox[3ex]{$^{ d}$}
\begin{minipage}[t]{14cm}
also at Agensi Nuklear Malaysia, 43000 Kajang, Bangi, Malaysia
\end{minipage}

\makebox[3ex]{$^{ e}$}
\begin{minipage}[t]{14cm}
also at Cracow University of Technology, Faculty of Physics, Mathematics and Applied Computer Science, Poland
\end{minipage}

\makebox[3ex]{$^{ f}$}
\begin{minipage}[t]{14cm}
partially supported by the Polish National Science Centre projects DEC-2011/01/B/ST2/03643 and DEC-2011/03/B/ST2/00220
\end{minipage}

\makebox[3ex]{$^{ g}$}
\begin{minipage}[t]{14cm}
now at Rockefeller University, New York, NY 10065, USA
\end{minipage}

\makebox[3ex]{$^{ h}$}
\begin{minipage}[t]{14cm}
now at University of Liverpool, United Kingdom
\end{minipage}

\makebox[3ex]{$^{ i}$}
\begin{minipage}[t]{14cm}
also affiliated with University College London, UK
\end{minipage}

\makebox[3ex]{$^{ j}$}
\begin{minipage}[t]{14cm}
supported by the Alexander von Humboldt Foundation
\end{minipage}

\makebox[3ex]{$^{ k}$}
\begin{minipage}[t]{14cm}
now at CERN, Geneva, Switzerland
\end{minipage}

\makebox[3ex]{$^{ l}$}
\begin{minipage}[t]{14cm}
also at Institute of Theoretical and Experimental Physics, Moscow, Russia
\end{minipage}

\makebox[3ex]{$^{ m}$}
\begin{minipage}[t]{14cm}
Alexander von Humboldt Professor; also at DESY and University of Oxford
\end{minipage}

\makebox[3ex]{$^{ n}$}
\begin{minipage}[t]{14cm}
also at University of Tokyo, Japan
\end{minipage}

\makebox[3ex]{$^{ o}$}
\begin{minipage}[t]{14cm}
now at Kobe University, Japan
\end{minipage}

\makebox[3ex]{$^{ p}$}
\begin{minipage}[t]{14cm}
supported by DESY, Germany
\end{minipage}

\makebox[3ex]{$^{ q}$}
\begin{minipage}[t]{14cm}
member of National Technical University of Ukraine, Kyiv Polytechnic Institute, Kyiv, Ukraine
\end{minipage}

\makebox[3ex]{$^{ r}$}
\begin{minipage}[t]{14cm}
now at DESY ATLAS group
\end{minipage}

\makebox[3ex]{$^{ s}$}
\begin{minipage}[t]{14cm}
member of National University of Kyiv - Mohyla Academy, Kyiv, Ukraine
\end{minipage}

\makebox[3ex]{$^{ t}$}
\begin{minipage}[t]{14cm}
now at BNL, USA
\end{minipage}

\makebox[3ex]{$^{ u}$}
\begin{minipage}[t]{14cm}
now at LNF, Frascati, Italy
\end{minipage}

\makebox[3ex]{$^{ v}$}
\begin{minipage}[t]{14cm}
also at Max Planck Institute for Physics, Munich, Germany, External Scientific Member
\end{minipage}

\makebox[3ex]{$^{ w}$}
\begin{minipage}[t]{14cm}
now at Tokyo Metropolitan University, Japan
\end{minipage}

\makebox[3ex]{$^{ x}$}
\begin{minipage}[t]{14cm}
also supported by DESY
\end{minipage}

\makebox[3ex]{$^{ y}$}
\begin{minipage}[t]{14cm}
also at \L\'{o}d\'{z} University, Poland
\end{minipage}

\makebox[3ex]{$^{ z}$}
\begin{minipage}[t]{14cm}
member of \L\'{o}d\'{z} University, Poland
\end{minipage}

\makebox[3ex]{$^{aa}$}
\begin{minipage}[t]{14cm}
now at Polish Air Force Academy in Deblin
\end{minipage}
}
}

%% file: zeusF2bc-txt.tex
\section{Introduction}
\label{sec:int}

The measurement of beauty and charm production in $ep$ collisions at
HERA is an important testing ground for perturbative Quantum
Chromodynamics (pQCD), since the heavy-quark masses provide a hard
scale that allows perturbative calculations to be made.
At leading
order, the dominant process for heavy-quark production at HERA is
boson-gluon fusion (BGF). 
In this process, a virtual photon emitted by
the incoming electron interacts with a gluon from the proton
forming a heavy quark--antiquark pair.
When the negative squared four-momentum of the virtual photon, \Qsq, is
large compared to the proton mass, the interaction is referred to as
deep inelastic scattering (DIS).  For heavy-quark transverse momenta
comparable to the quark mass, next-to-leading-order (NLO) QCD
calculations based on the dynamical generation of the massive
quarks~\cite{np:b452:109, pl:b353:535, pl:b359:423, np:b392:162, np:b392:229} are
expected to provide reliable predictions.

Beauty and charm production in DIS has been measured using several
methods by the
H1~\cite{zfp:c72:593, np:b545:21, pl:b528:199, epj:c38:447, epj:c40:349, epj:c41:453, epj:c45:23, epj:c51:271, Aaron:2009jy, Aaron:2009ut, Aaron:2010ib, Aaron:2011gp, Aaron:2011gp2}
and
ZEUS~\cite{pl:b407:402, epj:c12:35, pr:d69:012004, pl:b599:173, epj:c50:1434, pl:b649:111, Chekanov:2007ch, Chekanov:2008zz, epj:c63:2009:2:171-188, Chekanov:2009kj, Abramowicz:2010zq, Abramowicz:2010aa, Abramowicz:2011rs, jhep05.2013.023, jhep05.2013.097}
collaborations.
All but the two most recent measurements of charm production~\cite{jhep05.2013.023, jhep05.2013.097} and older data~\cite{pl:b649:111} have been
combined~\cite{epj.c73.2311}.
Predictions from NLO QCD describe all results reasonably well.

Inclusive jet cross sections in beauty and charm events are used in the analysis presented here to extract
the heavy-quark contribution to the proton structure function $F_{2}$
with high precision, and to measure the $b$-quark mass.
For this purpose, the long lifetimes of the
weakly decaying $b$ and $c$ hadrons, 
which make the reconstruction of their decay vertices possible,
as well as their large masses were exploited. 
Two
discriminating variables, the significance of the reconstructed decay
length and the invariant mass of the charged tracks associated with
the decay vertex (secondary vertex), were used.
This inclusive tagging method leads to a substantial increase in statistics with
respect to previous ZEUS measurements.

Differential cross sections as a function of \Qsq,
the Bjorken scaling variable, $x$, jet transverse energy, \ETjet, and
pseudorapidity, \etajet, were measured. They are compared to a
leading-order (LO) plus parton-shower (PS) Monte Carlo prediction and
to NLO QCD calculations. The beauty and charm contributions to the
proton structure function $F_{2}$, denoted as \Ftwob and \Ftwoc,
respectively, as well as beauty and charm reduced cross sections
(\sigredb and \sigredc, respectively) were extracted from the
double-differential cross section as a function of \Qsq and $x$.  The
results are compared to previous measurements and to predictions from
perturbative QCD.

The running \MSbar beauty-quark mass, $m_{b}$ at the scale $m_{b}$, denoted $m_b(m_b)$, 
is measured using \sigredb,
following a procedure similar to that used for a recent extraction
of the charm-quark mass~\cite{epj.c73.2311}.
This represents the
first measurement of the $b$-quark mass using HERA or any other hadron collider
data.

\section{Experimental set-up}
\label{sec:exp}

This analysis was performed with data taken with the ZEUS detector
from 2004 to 2007, when HERA collided electrons\footnote{%
  In this paper ``electron'' is used to denote both electron and positron.}
with energy $E_e=\SI{27.5}{\GeV}$ with protons of energy
\SI{920}{\GeV}, corresponding to a centre-of-mass energy $\sqrt{s} =
\SI{318}{\GeV}$. This data-taking period is denoted as HERA~II.  The
corresponding integrated luminosity is \SI{354 \pm 7}{\per\pb}.

\Zdetdesc

\Zctdmvddesc{\ZcoosysfnBEeta}

\Zcaldesc

\Zlumidesc

\section{Monte Carlo simulations}
\label{sec:montecarlo}

To evaluate the detector acceptance and to provide predictions of
the signal and background distributions, Monte Carlo (MC) samples of
beauty, charm and light-flavour events were generated, corresponding
to eighteen, three and one times the integrated luminosity of the
data, respectively.  The \RAPGAP~3.00 MC
program~\cite{cpc:86:147} in the massive mode ($m_{b} = \SI{4.75}{\GeV},
m_{c} = \SI{1.5}{\GeV}$) was used to generate the beauty and charm samples,
where the CTEQ5L~\cite{epj:c12:375} parameterisation for the proton
parton density functions (PDFs) was used. In \RAPGAP, LO matrix
elements are combined with higher-order QCD radiation simulated in the
leading-logarithmic approximation.  Higher-order QED effects are
included through \HERACLES~4.6~\cite{cpc:69:155}.
Light-flavour MC
events were extracted from an inclusive DIS sample generated with
\DJANGOH~1.6~\cite{proc:hera:1991:1419} interfaced to \ARIADNE{}
4.12~\cite{cpc:71:15}.
The CTEQ5D~\cite{epj:c12:375} PDFs were used and quarks were taken to
be massless.

Fragmentation and particle decays were simulated using the
\JETSET{}/\PYTHIA{} model~\cite{cpc:82:74, cpc:135:238}. 
The Bowler parameterisation~\cite{zfp:c11:169} of the fragmentation
function, as implemented in \PYTHIA~\cite{hep-ph-0108264} (with $r_{Q} = 1$), was used for the heavy-flavour samples.
The generated
events were passed through a full simulation of the ZEUS detector
based on \textsc{Geant 3.21}~\cite{tech:cern-dd-ee-84-1}.  The final
MC events were then subjected to the same trigger requirements and
processed by the same reconstruction program as the data.

For the acceptance determination, the \ETjet and \etajet distributions
in the charm MC, as well as the \Qsq distributions in both the beauty and charm
MCs, were reweighted in order to give a good description of the data.
The charm branching fractions and fragmentation fractions were adjusted to the world-average
values~\cite{Nakamura:2010zzi,arXiv:1112.3757,thesis:viazlo:2012}.

\section{Theoretical predictions and uncertainties}
\label{sec:theory}

Next-to-leading-order QCD predictions for differential cross sections
were obtained from the \HVQDIS program~\cite{pr:d57:2806}. 
The calculations were used to extrapolate the visible cross
sections to extract \Ftwob, \Ftwoc, \sigredb and \sigredc (see
Section~\ref{sec:F2q}).
The calculations are based on the
fixed-flavour-number scheme (FFNS) in which only light flavours are
present in the proton and heavy quarks are produced in the
interaction~\cite{np:b374:36}.
Therefore, the 3-flavour (4-flavour) FFNS variant of the ZEUS-S NLO QCD
fit~\cite{pr:d67:012007} was used for the proton PDF for the predictions of the 
charm (beauty) cross sections.
As in the PDF fit, the value of
$\alphaS(M_{Z})$ was set to 0.118 and the heavy-quark masses (pole
masses) were set to $m_{b} = \SI{4.75}{\GeV}$ and $m_{c} = \SI{1.5}{\GeV}$.  The
renormalisation and factorisation scales, $\mu_{R}$ and $\mu_{F}$,
were chosen to be equal and set to
$\mu_{R}=\mu_{F}=\sqrt{\Qsq+4m_{b(c)}^{2}}$.

The systematic uncertainty on the theoretical predictions with the
ZEUS-S PDFs were estimated by varying the quark masses and the
renormalisation and factorisation scales. Quark masses of $m_{b} = 4.5$
and \SI{5.0}{\GeV}, $m_{c} = 1.3$ and \SI{1.7}{\GeV} were used.  The
scales $\mu_{R}$, $\mu_{F}$ were varied independently by a factor of
two up and down.  Additionally, the experimental uncertainties of the
data used in the PDF fit were propagated to the predicted cross
sections.  The total uncertainties were obtained by adding positive
and negative changes to the cross sections in quadrature.  This
results in total uncertainties of \SIrange{10}{20}{\%} for beauty and
\SIrange{10}{50}{\%} for charm.

Predictions were also obtained using the 3- and 4-flavour variants of the
ABKM NLO PDFs~\cite{PhysRevD.81.014032} for the proton.
The pole masses of heavy quarks were set to $m_{b} = \SI{4.5}{\GeV}$ and
$m_{c} = \SI{1.5}{\GeV}$, both in the PDF fit and in the \HVQDIS calculation.
The values of $\alphaS(\mu_{R})$ were provided by
LHAPDF~\cite{Whalley:2005nh,LHAPDF} to
ensure that the same function was used as in the PDF fit. The
renormalisation and factorisation scales were both set to
$\mu_{R}=\mu_{F}=\sqrt{\Qsq+4m_{b(c)}^{2}}$.

The NLO QCD predictions are given for parton-level jets. These  
were reconstructed using the $k_{T}$ clustering
algorithm~\cite{pr:d48:3160} with a radius parameter $R = 1.0$ in the longitudinally invariant
mode~\cite{np:b406:187}. The $E$-recombination scheme, which produces
massive jets whose four-momenta are the sum of the four-momenta of the
clustered objects, was used.
The parton-level cross sections
were corrected for jet hadronisation effects to allow a
direct comparison with the measured hadron-level cross sections:
\begin{equation}
\sigma^{\text{had, NLO}} = \chad \sigma^{\text{parton,NLO}}\,,
\label{eq:chad}
\end{equation}
where the correction factors, $\chad = 1 + \deltahad$, were derived from the \RAPGAP MC
simulation. The factors \chad are defined as the ratio of the
hadron-level jet to the parton-level jet cross sections, and the parton
level is defined as the result of the parton-showering stage of the
simulation.

Since \chad were derived from an LO plus parton shower MC,
but are applied to an NLO prediction,
the uncertainty on \chad cannot be estimated in a straightforward 
way.
Within the framework of parton showering, MC subsets with different numbers of 
radiated partons were investigated using \RAPGAP and \PYTHIA samples.
These studies indicated that different approaches yield variations of
\deltahad of typically a factor of two.
Since it is not clear if the variations can be interpreted as uncertainties
on \chad, no such uncertainties were included in the cross-section ($F_{2}$) predictions.
However, for the extraction of the $b$-quark mass,
such a theoretical uncertainty needs to be included.

\section{Data selection}
\label{sec:data}

Events containing a scattered electron were selected online by means
of a three-level trigger
system~\cite{zeus:1993:bluebook, proc:chep:1992:222}.
The trigger~\cite{thesis:roloff:2011} did not require the presence of a secondary vertex nor of a jet.

Offline, the scattered electron was reconstructed using an electron finder based on a neural network~\cite{nim:a365:508}. The hadronic system
was reconstructed from energy-flow objects
(EFOs)~\cite{epj:c1:81, thesis:briskin:1998} which combine the
information from calorimetry and tracking, corrected for energy
loss in the detector material.
The kinematic variables used in the cross-section measurements, \Qsq
and $x$,  were reconstructed using the double-angle method~\cite{proc:hera:1991:23}.

The following cuts were applied to select a clean DIS sample:
\begin{itemize}
\item the reconstructed
  scattered electron~\cite{nim:a365:508, nim:a391:360} was required to
  have an energy $E'_{e} > \SI{10}{\GeV}$;
\item the impact position of the scattered electron on the face of the
  RCAL had to be outside the region \SI[parse-numbers=false]{26 \times 26}{\cm^{2}} centred on
  $X = Y = 0$;
\item the primary vertex had to be within
  \SI{\pm 30}{\cm} in $Z$ of the nominal interaction point;
\item the photon virtuality, \Qsq, had to be within $5 < \Qsq <
  \SI{1000}{\GeV^{2}}$;
\item $ y_{\text{JB}} > 0.02$, where $y_{\text{JB}}$ is the
  inelasticity reconstructed using the Jacquet-Blondel
  method~\cite{proc:epfacility:1979:391};
\item $ y_{e} < 0.7$, where $y_{e}$ is the inelasticity reconstructed
  using the electron method~\cite{proc:hera:1991:23};
\item $44 < (E-p_{Z}) < \SI{65}{\GeV}$, 
  where $(E-p_{Z}) = \sum_{i} (E_{i} - p_{Z,i})$ and $i$ runs over all final-state particles
  with energy $E_{i}$ and 
  $Z$-component of momentum $p_{{Z,i}}$; 
  this selects fully
  contained neutral-current $ep$ events for which
  $E-p_{Z}=2 E_{e}$.
\end{itemize}

Jets were reconstructed from EFOs using the $k_{T}$ clustering
algorithm~\cite{pr:d48:3160} as was described for parton-level jets 
in Section~\ref{sec:theory}.
Jets containing the identified scattered electron were not considered further.
Events were selected if
they contained at least one jet within the pseudorapidity range $-1.6
< \etajet < 2.2$ and with transverse energy, \ETjet,
of
\begin{equation*}
  \ETjet = \pTjet\frac{E^{\text{jet}}}{p^{\text{jet}}} > \SI[parse-numbers=false]{5\,(4.2)}{\GeV}
\end{equation*}
for beauty (charm), where $E^{\text{jet}}$, $p^{\text{jet}}$ and
\pTjet are the jet energy, momentum and transverse momentum. The cut
on \ETjet was optimised separately for beauty and charm measurements.
For beauty, a cut of $\ETjet > \SI{5}{\GeV}$ ensures a good correlation between momentum and angle for
reconstructed and hadron-level jets~\cite{thesis:kahle:2005}; for charm this cut was
$\SI{4.2}{\GeV}$ to reduce the extrapolation uncertainties for the
\Ftwoc and \sigredc measurements at low \Qsq.

In order to reconstruct potential secondary vertices related to $b$- and
$c$-hadron decays, tracks were selected if:
\begin{itemize}
\item they had a transverse momentum $\pT > \SI{0.5}{\GeV}$;
\item the total number of hits\footnote{Each MVD
  layer provided two coordinate measurements.} on the track in the MVD was $\ge 4$.
\item if the track was inside the CTD acceptance, track recognition in the CTD was required;
 the percentage of the tracks outside the CTD acceptance, and hence reconstructed using MVD hits only, was \SI{2.5}{\percent}.
\end{itemize}
  
Tracks were associated with the closest jet if they fulfilled the
criterion $\Delta R < 1$ with $\Delta R =
\sqrt{(\eta^{\text{trk}}-\eta^{\text{jet}})^{2}+(\phi^{\text{trk}}-\phi^{\text{jet}})^{2}}
$. If two or more of such tracks were associated with the jet, a
candidate vertex was fitted from the selected tracks using a
deterministic annealing
filter~\cite{RosePhysRevLett.65.945, Rose726788, Didierjean2010188}.
This fit provided the vertex position and its error matrix as
well as the invariant mass, \mvtx, of the charged tracks associated
with the reconstructed vertex. The charged-pion mass was assumed for all tracks
when calculating the vertex mass.
Vertices with $\chi^{2}/\ndf < 6$, a
distance from the interaction point within \SI{\pm 1}{\cm} in the
$X$--$Y$ plane, \SI{\pm 30}{\cm} in the $Z$ direction, and
$\num{1} < \mvtx < \SI{6}{\GeV}$ were kept for further
analysis.

The MC gives a good description of the track efficiencies, except for a
small fraction of tracks that are affected by hadronic interactions
in the detector material between the interaction point and the CTD.
Efficiency corrections for
this effect were determined from a study of exclusive $ep\to e\rho^0p$
events~\cite{thesis:libov:2012}, using a special track reconstruction.
The number of the pions
from the $\rho^0$ decay that were reconstructed in the MVD alone and had no
extension in the CTD was measured.  
The resulting track efficiency correction in the MC was applied by randomly rejecting
selected vertex tracks before the vertex fit, with a probability
that depends on the track parameters (around \SI{3}{\%} at $\eta=0$
and $\pT = \SI{1}{\GeV}$).

\section{Extraction of the heavy-flavour cross sections}
\label{sec:bc-extr}

Using the secondary-vertex candidates associated with jets, the decay length,
$d$, was defined as the vector in $X$--$Y$ between the secondary
vertex and the interaction point\footnote{%
  In the $X$--$Y$ plane, the interaction point was defined as the
  centre of the beam ellipse, determined using the average primary
  vertex position for groups of a few thousand events, taking into
  account the difference in angle between the beam direction and the
  $Z$ direction. The $Z$ coordinate was taken as the $Z$ position of
  the primary vertex of the event.}
projected onto the jet axis in
the $X$--$Y$ plane.  The sign of the decay length was assigned using
the axis of the jet to which the vertex was associated; if the decay-length vector
was in the same hemisphere as the jet axis, a
positive sign was assigned to it, otherwise the sign of the decay
length was negative. Negative decay lengths, which originate from
secondary vertices reconstructed on the wrong side of the interaction
point with respect to the direction of the associated jets, are
unphysical and caused by detector resolution effects. A small smearing
correction~\cite{thesis:libov:2012} to the MC decay-length
distribution was applied in order to reproduce the data with negative
values of decay length.

The beauty and charm content in the selected sample was determined
using the shape of the decay-length significance distribution together
with the secondary-vertex mass distribution, \mvtx.  The
decay-length significance, $S$, is defined as $d/\delta{d}$, where
$\delta{d}$ is the uncertainty on $d$. The invariant mass of the
tracks fitted to the secondary vertex provides a distinguishing
variable for jets from $b$ and $c$ quarks, reflecting the different
masses of the $b$ and $c$ hadrons. Figure~\ref{fig:significance} shows
the decay-length significance, $S$, for $\ETjet > \SI{4.2}{\GeV}$
divided into four bins: $1 < \mvtx < \SI{1.4}{\GeV}$, $1.4 <
\mvtx < \SI{2}{\GeV}$, $2 < \mvtx < \SI{6}{\GeV}$ and 
no restriction on \mvtx.  The MC
simulation provides a good description of the data.
The separation into subsamples is described below.

The contents of the negative bins of the significance distribution,
$N(S^{-})$, were subtracted from the contents of the corresponding
positive bins, $N(S^{+})$, yielding a subtracted decay-length
significance distribution.  In this way, the contribution from
light-flavour quarks is minimised.  
An additional advantage of this subtraction is that symmetric systematic 
effects, which might arise from discrepancies between the data and the MC, are 
removed. 
In order to reduce the
contamination of tracks originating from the primary vertex, a cut of
$|S| >4$ was applied.

To extract the contributions from beauty, charm and light flavours in
the data sample, a binned $\chi^{2}$ fit of
the subtracted significance distribution in the region $4 < |S| < 20$ was performed simultaneously
for three mass bins~\cite{thesis:roloff:2011}:
$1 < \mvtx < \SI{1.4}{\GeV}$; $1.4 < \mvtx < \SI{2}{\GeV}$;
$2 < \mvtx < \SI{6}{\GeV}$.
All MC distributions were normalised to the
integrated luminosity of the data before the fit. The overall MC
normalisation was constrained by requiring it to be consistent with
the normalisation of the data in the significance distribution with
$|S| < 20$ and $1 < \mvtx < \SI{6}{\GeV}$. The fit yielded scaling
factors $k_{b}$, $k_{c}$ and $k_{\text{lf}}$ for the beauty, charm and
light-flavour contributions, respectively, to obtain the best
description of the data.
The correlation coefficients were as follows: $\rho_{b,c}=-0.68(-0.67)$, 
$\rho_{b,\text{lf}}=0.58(0.57)$ and $\rho_{c,\text{lf}}=-0.98(-0.98)$ for 
$\ETjet > \SI[parse-numbers=false]{4.2(5.0)}{\GeV}$.
The subtracted and fitted distributions
for $\ETjet > \SI{4.2}{\GeV}$ are shown in 
Fig.~\ref{fig:subtracted_sig}.
A good agreement between data and MC is observed.
The first
two mass bins corresponding to the region $1 < \mvtx < \SI{2}{\GeV}$
are dominated by charm events. In the third mass bin,
beauty events are dominant at high values of significance.
The fit procedure was
repeated for every bin of a given observable to obtain differential
cross sections.  For the beauty cross-section extraction, the fit
procedure was repeated with the higher cut on \ETjet, $\ETjet > \SI{5}{\GeV}$.

Control distributions of \ETjet, \etajet,
$\log_{10}\Qsq$ and
$\log_{10}x$ are shown in Fig.~\ref{fig:beauty_enriched} after beauty
enrichment cuts ($2 < \mvtx < \SI{6}{\GeV}$ and $|S| > 8$) for 
$\ETjet > \SI{5.0}{\GeV}$ and in
Fig.~\ref{fig:charm_enriched} after charm enrichment cuts ($1 <
\mvtx < \SI{2}{\GeV}$ and $|S| > 4$) for $\ETjet > \SI{4.2}{\GeV}$. All 
data distributions are reasonably well
described by the MC.

The differential cross sections for jet production in beauty or charm events,
$q = b,c$, corrected to QED Born level, in a bin
$i$ of a given observable, $Y$, are given by:
\begin{equation}
  \label{eq:xsec4}
  \frac{\dif \sigma_{q}^{\text{jet}}}{\dif Y_{i}} =
  k_{q}(Y_{i}) \frac{N_{q}^{\text{had,MC}}(Y_{i})}{%
    \mathcal{L} \cdot \Delta Y_{i}} \frac{1}{\crad},
\end{equation}
where $\Delta Y_{i}$ is the width of the bin, $k_{q}$ denotes the
scaling factor obtained from the fit,
$N_{q}^{\text{had,MC}}$ is the number of generated jets in beauty or charm events
at the MC hadron level, $\crad$ is the QED radiative correction
and $\mathcal{L}$ is the corresponding integrated luminosity.

Hadron-level jets were obtained by running the
$k_{T}$ clustering algorithm
on all stable final-state particles,
in the same mode as for the data. 
Weakly decaying $b$ and $c$ hadrons were treated as stable particles
and were decayed only after the application of the jet algorithm.

The predictions from the \HVQDIS program are given at the QED Born
level with a running coupling, \alphaem. Hence, a correction of the
measured cross sections for QED radiative effects is necessary in
order to be able to compare them directly to the \HVQDIS predictions.
The corrections were obtained using the \RAPGAP Monte Carlo as
$\crad=\sigma_{\text{rad}}/\sigma_{\text{Born}}$, where
$\sigma_{\text{rad}}$ is the cross section with full QED corrections,
as used in the standard MC samples, and $\sigma_{\text{Born}}$ was
obtained with the QED corrections turned off but keeping \alphaem
running. Both cross sections, $\sigma_{\text{rad}}$ and
$\sigma_{\text{Born}}$, were obtained at the hadron level.

\section{Systematic uncertainties}
\label{sec:syst}

The systematic uncertainties were evaluated by varying the analysis
procedure or by changing the selection cuts
and repeating the extraction of the cross section. 
The following sources of experimental systematic uncertainties were
identified~\cite{thesis:libov:2012, thesis:roloff:2011};
the uncertainties on the integrated cross sections determined for each
source are summarised in Table~\ref{tab:syst} to indicate the sizes of the 
different effects:

\begin{enumerate}[$\delta_1$]
\item DIS selection -- the cuts for DIS event selection were varied in
  both data and MC. 
  The cut on the scattered electron energy was varied
  between $9 < E'_{e} < \SI{11}{\GeV}$ ($\delta_{1}^{E_{e}}$), the cut on the inelasticity was varied
  between $0.01 < \yJB < 0.03$ ($\delta_{1}^{y}$), 
  and the lower cut on $E-p_{Z}$ 
  was changed by $\pm \SI{2}{\GeV}$ ($\delta_{1}^{E-p_{Z}}$);

\item trigger efficiency -- the uncertainty on the trigger efficiency
  was evaluated by comparing events taken with independent triggers;

\item tracking efficiency correction -- the size of the correction was varied 
  by its estimated uncertainty of $\pm \SI{50}{\%}$;

\item decay-length smearing -- the fraction of secondary vertices for
  which the decay length was smeared was varied separately in the 
  core ($\delta_{4}^{\text{core}}$) and the tails ($\delta_{4}^{\text{tail}}$) of the distribution
  such that the agreement between data and MC remained reasonable;

\item signal extraction procedure -- the systematic uncertainty on the
  signal extraction procedure was estimated by changing the lower
  $|S|$ cut from $|S| > 4$ to $|S| > 3$ and $|S| > 5$;

\item jet energy scale -- the calorimetric part of the transverse jet
  energy in the MC was varied by its estimated uncertainty of $\pm \SI{3}{\%}$;

\item electron energy scale -- the reconstructed energy of the scattered 
  electron was varied in the MC by its estimated uncertainty of $\pm \SI{2}{\%}$;

\item \label{syst:MCc}MC model dependence -- the 
  \Qsq ($\delta_{8}^{\Qsq}$), 
  \etajet ($\delta_{8}^{\etajet}$) and 
  \ETjet  ($\delta_{8}^{\ETjet}$) reweighting
  corrections in the charm MC were varied in a range for which the description of data by MC remained reasonable. 
  The same relative variations were applied to the beauty MC;

\item light-flavour background -- the light-flavour contribution to
  the subtracted significance distribution includes a contribution
  from long-lifetime strange-hadron decays.  To estimate the
  uncertainty due to modelling of this effect, the MC light-flavour
  distribution of $N(S^{+}) - N(S^{-})$ was scaled by $\pm
  \SI{30}{\%}$~\cite{Aaron:2009ut} and the fit was repeated;

\item[$\delta_{10}$] charm fragmentation function -- to estimate the sensitivity to
  the charm fragmentation function, it was changed in the MC from the
  Bowler
  to the Peterson~\cite{pr:d27:105} parameterisation with
  $\epsilon=0.062$~\cite{1126-6708-2009-04-082};

\item[$\delta_{11}$] beauty fragmentation function -- to estimate the sensitivity to
  the beauty fragmentation function, it was changed in the MC from the
  Bowler
  to the Peterson parameterisation with
  $\epsilon=0.0041$~\cite{Abbiendi:2002vt};

\item[$\delta_{12}$] charm branching fractions ($\delta^{\text{BR}}_{12}$) and 
  fragmentation fractions ($\delta^{\text{frag}}_{12}$) -- these
  were varied
  within the uncertainties of the world-average
  values~\cite{Nakamura:2010zzi,arXiv:1112.3757,thesis:viazlo:2012};

\item[$\delta_{13}$] luminosity measurement -- a $\SI{1.9}{\%}$ overall
  normalisation uncertainty was associated with the luminosity
  measurement.
\end{enumerate}

To evaluate the total
systematic uncertainty on the integrated cross sections, the contributions from the different
systematic uncertainties were added in quadrature, separately for the
negative and the positive variations.
The same procedure was applied
to each bin for the differential cross sections. However,
the luminosity measurement uncertainty was not included.
In the case of beauty, the dominant effects arise from the
uncertainties on the track-finding inefficiencies, the beauty fragmentation function
and MC modelling. For charm, the uncertainties on the branching
fractions, the light-flavour asymmetry as well as on the MC modelling
contribute most to the total systematic uncertainty.

\section{Cross sections}
\label{sec:results}

Cross sections for inclusive jet production in beauty (charm) events were measured 
in the range $\ETjet > \SI[parse-numbers=false]{5 (4.2)}{\GeV}$,
$-1.6 < \etajet < 2.2$
for DIS events with $5 < \Qsq < \SI{1000}{\GeV^{2}}$ and 
$0.02 < y < 0.7$, where the jets are defined as in Section~\ref{sec:bc-extr}.
The single-differential cross sections for jet production in beauty
and charm events were measured as a function of $\ETjet, \etajet,
\Qsq$ and $x$. 
The results of the measured cross
sections are given in Tables~\ref{tab:diffet}--\ref{tab:diffx} and
shown in Figs.~\ref{fig:diff_et}--\ref{fig:diff_x}. 
The measurements
are compared to the \HVQDIS NLO QCD predictions obtained using ZEUS-S and ABKM
as proton PDFs, and to the \RAPGAP predictions scaled by a factor of
1.49 for beauty and 1.40 for charm. The scale factors correspond to
the ratio of the measured integrated visible cross section to the \RAPGAP
prediction.  The shapes of all measured beauty cross sections are reasonably
well described by \HVQDIS and the \RAPGAP MC. \RAPGAP provides a worse 
description of the shape of the charm cross sections than \HVQDIS.\footnote{For 
the acceptance corrections, the Monte Carlo was reweighted as discussed in 
Section~\ref{sec:montecarlo}.} For charm, the data are
typically \SIrange{20}{30}{\%} above the \HVQDIS NLO prediction, but
in reasonable agreement within uncertainties. Differences between the
NLO predictions using the different proton PDFs are mostly very small.

Double-differential cross sections 
as a function of $x$ for different
ranges of \Qsq for inclusive jet production in beauty and charm events are 
listed in Tables~\ref{tab:ddiffb} and~\ref{tab:ddiffc}, respectively.

\section{\boldmath Extraction of \Ftwoq and \sigredq}
\label{sec:F2q}

The heavy-quark contribution to the proton structure function $F_{2}$,
\Ftwoq with $q = b, c$, can be defined in terms of
the inclusive heavy flavour double-differential cross section as a function of $x$ and \Qsq,
\begin{equation*}
  \frac{\dif^{2}\sigma_{q\bar{q}}}{\dif x\,\dif \Qsq} =
  \frac{2\pi \alphaem^{2}}{xQ^{4}} {\Big\{[1 + (1 - y^{2})] \Ftwoq(x,\Qsq)-y^{2} \Flq(x
      ,\Qsq)\Big\}} \,,
\end{equation*}
where \Flq is the heavy-quark contribution to the
structure function $F_{L}$.

To extract \Ftwoq from the visible jet production cross sections in heavy-quark events,
measured in bins of $x$ and \Qsq, an
extrapolation from the measured range in \ETjet and \etajet to the
full kinematic phase space was performed. 
This implicitly takes into account the jet multiplicity.
The measured values of \Ftwoq at a
reference point in the $x$--\Qsq plane were calculated using
\begin{equation}
  \Ftwoq(x,\Qsq) =
  \frac{\dif^{2}\sigma_{q}^{\text{jet}} /\dif x\,\dif \Qsq}
       {\dif^{2}\sigma_{q}^{\text{had,NLO}}/\dif x\,\dif \Qsq}
  F_{2}^{q\bar{q},\text{NLO}}(x,\Qsq)\,,
  \label{eq:F2q}
\end{equation}
where
$\dif^{2}\sigma_{q}^{\text{jet}}/\dif x\,\dif \Qsq$ is determined in analogy to
Eq.~\eqref{eq:xsec4}, and $F_{2}^{q\bar{q},\text{NLO}}$ and
$\dif^{2}\sigma_{q}^{\text{had,NLO}}/\dif x\,\dif \Qsq$ were
calculated at NLO in the FFNS using the \HVQDIS program with the factor \chad applied as in Eq.~\eqref{eq:chad}.  The proton
PDFs were obtained from the FFNS variant of the HERAPDF~1.0 NLO QCD
fit~\cite{epj.c73.2311}.
This PDF was used in order to be consistent with the HERA combined results~\cite{epj.c73.2311}.
The strong coupling constant $\alphaS(M_Z)$ was set to 0.105 as in
the PDF fit.  Other settings were as described in
Section~\ref{sec:theory} for the ZEUS-S variant.  
As discussed in
Section~\ref{sec:bc-extr}, $\dif^{2}\sigma_{q}^{\text{jet}}/\dif x\,\dif \Qsq$
was multiplied by $1/\cqrad$, hence \Ftwoq is given at QED Born level,
consistent with the usual convention.  
The procedure of Eq.~\eqref{eq:F2q} also
corrects for the \Flq contribution to the cross section. This assumes that the
calculation correctly predicts the ratio $\Flq/\Ftwoq$.

The
extrapolation factors for beauty due to cuts on \ETjet and \etajet
typically range from 1.3 to 1.0, decreasing with increasing \Qsq. The factor
is up to 1.7 at high values of $x$. 
For charm, the extrapolation
factors are typically about 4 in the region $5 < \Qsq < \SI{20}{\GeV^{2}}$
and about 2 in the region $20 < \Qsq < \SI{60}{\GeV^{2}}$. The uncertainty on
the extrapolation from the measured range to the full kinematic phase
space was estimated by varying the parameters of the calculation for the
extrapolation factors and adding the resulting uncertainties in
quadrature.  For charm, the same variations were performed as for
the HERA combined results~\cite{epj.c73.2311}:
the charm mass was varied by
$\pm \SI{0.15}{\GeV}$; the strong coupling constant $\alphaS(M_Z)$ was changed by $\pm 0.002$;
renormalisation and factorisation scales were multiplied
simultaneously by 0.5 or 2.
Uncertainties resulting from the proton PDF uncertainty
are small~\cite{zeush1.web} 
and were neglected.
For beauty, the same variations of \alphaS and scales were made and
the beauty mass was varied by $\pm \SI{0.25}{\GeV}$.
For each bin, a reference
point in $x$ and \Qsq was defined (see Table~\ref{tab:f2b}) to
calculate the structure function.

In addition, beauty and charm reduced cross sections were determined.
They are defined as
\begin{equation*}
  \sigredq = \frac{\dif^{2}\sigma_{q\bar{q}}}{\dif x\,\dif \Qsq} \cdot
  \frac{xQ^{4}}{2\pi \alphaem^{2}[1 + (1 - y^{2})]} = F_{2}^{q\bar{q}}(x,\Qsq)-\frac{y^{2}}{1 + (1 - y^{2})} F_{L}^{q\bar{q}}(x
      ,\Qsq) \,,
\end{equation*}
and are extracted in analogy to \Ftwoq as described above except that
no assumption on \Flq is required.

The extracted values of \Ftwob and \Ftwoc are given in
Tables~\ref{tab:f2b} and~\ref{tab:f2c}, respectively, while \sigredb and
\sigredc are shown in Tables~\ref{tab:reduced_beauty}
and~\ref{tab:reduced_charm}.  The total uncertainties of the
measurements were calculated from the statistical and systematic
uncertainties of the measured cross sections
(Tables~\ref{tab:ddiffb},~\ref{tab:ddiffc},~\ref{tab:beauty_systematics_part1}--\ref{tab:charm_systematics_part2}) and of the extrapolation uncertainty
(Tables~\ref{tab:f2b_extrapolation}--\ref{tab:redc_extrapolation}),
added in quadrature.

The structure function \Ftwoc is shown in Fig.~\ref{fig:f2c_x} as a
function of $x$ for different values of \Qsq. The measurements are
compared to the NLO QCD HERAPDF~1.5~\cite{herapdf.web} predictions,
the most recent official release of the HERAPDF, based on 
the RT~\cite{springerlink:10.1140/epjc/s10052-009-1072-5} general-mass
variable-flavour-number scheme (GMVFNS).
The predictions are consistent with the measurements.

In Fig.~\ref{fig:sigma_red_c}, the measured \sigredc
values are compared to the HERA combined results~\cite{epj.c73.2311} as well as to the two recent results from
ZEUS~\cite{jhep05.2013.097, jhep05.2013.023} which are not yet included
in the combination.  For the comparison, some of the measured values of
this analysis were swum in \Qsq and $x$ using \HVQDIS. This measurement
is competitive, especially at high \Qsq, where the extrapolation
uncertainty is low, and is in agreement with the HERA combined
measurements.

The structure function \Ftwob is shown in Fig.~\ref{fig:f2b_x} as a
function of $x$ for different values of \Qsq. The measurements
are compared to HERAPDF~1.5 GMVFNS predictions. The increase in the uncertainty on the prediction 
around $\Qsq = m_{b}^{2}$ is a feature of the GMVFNS scheme used. 
The predictions
are consistent with the measurements.

The \Ftwob measurement is also shown as a function of \Qsq for fixed
$x$ in Fig.~\ref{fig:f2b_q2}, and is compared to previous ZEUS and H1
measurements. Again, \HVQDIS was used to swim the measured values in
\Qsq and $x$ to match the previous measurements.
In a wide range of \Qsq, this measurement represents the most precise
determination of \Ftwob at HERA. It is in good agreement with the
previous ZEUS analyses and H1 measurements. Several NLO and NNLO QCD
predictions based on the fixed- or variable-flavour-number
schemes~\cite{herapdf.web,PhysRevD.78.013004,
  springerlink:10.1140/epjc/s10052-009-1072-5,
  Thorne:2008xf, jimenez:2009,
  Alekhin2009166, Alekhin:2009ni, Alekhin:2010iu}
are also compared to the measurements. 
Predictions from different theoretical approaches agree
well with each other. 
All predictions provide a reasonable description of the data.

\section{Measurement of the running beauty-quark mass}
\label{sec:mb}

The reduced beauty cross sections, \sigredb,
(Fig.~\ref{fig:sigma_red_b} and Table~\ref{tab:reduced_beauty})
together with inclusive DIS data were used
to determine
the beauty-quark mass, in a simultaneous fit of the
mass and the parton densities.

The measurement procedure follows closely the method presented in a
recent H1-ZEUS publication~\cite{epj.c73.2311}, where the running
charm-quark mass in the \MSbar scheme was extracted using a
simultaneous QCD fit of the combined HERA~I inclusive DIS
data~\cite{Aaron:2009aa} and the HERA combined charm DIS
data~\cite{epj.c73.2311}.
This approach 
was also used and extended
by a similar independent analysis~\cite{adlm2},
and was preceded by a
similar analysis of a partial charm data set~\cite{adlm}.

The fit for the running beauty-quark mass was performed within the
\HERAFITTER~\cite{herafitter} framework choosing the ABM implementation 
of the fixed-flavour-number
scheme at next-to-leading order~\cite{ABKMMSbar, ABKMMSbar2,
  Alekhin:2009ni, np:b392:162, np:b392:229}. The
OPENQCDRAD~\cite{openqcdrad} option in \HERAFITTER was used 
in the \MSbar running-mass mode.
The fit was applied to the beauty data listed in Table~\ref{tab:reduced_beauty} 
and to the same inclusive DIS data~\cite{Aaron:2009aa} as in
the charm-quark mass fit~\cite{epj.c73.2311}.
A fit to the inclusive data only shows a very weak dependence on $m_{b}$.
In order to avoid technical complications,
no charm data were included in the simultaneous fit and only $m_b$ was extracted.

The PDFs resulting from the simultaneous fit changed only marginally with respect to 
the nominal PDFs obtained from the fit to the inclusive DIS data only.
The $\chi^2$ of the QCD fit, including the beauty data, shows a clear dependence on the
beauty-quark mass, $m_b$, as can be seen in Fig.~\ref{fig:chi2}.
The total $\chi^2$ for the best fit
is 587 for 596 degrees of freedom, and the partial contribution from
the beauty data is 11.4 for 17 points. The beauty-quark mass and its
uncertainty are determined from a parabolic parameterisation. The best fit yields
\begin{equation*}
  \mbMSbarmeas
\end{equation*}
for the \MSbar running beauty-quark mass at NLO.
The fit uncertainty\footnote{For the charm-quark mass fit~\cite{epj.c73.2311} this uncertainty was denoted 
  ``exp''.} (fit) is determined from
$\Delta\chi^2=1$. It contains the experimental uncertainties, 
the extrapolation uncertainties, the uncertainties of the standard PDF 
parameterisation, as well as an estimate of the uncertainty on the hadronisation
corrections, as detailed below. 
In addition, the result has uncertainties attributed to the
choices of some extra model parameters (mod.), some additional
variations of the PDF parameterisation (param.) and uncertainties on the
perturbative QCD parameters
(theo.). Details of the uncertainty evaluation include:

\begin{description}
\item[Fit uncertainty:]
For the beauty data, all uncertainties from
Tables~\ref{tab:beauty_systematics_part1}, 
\ref{tab:beauty_systematics_part2} (experimental) and
\ref{tab:redb_extrapolation} (extrapolation), and the statistical uncertainty, 
as summarised in 
Table~\ref{tab:reduced_beauty}, were accounted for in the fit.
Following the discussion in Section~\ref{sec:theory}, an uncertainty of \SI{100}{\percent}
on $\deltahad = \chad - 1$ (Table~\ref{tab:ddiffb}) was introduced 
as an additional uncorrelated uncertainty.
The uncertainties arising from the default PDF parameterisation~\cite{epj.c73.2311}, 
including the so-called ``flexible'' gluon 
parameterisation, are implicitly part of the fit uncertainty. 

The statistical uncertainties and the uncertainties $\delta_1,
\delta_2, \delta_4^{\text{core}}, \delta_5$ and $\delta_{12}$ from
Tables~\ref{tab:beauty_systematics_part1} and~\ref{tab:beauty_systematics_part2} were treated as uncorrelated, while
all other uncertainties, including those from luminosity and 
from Table~\ref{tab:redb_extrapolation}, were treated as
point-to-point correlated.
The ``multiplicative'' uncertainty option~\cite{herafitter} from \HERAFITTER was
used.  In the case of asymmetric uncertainties, the larger was used
in both directions.
The uncertainties of the inclusive data were used as published.
Since the inclusive data were taken during the HERA~I phase
and the beauty data during the HERA~II phase, the two sets of
data were treated as  uncorrelated.

\item[Model uncertainty:]
The model choices include an assumption on the strangeness fraction, $f_s$,
the minimum $Q^2$ used in the data selection, $Q^2_{\text{min}}$, and
$Q^2_0$, the starting value for the QCD evolution.
These were treated exactly as in the charm-quark mass fit~\cite{epj.c73.2311}.
Table~\ref{tab:result} lists the choices and variations and their
individual contributions to the model uncertainty attributed to the
model choices.

Another source of uncertainty is that the $b$-quark mass was
used earlier to extrapolate the measured visible cross sections to
the reduced cross sections. The corresponding uncertainty is
parameterised in Table~\ref{tab:redb_extrapolation} and used in the
fit, but the correlation of this uncertainty with the mass used in
the QCD fit is lost.  Since the \HVQDIS~\cite{pr:d57:2806} program
used for the extrapolation uses the pole-mass scheme, and no
differential calculations are available in the running-mass scheme,
no fully consistent treatment of this correlation is possible.
However, the equivalent uncertainty when using the pole-mass scheme
can be consistently estimated.
For this purpose, the fit was repeated with
the pole-mass option of OPENQCDRAD, which was checked to yield
results consistent with the \HVQDIS predictions for \sigredb.

The result,
$m_b(\text{pole}) = \SI[parse-numbers=false]{4.35 \pm 0.14\,(\text{fit})}{\GeV}$,
has a fit uncertainty which is the same as
the fit uncertainty for the running-mass fit. However, since the pole-mass definition
includes an unavoidable additional theoretical uncertainty due to a
nonperturbative contribution, 
no attempt to extract a pole-mass measurement with full systematic uncertainties was made.
To recover the correlation between the extrapolation and the mass fit,
the extrapolated cross sections were iteratively modified
using the predictions from the mass values obtained in each fit.
This removes the uncertainty on $m_{b}$ in the extrapolation
and takes the full correlations into account.  
The resulting mass
$m_b(\text{pole}) = \SI[parse-numbers=false]{4.28 \pm
  0.13\,(\text{fit})}{\GeV}$ is slightly lower.  
The difference between the results from the two procedures 
($\delta m_{\text{ext}} = \SI{-0.07}{\GeV}$) was treated as an additional
model uncertainty.

\item[PDF parameterisation uncertainty:]
The parameterisation of the PDFs is chosen as for the charm-quark mass
fit~\cite{epj.c73.2311}, including the ``flexible'' parameterisation
of the gluon distribution. An additional
uncertainty is estimated by freeing three extra PDF
parameters $D_{u_v}$, $D_{\bar D}$ and $D_{\bar U}$ in the fit which allow
for small shape variations in the $u_v$, $\bar{U}$ and $\bar{D}$ parton
distributions~\cite{epj.c73.2311}.
The change in $m_{b}$ due to leaving the parameters free is given as the systematic uncertainty in Table~\ref{tab:result}.

\item[Perturbative scheme and related theory uncertainty:]
The parameters used for the perturbative part of the QCD calculations
also introduce uncertainties;
the effects are listed in Table~\ref{tab:result}.
As in previous analyses~\cite{adlm, epj.c73.2311, adlm2}, the \MSbar running-mass
scheme~\cite{mc:run, mc:run2, mc:run3}
was chosen for all calculations of the reduced cross sections and
the fit because it
shows better perturbative convergence behaviour than the pole-mass
scheme.
In order to allow the low-\Qsq points of the inclusive DIS
measurement to be included without the need of additional charm-quark mass
corrections, the number of active flavours (NF) was set to three, i.e.\ the charm
contribution was also treated in the fixed-flavour-number scheme.
Accordingly, the strong coupling constant was set to
$\alphaS(M_Z)^{\text{NF}=3} = 0.105 \pm 0.002$, corresponding to
$\alphaS(M_Z)^{\text{NF}=5} = 0.116 \pm 0.002$.

The theoretical prediction
of the charm contribution to the inclusive DIS data is obtained using the
running charm-quark mass obtained from the fit to the combined HERA charm data~\cite{epj.c73.2311},
i.e.\  $m_c(m_c) = \SI{1.26(6)}{\GeV}$.
It was checked that, as expected, using this mass together with the central 
PDF from the $m_{b}$ fit,
a good description of the combined HERA charm data~\cite{epj.c73.2311} was obtained.
Thus, the charm contribution to the inclusive data should be well described.

The renormalisation and factorisation scales were set to $\mu =
\mu_R = \mu_F = \sqrt{\Qsq+4m^2}$ with $m=0, m_c, m_b$ for the light quark,
charm, and beauty contributions, respectively, and varied simultaneously by a factor two as in 
previous analyses~\cite{adlm, adlm2}.
\end{description}

The measured beauty-quark mass is in
very good agreement with the world average $m_b(m_b)= \SI{4.18 \pm
  0.03}{\GeV}$~\cite{PDG2012}.
The resulting predictions for the beauty cross
sections are shown together with the data in Fig.~\ref{fig:sigma_red_b}.
Figure~\ref{fig:sigma_red_b} also shows the change in the predictions
resulting from the fit when different $m_b$ values are assumed.
The largest sensitivity to
$m_b$ arises from the low-\Qsq region, while at high \Qsq the
impact of the beauty-quark mass is small.

\section{Conclusions}
\label{sec:conclusions}

Inclusive jet production cross sections in events
containing beauty or charm quarks have been
measured in DIS at HERA, exploiting the long lifetimes and large
masses of $b$ and $c$ hadrons. In contrast to previous analyses at ZEUS,
the measurement was not restricted to any particular final state. This
resulted in substantially increased statistics.

Differential cross sections as functions of \ETjet, \etajet, \Qsq and
$x$ were determined.  Next-to-leading-order QCD predictions calculated using the \HVQDIS
program using two different sets of proton PDFs are consistent with
the measurements.  

The heavy-quark contributions to the proton structure function $F_{2}$
as well as beauty and charm reduced cross sections were extracted from
the double-differential cross sections as a function of $x$ and \Qsq.
The \Ftwob, \Ftwoc and \sigredb, \sigredc values are in agreement with previous
measurements at HERA.
The results were also compared to several NLO and NNLO QCD
calculations, which provide a good description of the data.  The
precision of the \Ftwoc measurement is competitive with other
analyses.  For a wide range of \Qsq, the \Ftwob measurement represents
the most precise determination of \Ftwob.

The running beauty-quark mass in the \MSbar scheme was determined from
an NLO QCD fit in the fixed-flavour-number scheme to the \sigredb 
cross sections from
this analysis and to HERA~I inclusive DIS data:
\begin{equation*}
  \mbMSbarmeas
\end{equation*}
This value agrees well with the world average.

\section*{Acknowledgements}
\label{sec:acknowledge}

We appreciate the contributions to the construction, maintenance and operation of
the ZEUS detector of many people who are not listed as authors.  The
HERA machine group and the DESY computing staff are especially
acknowledged for their success in providing excellent operation of the
collider and the data-analysis environment. We thank the DESY
directorate for their strong support and encouragement.
It is a pleasure to thank the ABKM, CTEQ, JR and MSTW groups that
provided the predictions for \Ftwob shown in
Fig.~\ref{fig:f2b_q2}.
We gratefully acknowledge the advice from S.~Alekhin
and R.~Pla\v{c}akyt\.{e} concerning the appropriate usage of OPENQCDRAD and
\HERAFITTER.


%% file: zeusF2bc.bbl
\providecommand{\urlprefix}{}
\providecommand{\etal}{et al.\xspace}
\providecommand{\coll}{collaboration\xspace}

%% file: zeusF2bc-tab.tex
\newcommand{\pho}{\phantom{0}}
\newcommand{\phd}{\phantom{.}}
\newcommand{\phdo}{\phantom{.0}}

\vspace*{\fill}
\begin{table}[!hp]
  \centering
  \begin{tabular}{ll|c |c}
    \toprule
        & Source & Beauty & Charm \\
        & & (\%) & (\%) \\\hline
    $\delta_1$  & Event and DIS selection & $\pm 1.4$ & $\pm 0.8$\\\hline
    $\delta_2$  & Trigger efficiency & $+2.0$ & $+1.0$\\\hline
    $\delta_3$  & Tracking efficiency & $\pm 2.0$ & $\pm 0.5$\\\hline
    $\delta_4$  & Decay-length smearing & $\pm 1.3$ & $\pm 1.2$\\\hline
    $\delta_5$  & Signal extraction procedure & $\pm 0.8$ & $\pm 0.8$\\\hline
    $\delta_6$  & Jet energy scale & $\pm 0.7$ & $\pm 0.9$ \\\hline
    $\delta_7$  & EM energy scale & $\pm  0.3$ &  $\pm  0.1$\\\hline
    $\delta_8$  & Charm \Qsq reweighting ($\delta_8^{\Qsq,c}$) & $\pm 1.7$ & $\pm 1.8$\\
        & Beauty \Qsq reweighting ($\delta_8^{\Qsq,b}$) & $\pm 2.9$ & $\pm 0.4$\\
        & Charm \etajet reweighting ($\delta_8^{\etajet,c}$) & $^{+ 0.3}_{-0.4}$ & $^{+ 1.5}_{-1.0}$\\
        & Beauty \etajet reweighting ($\delta_8^{\etajet,b}$) & $^{+0.7}_{-0.4}$ & $^{+0.0}_{-0.1}$ \\
        & Charm \ETjet reweighting ($\delta_8^{\ETjet,c}$) & $^{+1.7}_{-1.3}$ & $^{+2.2}_{-1.7}$\\
        & Beauty \ETjet reweighting ($\delta_8^{\ETjet,b}$) & $^{+5.4}_{-4.2}$ & $^{+0.5}_{-0.6}$ \\ \hline
    $\delta_9$  & Light-flavour asymmetry & $\pm 0.4$ & $\pm 2.0$\\\hline
    $\delta_{10}$ & Charm fragmentation function& $- 0.9$ & $+ 1.0$\\\hline
    $\delta_{11}$ & Beauty fragmentation function & $- 3.1$ & $+0.0$\\\hline
    $\delta_{12}$ & BR and fragmentation fractions & $^{+1.8}_{-2.1}$ & $^{+3.5}_{-2.6}$\\\hline
    $\delta_{13}$ & Luminosity measurement & $\pm 1.9$ &  $\pm 1.9$\\
    \midrule
        & Total & $^{+8.0}_{-7.6}$ & $^{+6.0}_{-5.1}$\\
    \bottomrule
    \end{tabular}
    \caption[Systematic uncertainties]{Effects of the systematic
      uncertainties on the integrated beauty- and charm-jet cross sections.}
    \label{tab:syst}
\end{table}
\vspace*{\fill}
\clearpage


\vspace*{\fill}
\sisetup{round-mode=figures, round-precision=2, retain-explicit-plus=true, group-digits = integer, group-minimum-digits=4}
\begin{table}[!hp]
  \centering
  \begin{tabular}{%
      S[table-format=2.1, table-number-alignment=right,
      round-mode=places, round-precision=1]@{$\,:\,$}
      S[table-format=2.0, table-number-alignment=left,
      round-mode=places, round-precision=0]|
      S[table-format=4.3, table-number-alignment=right,
      round-mode=figures, round-precision=3]@{$\,\pm\,$}
      S[table-format=3.3, table-number-alignment=right,
      round-mode=figures, round-precision=2]@{$\,$}l|
      S[table-format=1.2, table-number-alignment=left,
      round-mode=places, round-precision=2]|
      S[table-format=1.2, table-number-alignment=left,
      round-mode=places, round-precision=2]} 
    \toprule

    \multicolumn{2}{c|}{\ETjet} & \multicolumn{3}{c|}{\diffEtb (\si{\pb/\GeV})} & \chad & \crad \\
    \multicolumn{2}{c|}{(\si{\GeV})} & 
    \multicolumn{1}{c}{\mbox{}} & \multicolumn{1}{c}{\stat} &\multicolumn{1}{c|}{\syst} & & \\\hline

      5 	& 8 	& 134.432 	& 26.259 	& \numpmerr{+23.6011}{-24.9878}{2} 	& 0.95	& 1.01 	 \\
      8 	& 11 	& 66.1404 	& 5.83552 	& \numpmerr{+5.13624}{-6.83356}{2} 	& 1.08	& 0.98 	 \\
      11 	& 14 	& 30.1461 	& 1.88766 	& \numpmerr{+1.3801}{-1.97175}{2} 	& 1.05	& 0.96 	 \\
      14 	& 17 	& {\pho\num[round-precision=4]{11.2732}} 	& 0.895181 	& \numpmerr{+0.523118}{-0.597212}{2} 	& 1.04	& 0.95 	 \\
      17 	& 20 	& 4.70856 	& 0.496915 	& \numpmerr{+0.335698}{-0.32315}{2} 	& 0.99	& 0.93 	 \\
      20 	& 25 	& 2.03698 	& 0.278281 	& \numpmerr{+0.275551}{-0.275902}{2} 	& 0.93	& 0.89 	 \\
      25 	& 35 	& 0.38001 	& 0.094178 	& \numpmerr{+0.0783403}{-0.076364}{2} 	& 0.8	& 0.89 	 \\
    \bottomrule
    
    \multicolumn{7}{c}{\mbox}{}\\
    
    \toprule
    \multicolumn{2}{c|}{\ETjet} & \multicolumn{3}{c|}{\diffEtc (pb/GeV)} & \chad & \crad \\
    \multicolumn{2}{c|}{(GeV)} &
    \multicolumn{1}{c}{\mbox{}} & \multicolumn{1}{c}{\stat} &\multicolumn{1}{c|}{\syst} & & \\\hline
    
      4.2 	& 8 	& 3663.71 	& 115.859 	& \numpmerr{+204.421}{-183.457}{2} 	& 1.06	& 0.98 	 \\
      8 	& 11 	& 748.158 	& 22.4793 	& \numpmerr{+44.8115}{-41.1266}{2} 	& 1.05	& 0.97 	 \\
      11 	& 14 	& 222.129 	& 10.1362 	& \numpmerr{+21.3014}{-20.3774}{2} 	& 1.03	& 0.96 	 \\
      14 	& 17 	& 91.4223 	& 6.46509 	& \numpmerr{+10.8208}{-9.66619}{2} 	& 0.99	& 0.93 	 \\
      17 	& 20 	& 38.9494 	& 4.36745 	& \numpmerr{+6.11693}{-5.97963}{2} 	& 0.96	& 0.93 	 \\
      20 	& 25 	& 16.3889 	& 3.21475 	& \numpmerr{+3.97609}{-3.69172}{2} 	& 0.95	& 0.85 	 \\
      25 	& 35 	& {\phdo\num[round-precision=2]{2.55918}}	& 1.12963 	& \numpmerr{+0.912201}{-0.918568}{1} 	& 0.86	& 0.88 	 \\
      \bottomrule
  \end{tabular}
  \caption{Differential cross sections for inclusive jet production in beauty
    events (top) and charm events (bottom) as a function of \ETjet.
    The beauty (charm) cross sections are given for $5 < Q^{2} < \SI{1000}{\GeV^{2}}$,
    $0.02 < y < 0.7$, $\ETjet > \SI[parse-numbers=false]{5 (4.2)}{\GeV}$ and $-1.6 < \etajet < 2.2$.
    The measurements are given together with their statistical and
    systematic uncertainties (not including the error on the integrated luminosity). Hadronisation and QED radiative corrections, \chad and \crad, respectively, are also shown.}
  \label{tab:diffet}
\end{table}
\vspace*{\fill}
\clearpage

\vspace*{\fill}
\begin{table}[!hp]
  \centering
  \begin{tabular}{%
      S[table-format=+1.1, table-number-alignment=right,
      round-mode=places, round-precision=1]@{$\,:\,$}
      S[table-format=+1.1, table-number-alignment=left,
      round-mode=places, round-precision=1]|
      S[table-format=4.0, table-number-alignment=right,
      round-mode=figures, round-precision=2]@{$\,\pm\,$}
      S[table-format=3.0, table-number-alignment=center,
      round-mode=figures, round-precision=2]@{$\,$}c|
      S[table-format=1.2, table-number-alignment=left,
      round-mode=places, round-precision=2]|
      S[table-format=1.2, table-number-alignment=left,
      round-mode=places, round-precision=2]}
   \toprule

    \multicolumn{2}{c|}{\etajet} & \multicolumn{3}{c|}{\diffetab (\si{\pb})} & \chad & \crad \\
    \multicolumn{2}{c|}{} &
    \multicolumn{1}{c}{\mbox{}} & \multicolumn{1}{c}{\stat} &\multicolumn{1}{c|}{\syst} & & \\\hline
    
        -1.6 	& -0.8 	& 88.7486 	& 30.8037 	& \numpmerr{+24.5925}{-40.9638}{2} 	& 0.96	& 0.98 	 \\
        -0.8 	& -0.5 	& 220.465 	& 29.8391 	& \numpmerr{+15.2221}{-22.7891}{2} 	& 0.98	& 0.98 	 \\
        -0.5 	& -0.2 	& 206.218 	& 24.0276 	& \numpmerr{+17.8843}{-20.9708}{2} 	& 0.93	& 0.99 	 \\
        -0.2 	& 0.1 	& 282.574 	& 21.9724 	& \numpmerr{+21.4662}{-22.8903}{2} 	& 0.91	& 0.99 	 \\
        0.1 	& 0.4 	& 261.219 	& 22.003 	  & \numpmerr{+20.6369}{-19.2677}{2} 	& 0.94	& 0.99 	 \\
        0.4 	& 0.7 	& 312.231 	& 23.3473 	& \numpmerr{+30.4042}{-28.5722}{2} 	& 1.01	& 0.99 	 \\
        0.7 	& 1.0 	& 265.416 	& 26.0359 	& \numpmerr{+26.1089}{-24.4162}{2} 	& 1.06	& 0.99 	 \\
        1.0 	& 1.3 	& 286.376 	& 30.6789 	& \numpmerr{+30.4274}{-30.3705}{2} 	& 1.07	& 0.99 	 \\
        1.3 	& 1.6 	& 217.711 	& 41.1839 	& \numpmerr{+23.7937}{-23.4561}{2} 	& 1.07	& 0.99 	 \\
        1.6 	& 2.2 	& 218.954 	& 71.3579 	& \numpmerr{+78.2865}{-78.6195}{2} 	& 1.07	& 0.98 	 \\
      \bottomrule

    \multicolumn{7}{c}{}\\
    
    \toprule
    \multicolumn{2}{c|}{\etajet} & \multicolumn{3}{c|}{\diffetac (\si{\pb})} & \chad & \crad \\
    \multicolumn{2}{c|}{} &
    \multicolumn{1}{c}{\mbox{}} & \multicolumn{1}{c}{\stat} &\multicolumn{1}{c|}{\syst} & & \\\hline

        -1.6 	& -1.1 	& 1862.99 	& 257.965 	& \numpmerr{+197.01}{-176.819}{2} 	& 0.89	& 0.99 	 \\
        -1.1 	& -0.8 	& 3582.42 	& 216.166 	& \numpmerr{+239.62}{-221.298}{2} 	& 0.97	& 0.98 	 \\
        -0.8 	& -0.5 	& 4196.36 	& 199.194 	& \numpmerr{+207.201}{-181.171}{2} 	& 1.02	& 0.98 	 \\
        -0.5 	& -0.2 	& 5236.62 	& 192.819 	& \numpmerr{+263.6}{-236.378}{2} 	& 1.05	& 0.98 	 \\
        -0.2 	& 0.1 	& 5418.12 	& 204.421 	& \numpmerr{+376.769}{-359.057}{2} 	& 1.07	& 0.98 	 \\
        0.1 	& 0.4 	& 5958.86 	& 212.74 	  & \numpmerr{+405.036}{-383.702}{2} 	& 1.1	& 0.98 	 \\
        0.4 	& 0.7 	& 5673.08 	& 215.872 	& \numpmerr{+341.669}{-321.991}{2} 	& 1.11	& 0.98 	 \\
        0.7 	& 1.0 	& 5674.09 	& 239.103 	& \numpmerr{+315.756}{-295.958}{2} 	& 1.1	& 0.98 	 \\
        1.0 	& 1.3 	& 4945.69 	& 267.456 	& \numpmerr{+319.029}{-300.982}{2} 	& 1.09	& 0.98 	 \\
        1.3 	& 1.6 	& 4856.08 	& 364.701 	& \numpmerr{+330.137}{-295.378}{2} 	& 1.07	& 0.97 	 \\
        1.6 	& 2.2 	& 4756.34 	& 626.922 	& \numpmerr{+639.32}{-638.604}{2} 	& 1.13	& 0.97 	 \\
        \bottomrule
  \end{tabular}
  \caption{Differential cross sections for inclusive jet production in beauty
    events (top) and charm events (bottom) as a function of \etajet.
    For details, see the caption of Table~\protect\ref{tab:diffet}.}
    \label{tab:diffeta}
\end{table}
\vspace*{\fill}
\clearpage

\vspace*{\fill}
\begin{table}[!hp]
  \centering
  \begin{tabular}{%
      S[table-format=3.0, table-number-alignment=right,
      round-mode=places, round-precision=1]@{$\,:\,$}
      S[table-format=4.0, table-number-alignment=left,
      round-mode=places, round-precision=0]|
      S[table-format=4.4, table-number-alignment=right,
      round-mode=figures, round-precision=3]@{$\,\pm\,$}
      S[table-format=3.4, table-number-alignment=right,
      round-mode=figures, round-precision=2]@{$\,$} l|
      S[table-format=1.2, table-number-alignment=left,
      round-mode=places, round-precision=2]|
      S[table-format=1.2, table-number-alignment=left,
      round-mode=places, round-precision=2]}
    \toprule

    \multicolumn{2}{c|}{\Qsq} & \multicolumn{3}{c|}{\diffQsqb (\si{\pb/\GeV^{2}})} & \chad & \crad \\
    \multicolumn{2}{c|}{(\si{\GeV^{2}})} &
    \multicolumn{1}{c}{\mbox{}} & \multicolumn{1}{c}{\stat} &\multicolumn{1}{c|}{\syst} & & \\\hline
        
      5 	& 10 	& 43.7925 	& 4.10641 	& \numpmerr{+3.96826}{-3.23861}{2} 	& 1.01	& 0.99 	 \\
      10 	& 20 	& 18.0256 	& 1.77665 	& \numpmerr{+1.67506}{-1.50422}{2} 	& 1.01	& 0.99 	 \\
      20 	& 40 	& 7.57717 	& 0.815906 	& \numpmerr{+0.736787}{-0.720338}{2} 	& 0.99	& 0.99 	 \\
      40 	& 70 	& 3.79767 	& 0.385252 	& \numpmerr{+0.300048}{-0.310328}{2} 	& 0.98	& 0.99 	 \\
      70 	& 120 	& 1.26479 	& 0.16202 	& \numpmerr{+0.143118}{-0.149595}{2} 	& 0.98	& 0.98 	 \\
      120 	& 200 	& 0.623354 	& 0.0664451 & \numpmerr{+0.0419258}{-0.0468477}{2} 	& 0.99	& 0.99 	 \\
      200 	& 400 	& 0.141922 	& 0.0175931 & \numpmerr{+0.0100339}{-0.0101913}{2} 	& 0.99	& 0.99 	 \\
      400 	& 1000 	& 0.0194489 & 0.00335429& \numpmerr{+0.00202807}{-0.00191172}{2} 	& 1.01	& 0.95 	 \\
    \bottomrule

    \multicolumn{7}{c}{}\\
    \toprule
    \multicolumn{2}{c|}{\Qsq} & \multicolumn{3}{c|}{\diffQsqc (\si{\pb/\GeV^{2}})} & \chad & \crad \\
    \multicolumn{2}{c|}{(\si{\GeV^{2}})} &
    \multicolumn{1}{c}{\mbox{}} & \multicolumn{1}{c}{\stat} &\multicolumn{1}{c|}{\syst} & & \\\hline
        
      5 	& 10 	& 835.049 	& 34.2112 	& \numpmerr{+45.9669}{-38.9157}{2} 	& 1.15	& 0.98 	 \\
      10 	& 20 	& 460.368 	& 15.0391 	& \numpmerr{+25.9113}{-22.013}{2} 	& 1.08	& 0.99 	 \\
      20 	& 40 	& 207.004 	& 6.35837 	& \numpmerr{+10.384}{-9.52269}{2} 	& 1.01	& 0.98 	 \\
      40 	& 70 	& 68.4758 	& 2.73395 	& \numpmerr{+3.79334}{-3.45069}{2} 	& 1.00	& 0.97 	 \\
      70 	& 120 	& 22.5422 	& 1.04134 	& \numpmerr{+1.35912}{-1.18263}{2} 	& 1.00	& 0.97 	 \\
      120 	& 200 	& 7.2829 	& 0.461989 	& \numpmerr{+0.482234}{-0.411741}{2} 	& 1.01	& 0.96 	 \\
      200 	& 400 	& 1.81963 	& 0.138055 	& \numpmerr{+0.0984157}{-0.0769446}{1} 	& 1.01	& 0.95 	 \\
      400 	& 1000 	& 0.218761 	& 0.0366047 & \numpmerr{+0.0323175}{-0.0291866}{2} 	& 1.02	& 0.87 	 \\
      \bottomrule
  \end{tabular}
  \caption[Differential cross sections as a function of \Qsq.]{Differential cross sections 
    for inclusive jet production in beauty
    events (top) and charm events (bottom) as a function of \Qsq.
    For details, see the caption of Table~\protect\ref{tab:diffet}.}
  \label{tab:diffq2}
\end{table}
\vspace*{\fill}
\clearpage

\vspace*{\fill}
\sisetup{round-mode=figures, round-precision=2,
  retain-explicit-plus=true, group-digits = integer, group-minimum-digits=4,
  scientific-notation=false}
\begin{table}[!hp]
  \centering
  \begin{tabular}{%
      S[table-format=1.5, table-number-alignment=right, round-mode=off]@{$\,:\,$}
      S[table-format=1.4, table-number-alignment=left,  round-mode=off]|
      S[table-format=8.0, table-number-alignment=right,
      round-mode=figures, round-precision=3]@{$\,\pm\,$}
      S[table-format=6.0, table-number-alignment=right,
      round-mode=figures, round-precision=2]@{$\,$} r|
      S[table-format=1.2, table-number-alignment=left,
      round-mode=places, round-precision=2]|
      S[table-format=1.2, table-number-alignment=left,
      round-mode=places, round-precision=2]}
    \toprule

    \multicolumn{2}{c|}{$x$} & \multicolumn{3}{c|}{\diffxb (\si{\pb})} & \chad & \crad \\
    \multicolumn{2}{c|}{} &
    \multicolumn{1}{c}{\mbox{}} & \multicolumn{1}{c}{\stat} &\multicolumn{1}{c|}{\syst} & & \\\hline
    
        0.00008	& 0.0002 	& 685820 	& 107987 	& \numpmerr{+85169.0}{-77780.6}{2} 	& 1.09	& 0.99 	 \\
        0.0002  & 0.0006 	& 614493 	& 46709.1 	& \numpmerr{+52256.5}{-45101.7}{2} 	& 1.05	& 0.99 	 \\
        0.0006 	& 0.0016 	& 218324 	& 14969.7 	& \numpmerr{+15778.4}{-14483}{2} 	& 0.99	& 0.99 	 \\
        0.0016 	& 0.005 	& 49784.9 	& 3471.98 	& \numpmerr{+3565.55}{-3544.64}{2} 	& 0.95	& 0.99 	 \\
        0.005 	& 0.01 	& 11170 	& 1329.6 	& \numpmerr{+952.614}{-918.569}{2} 	& 0.93	& 1.00 	 \\
        0.01 	& 0.1 	& 374.016 	& 79.2837 	& \numpmerr{+50.7126}{-50.3992}{2} 	& 0.92	& 0.95 	 \\
    \bottomrule

    \multicolumn{7}{c}{}\\
    \toprule
    \multicolumn{2}{c|}{$x$} & \multicolumn{3}{c|}{\diffxc (\si{\pb})} & \chad & \crad \\
    \multicolumn{2}{c|}{} &
    \multicolumn{1}{c}{\mbox{}} & \multicolumn{1}{c}{\stat} &\multicolumn{1}{c|}{\syst} & & \\\hline
    
        0.00008	& 0.0002 	& 10708000 	& 868147 	& \numpmerr{+758442}{-649688}{2} 	& 1.19	& 0.96 	 \\
        0.0002 	& 0.0006 	& 10270100 	& 385950 	& \numpmerr{+539076}{-423537}{2} 	& 1.20	& 0.98 	 \\
        0.0006 	& 0.0016 	& 4987690 	& 135403 	& \numpmerr{+259149}{-237028}{2} 	& 1.09	& 0.99 	 \\
        0.0016 	& 0.005 	& 1245090 	& 31513.7 	& \numpmerr{+70733.1}{-64016.7}{2} 	& 0.97	& 0.99 	 \\
        0.005 	& 0.01 	& 264303 	& 12116.9 	& \numpmerr{+18672.9}{-16968.5}{2} 	& 0.91	& 1.00 	 \\
        0.01 	& 0.1 	& 12473.8 	& 896.722 	& \numpmerr{+1049.92}{-965.189}{2} 	& 0.88	& 0.88 	 \\
        \bottomrule
  \end{tabular}
  \caption[Differential cross sections as a function of $x$]{Differential cross sections 
    for inclusive jet production in beauty
    events (top) and charm events (bottom) as a function of $x$.
    For details, see the caption of Table~\protect\ref{tab:diffet}.}
    \label{tab:diffx}
\end{table}
\vspace*{\fill}
\clearpage

\vspace*{\fill}
\sisetup{round-mode=figures, round-precision=2,
  retain-explicit-plus=true, group-digits = integer, group-minimum-digits=4,
  scientific-notation=false}
\begin{table}[!hp]
  \renewcommand{\arraystretch}{1.3}
  \centering
  \begin{tabular}{%
      S[table-format=3.0, table-number-alignment=right,
      round-mode=figures, round-precision=2]@{$\,:\,$}
      S[table-format=4.0, table-number-alignment=left,
      round-mode=figures, round-precision=2]|
      S[table-format=1.5, table-number-alignment=right, round-mode=off]@{$\,:\,$}
      S[table-format=1.4, table-number-alignment=left,  round-mode=off]|
      S[table-format=6.0, table-number-alignment=right,
      round-mode=figures, round-precision=3]@{$\,\pm\,$}
      S[table-format=6.0, table-number-alignment=right,
      round-mode=figures, round-precision=2]@{$\,$}r|
      S[table-format=1.2, table-number-alignment=left,
      round-mode=places, round-precision=2]|
      S[table-format=1.2, table-number-alignment=left,
      round-mode=places, round-precision=2]}
   \toprule

    \multicolumn{2}{c|}{\Qsq} & \multicolumn{2}{c|}{$x$} & \multicolumn{3}{c|}{\ddiffQsqxb (\si{pb/\GeV^{2}})} &
    \chad & \crad \\
    \multicolumn{2}{c|}{(\si{\GeV^{2}})} & \multicolumn{2}{c|}{} &
    \multicolumn{1}{c}{\mbox{}} & \multicolumn{1}{c}{\stat} &\multicolumn{1}{c|}{\syst} & & \\\hline
    
      5 	& 20 	& 0.00008& 0.0002 & {\num[round-precision=2]{685820}} 	& 107987 	& \numpmerr{+85211.1}{-77138.1}{1} 	& 1.09	& 0.99 	 \\
      5 	& 20 	& 0.0002 & 0.0003 & {\num[round-precision=2]{834185}} 	& 120239 	& \numpmerr{+84402.7}{-76794.5}{1} 	& 1.07	& 0.98 	 \\
      5 	& 20 	& 0.0003 & 0.0005 & 501097 	& 54637.3 	& \numpmerr{+48815.9}{-41434.7}{2} 	& 1.04	& 0.99 	 \\
      5 	& 20 	& 0.0005 & 0.003 & 48223.7 	& 5807.02 	& \numpmerr{+5126.01}{-4681.3}{2} 	& 0.91	& 0.99 	 \\ \hline
      
      20 	& 60 	& 0.0003 & 0.0005 & {\pho\num[round-precision=2]{81820.1}} & 24429.1 	& \numpmerr{+10743.2}{-10825.1}{2} 	& 1.07	& 0.98 	 \\
      20 	& 60 	& 0.0005 & 0.0012 & 133768 	& 13629.7 	& \numpmerr{+9990.18}{-9540.24}{1} 	& 1.05	& 0.99 	 \\
      20 	& 60 	& 0.0012 & 0.002 & 73427.2 	& 8527.85 	& \numpmerr{+6932.81}{-6852.47}{2} 	& 1.00	& 1.00 	 \\
      20 	& 60 	& 0.002  & 0.0035 & 25813.2 & 4558.39 	& \numpmerr{+3493.33}{-3438.41}{2} 	& 0.94	& 1.01 	 \\
      20 	& 60 	& 0.0035 & 0.01 & {\pho\pho\num[round-precision=2]{3607.94}} 	& 2017.9 	& \numpmerr{+998.454}{-1030.22}{2} 	& 0.81	& 0.99 	 \\ \hline
      
      60 	& 120 	& 0.0008 & 0.0018 & 33366.7 & 4451.86 	& \numpmerr{+3175.97}{-3092.36}{2} 	& 1.03	& 0.98 	 \\
      60 	& 120 	& 0.0018 & 0.003 & 22466.7 	& 2895.92 	& \numpmerr{+1979.28}{-2081.55}{2} 	& 1.02	& 0.99 	 \\
      60 	& 120 	& 0.003  & 0.006 & {\pho\pho\num[round-precision=2]{7429.65}} 	& 1331.49 	& \numpmerr{+848.402}{-893.717}{1} 	& 0.98	& 0.98 	 \\ \hline
      
      120 	& 400 	& 0.0016 & 0.005 & {\pho\pho\num[round-precision=2]{6711.61}} 	& 928.434 	& \numpmerr{+454.966}{-496.86}{2} 	& 1.01	& 0.99 	 \\
      120 	& 400 	& 0.005  & 0.016 & 3815.42 	& 337.655 	& \numpmerr{+172.05}{-197.342}{2} 	& 0.99	& 1.02 	 \\
      120 	& 400 	& 0.016  & 0.06 & 268.922 	& 133.471 	& \numpmerr{+73.5295}{-75.648}{1} 	& 0.92	& 0.87 	 \\ \hline
      
      400 	& 1000 	& 0.005  & 0.02 & 401.028 	& 87.7049 	& \numpmerr{+56.2695}{-52.5086}{2} 	& 1.01	& 0.95 	 \\
      400 	& 1000 	& 0.02   & 0.1 & {\pho\pho\pho\pho\num[round-precision=2]{69.5602}} 	& 20.7294 	& \numpmerr{+15.0483}{-16.0772}{2} 	& 1.00	& 0.95 	 \\
    \bottomrule
  \end{tabular}
  \caption[Beauty double differential cross sections as a function of $x$ and \Qsq]{%
      Double-differential cross sections for inclusive jet production in beauty
    events as a function of $x$ for different ranges of \Qsq. The
    cross sections are given for $5 < \Qsq < \SI{1000}{\GeV^{2}}$,
    $0.02 < y < 0.7$, $\ETjet > \SI{5}{\GeV}$ and $-1.6 < \etajet < 2.2$. The
    measurements are given together with their statistical and
    systematic uncertainties (not including the error on the integrated luminosity). Hadronisation and QED radiative corrections, \chad and \crad, respectively, are also shown.}
  \label{tab:ddiffb}
\end{table}
\vspace*{\fill}
\clearpage

\vspace*{\fill}
\begin{table}[!hp]
  \renewcommand{\arraystretch}{1.3}
  \centering
  \begin{tabular}{%
      S[table-format=3.0, table-number-alignment=right,
      round-mode=figures, round-precision=2]@{$\,:\,$}
      S[table-format=4.0, table-number-alignment=left,
      round-mode=figures, round-precision=2]|
      S[table-format=1.5, table-number-alignment=right, round-mode=off]@{$\,:\,$}
      S[table-format=1.4, table-number-alignment=left,  round-mode=off]|
      S[table-format=8.0, table-number-alignment=right,
      round-mode=figures, round-precision=3]@{$\,\pm\,$}
      S[table-format=6.0, table-number-alignment=right,
      round-mode=figures, round-precision=2]@{$\,$}r|
      S[table-format=1.2, table-number-alignment=left,
      round-mode=places, round-precision=2]|
      S[table-format=1.2, table-number-alignment=left,
      round-mode=places, round-precision=2]}
   \toprule

    \multicolumn{2}{c|}{\Qsq} & \multicolumn{2}{c|}{$x$} & \multicolumn{3}{c|}{\ddiffQsqxc (\si{pb/\GeV^{2}})} &
    \chad & \crad \\
    \multicolumn{2}{c|}{(\si{\GeV^{2}})} & \multicolumn{2}{c|}{} &
    \multicolumn{1}{c}{\mbox{}} & \multicolumn{1}{c}{\stat} &\multicolumn{1}{c|}{\syst} & & \\\hline

      5 	& 20 	& 0.00008& 0.0002 & 10708000 	& 868147 	& \numpmerr{+741266}{-597879}{2} 	& 1.19	& 0.96 	 \\
      5 	& 20 	& 0.0002 & 0.0003 & 13460200 & 949032 	& \numpmerr{+892710}{-731810}{2} 	& 1.21	& 0.98 	 \\
      5 	& 20 	& 0.0003 & 0.0005 & 8218150 & 473487 	& \numpmerr{+539562}{-470875}{2} 	& 1.23	& 0.98 	 \\
      5 	& 20 	& 0.0005 & 0.003  & {\phdo\num[round-precision=4]{1619500}} 	& 55794.8 	& \numpmerr{+99889.8}{-86812.1}{2} 	& 1.07	& 1.00 	 \\ \hline
      
      20 	& 60 	& 0.0003 & 0.0005 & 1574480 & 227228 	& \numpmerr{+141212}{-115194}{2} 	& 1.13	& 0.97 	 \\
      20 	& 60 	& 0.0005 & 0.0012 & 2603750 & 112795 	& \numpmerr{+137492}{-122519}{2} 	& 1.09	& 0.97 	 \\
      20 	& 60 	& 0.0012 & 0.002  & 1392860 & 68090.6 	& \numpmerr{+72583.6}{-68658.4}{2} 	& 1.05	& 0.98 	 \\
      20 	& 60 	& 0.002  & 0.0035 & 649576 	    & 34041.3 	& \numpmerr{+48333}{-46006.6}{2} 	& 1.01	& 0.99 	 \\
      20 	& 60 	& 0.0035 & 0.01   & 190150 	    & 13261.3 	& \numpmerr{+14257.8}{-13392.8}{2} 	& 0.91	& 0.99 	 \\ \hline
      
      60 	& 120 	& 0.0008 & 0.0018 & 251004	    & 33152.4 	& \numpmerr{+33220.4}{-33503.3}{2} 	& 1.07	& 0.97 	 \\
      60 	& 120 	& 0.0018 & 0.003  & 283017 	    & 21232.1 	& \numpmerr{+21890.1}{-21031}{2} 	& 1.03	& 0.99 	 \\
      60 	& 120 	& 0.003  & 0.006  & 135838 	    & 8092.47 	& \numpmerr{+10153}{-9738.35}{2} 	& 1.01	& 0.98 	 \\
      60 	& 120 	& 0.006  & 0.04   & 17107.1 	& 1612.27 	& \numpmerr{+1377.53}{-1238.1}{2} 	& 0.93	& 0.93 	 \\ \hline
      
      120 	& 400 	& 0.0016 & 0.005  & 109820 	    & 7792.75 	& \numpmerr{+7418.08}{-5623.43}{2} 	& 1.05	& 0.97 	 \\
      120 	& 400 	& 0.005  & 0.016  & 34508.3 	& 2165.68 	& \numpmerr{+1936.4}{-1722.26}{2} 	& 1.01	& 1.00 	 \\
      120 	& 400 	& 0.016  & 0.06   & {\pho\pho\pho\phdo\num[round-precision=2]{5310.5}} 	    & 1083.32 	& \numpmerr{+786.217}{-764.99}{1} 	& 0.96	& 0.8 	 \\ \hline
      
      400 	& 1000 	& 0.005  & 0.02   & 5785.28 	& 898.61 	& \numpmerr{+903.841}{-847.828}{2} 	& 1.02	& 0.88 	 \\
      400 	& 1000 	& 0.02   & 0.1    & 540.021 	& 281.07 	& \numpmerr{+156.328}{-157.228}{2} 	& 1.01	& 0.84 	 \\
      \bottomrule
  \end{tabular}
  \caption[Charm double-differential cross sections as a function of $x$ and \Qsq.]{%
    Double-differential cross sections for inclusive jet production in charm
    events as a function of $x$ for different ranges of \Qsq. The
    cross sections are given for $5 < \Qsq < \SI{1000}{\GeV^{2}}$, 
    $0.02 < y < 0.7$, $\ETjet > \SI{4.2}{\GeV}$ and $-1.6 < \etajet < 2.2$. The
    measurements are given together with their statistical and
    systematic uncertainties (not including the error on the integrated luminosity). Hadronisation and QED radiative
    corrections, \chad and \crad, respectively, are also shown.}
  \label{tab:ddiffc}
\end{table}
\vspace*{\fill}
\clearpage

\vspace*{\fill}
\begin{table}[!hp]
  \centering
  \renewcommand{\arraystretch}{1.3}
\begin{tabular}{%
      S[table-format=3.1, table-number-alignment=center,
      round-mode=places, round-precision=1]|
      S[table-format=1.5, table-number-alignment=center,
      round-mode=figures, round-precision=2]|
      S[table-format=1.5, table-number-alignment=left,
      round-mode=figures, round-precision=3]@{$\,\pm\,$}
      S[table-format=1.5, table-number-alignment=left,
      round-mode=figures, round-precision=2]@{\ }l@{\ }l} 
    \toprule

    {\Qsq} & {$x$} & \multicolumn{4}{c}{\Ftwob}\\
    {(\si{\GeV^2})} & &
    \multicolumn{1}{c}{\mbox{}} & \multicolumn{1}{c}{\stat} & \multicolumn{1}{c}{\syst} & 
    \multicolumn{1}{c}{\extr}\\\hline
    
        6.5 & 0.00015 & 0.00430744 & 0.000678236 & \numpmerrp{+0.000535187}{-0.000484483}{5} & \numpmerrp{+0.000340672}{-0.000292465}{5} \\
        6.5 & 0.00028 & 0.00357393 & 0.000515144 & \numpmerrp{+0.000361609}{-0.000329013}{5} & \numpmerrp{+0.000290839}{-0.000250723}{5} \\
        12  & 0.00043 & 0.00632397 & 0.000689536 & \numpmerrp{+0.000616069}{-0.000522916}{5} & \numpmerrp{+0.000440387}{-0.000343815}{5} \\
        12  & 0.00065 & 0.00438118 & 0.000527575 & \numpmerrp{+0.000465704}{-0.000425302}{5} & \numpmerrp{+0.000201543}{-0.000123445}{5} \\ 
        \hline
        
        25  & 0.00043 & 0.0118218  & 0.00352965  & \numpmerrp{+0.00155224}{-0.00156407}{4}   & \numpmerrp{+0.000903002}{-0.0006563}{4} \\
        25  & 0.0008  & 0.0104674  & 0.00106653  & \numpmerrp{+0.000781736}{-0.000746528}{4} & \numpmerrp{+0.00060856}{-0.000491746}{4} \\
\bottomrule
  \end{tabular}
  \caption[The structure function \Ftwob as a function of $x$]{%
      The structure function \Ftwob as a function of $x$ for seven different
    values of \Qsq. The first error is statistical, the
    second systematic (not including the error on the integrated luminosity) and the last is the extrapolation
    uncertainty.
    The horizontal lines correspond to the bins in \Qsq in Table~\protect\ref{tab:ddiffb}.}
  \label{tab:f2b}
\end{table}
\vspace*{\fill}
\clearpage

\vspace*{\fill}
\begin{table}[!hp]
  \centering
  \renewcommand{\arraystretch}{1.3}
  \begin{tabular}{%
      S[table-format=3.1, table-number-alignment=center,
      round-mode=places, round-precision=1]|
      S[table-format=1.5, table-number-alignment=center,
      round-mode=figures, round-precision=2]|
      S[table-format=1.3, table-number-alignment=left,
      round-mode=figures, round-precision=3]@{$\,\pm\,$}
      S[table-format=1.3, table-number-alignment=left,
      round-mode=figures, round-precision=2]@{\ }l@{\ }l}
    \toprule

    {\Qsq} & {$x$} & \multicolumn{4}{c}{\Ftwoc}\\
    {(\si{\GeV^2})} & &
    \multicolumn{1}{c}{\mbox{}} & \multicolumn{1}{c}{\stat} & \multicolumn{1}{c}{\syst} & 
    \multicolumn{1}{c}{\extr}\\\hline
    
        6.5 & 0.00015 & 0.202358 & 0.0164061 & \numpmerrp{+0.0140083}{-0.0112986}{3} & \numpmerrp{+0.0464721}{-0.0420341}{3} \\
        6.5 & 0.00028 & 0.188867 & 0.0133164 & \numpmerrp{+0.0125261}{-0.0102684}{3} & \numpmerrp{+0.0389445}{-0.0385403}{3} \\
        12  & 0.00043 & 0.230733 & 0.0132937 & \numpmerrp{+0.0151488}{-0.0132203}{3} & \numpmerrp{+0.0388088}{-0.0440923}{3} \\
        12  & 0.00065 & 0.224027 & {\num[round-precision=1]{0.00771815}}& \numpmerrp{+0.0138179}{-0.0120088}{3} & \numpmerrp{+0.0284544}{-0.0282945}{3} \\
        \hline
        
        25  & 0.00043 & 0.492397 & 0.0710624 & \numpmerrp{+0.0441621}{-0.0360253}{3} & \numpmerrp{+0.0854343}{-0.0732749}{3} \\
        25  & 0.0008  & 0.417956 & 0.018106  & \numpmerrp{+0.0220703}{-0.0196669}{3} & \numpmerrp{+0.0295934}{-0.0340291}{3} \\
\bottomrule
  \end{tabular}
  \caption[The structure function \Ftwoc as a function of $x$]{The structure function \Ftwoc as a function of $x$ for seven different
    values of \Qsq. The first error is statistical, the
    second systematic (not including the error on the integrated luminosity) and the last is the extrapolation
    uncertainty.
    The horizontal lines correspond to the bins in \Qsq in Table~\protect\ref{tab:ddiffc}.}      
  \label{tab:f2c}
\end{table}
\vspace*{\fill}
\clearpage

\vspace*{\fill}
\begin{table}[!hp]
  \centering
  \renewcommand{\arraystretch}{1.3}
  \begin{tabular}{%
      S[table-format=3.1, table-number-alignment=center,
      round-mode=places, round-precision=1]|
      S[table-format=1.5, table-number-alignment=center,
      round-mode=figures, round-precision=2]|
      S[table-format=1.5, table-number-alignment=left,
      round-mode=figures, round-precision=3]@{$\,\pm\,$}
      S[table-format=1.5, table-number-alignment=left,
      round-mode=figures, round-precision=2]@{\ }l@{\ }l}
    \toprule

    {\Qsq} & {$x$} & \multicolumn{4}{c}{\sigredb}\\
    {(\si{\GeV^2})} & &
    \multicolumn{1}{c}{\mbox{}} & \multicolumn{1}{c}{\stat} & \multicolumn{1}{c}{\syst} & 
    \multicolumn{1}{c}{\extr}\\\hline
    
        6.5 & 0.00015 & 0.00429463 & 0.000676218 & \numpmerrp{+0.000533595}{-0.000483041}{5} & \numpmerrp{+0.000335938}{-0.000290441}{5} \\
        6.5 & 0.00028 & 0.00357045 & 0.000514642 & \numpmerrp{+0.000361257}{-0.000328693}{5} & \numpmerrp{+0.000289854}{-0.000248819}{5} \\
        12 & 0.00043 & 0.00631148 & 0.000688174 & \numpmerrp{+0.000614852}{-0.000521883}{5} & \numpmerrp{+0.00043273}{-0.000340334}{5} \\
        12 & 0.00065 & 0.00435843 & 0.000524836 & \numpmerrp{+0.000463286}{-0.000423094}{5} & \numpmerrp{+0.000229076}{-0.0000994433}{5} \\
        \hline
        
        25 & 0.00043 & 0.0115863 & 0.00345935 & \numpmerrp{+0.00152132}{-0.00153292}{4} & \numpmerrp{+0.00087818}{-0.000639322}{4} \\
        25 & 0.0008 & 0.0104247 & 0.00106218 & \numpmerrp{+0.000778544}{-0.00074348}{4} & \numpmerrp{+0.000595703}{-0.000504791}{4} \\
\bottomrule
  \end{tabular}
  \caption{Reduced beauty cross sections, \sigredb, as a function of $x$ for seven different
    values of \Qsq. 
    For more details, see the caption of Table~\protect\ref{tab:f2b}.}
  \label{tab:reduced_beauty}
\end{table}
\vspace*{\fill}
\clearpage

\vspace*{\fill}
\begin{table}[!hp]
  \centering
  \renewcommand{\arraystretch}{1.3}
  \begin{tabular}{%
      S[table-format=3.1, table-number-alignment=center,
      round-mode=places, round-precision=1]|
      S[table-format=1.5, table-number-alignment=center,
      round-mode=figures, round-precision=2]|
      S[table-format=1.3, table-number-alignment=left,
      round-mode=figures, round-precision=3]@{$\,\pm\,$}
      S[table-format=1.3, table-number-alignment=left,
      round-mode=figures, round-precision=2]@{\ }l@{\ }l}
    \toprule

    {\Qsq} & {$x$} & \multicolumn{4}{c}{\sigredc}\\
    {(\si{\GeV^2})} & &
    \multicolumn{1}{c}{\mbox{}} & \multicolumn{1}{c}{\stat} & \multicolumn{1}{c}{\syst} & 
    \multicolumn{1}{c}{\extr}\\\hline

        6.5 & 0.00015 & 0.201019 & 0.0162976 & \numpmerrp{+0.0139156}{-0.0112239}{3} & \numpmerrp{+0.0408899}{-0.041564}{3} \\
        6.5 & 0.00028 & 0.188081 & 0.013261  & \numpmerrp{+0.012474}{-0.0102257}{3}  & \numpmerrp{+0.03988}{-0.0420303}{3} \\
        12  & 0.00043 & 0.229988 & 0.0132507 & \numpmerrp{+0.0150998}{-0.0131776}{3} & \numpmerrp{+0.0368009}{-0.0466139}{3} \\
        12  & 0.00065 & 0.22384  & {\num[round-precision=1]{0.00771169}}& \numpmerrp{+0.0138063}{-0.0119988}{3} & \numpmerrp{+0.028382}{-0.0339199}{3} \\
        \hline
        
        25  & 0.00043 & 0.465047 & 0.0671153 & \numpmerrp{+0.0417092}{-0.0340243}{3} & \numpmerrp{+0.0806689}{-0.0674669}{3} \\
        25  & 0.0008  & 0.413074 & 0.0178945 & \numpmerrp{+0.0218125}{-0.0194371}{3} & \numpmerrp{+0.0264627}{-0.0348912}{3} \\
\bottomrule
  \end{tabular}
  \caption{Reduced charm cross section, \sigredc, as a function of $x$
    for seven different values of \Qsq. 
    For more details, see the caption of Table~\protect\ref{tab:f2c}.}
  \label{tab:reduced_charm}
\end{table}
\vspace*{\fill}
\clearpage

\begin{sidewaystable}[!hp]
  \renewcommand{\arraystretch}{1.3}
  \centering
  \footnotesize
  \begin{tabular}{%
    S[table-format=3.0, table-number-alignment=right,
    round-mode=figures, round-precision=2]@{$\,:\,$}
    S[table-format=4.0, table-number-alignment=left,
    round-mode=figures, round-precision=2]|
    S[table-format=1.5, table-number-alignment=right,round-mode=off]@{$\,:\,$}
    S[table-format=1.4, table-number-alignment=left, round-mode=off]|
    S[table-format=+2.1, table-number-alignment=center,
    round-mode=places, round-precision=1]|
    S[table-format=+2.1, table-number-alignment=center,
    round-mode=places, round-precision=1]|
    S[table-format=+2.1, table-number-alignment=center,
    round-mode=places, round-precision=1]|
    S[table-format=+2.1, table-number-alignment=center,
    round-mode=places, round-precision=1]|
    S[table-format=+2.1, table-number-alignment=center,
    round-mode=places, round-precision=1]|
    S[table-format=+2.1, table-number-alignment=center,
    round-mode=places, round-precision=1]|
    S[table-format=+2.1, table-number-alignment=center,
    round-mode=places, round-precision=1]|
    S[table-format=+2.1, table-number-alignment=center,
    round-mode=places, round-precision=1]|
    S[table-format=+2.1, table-number-alignment=center,
    round-mode=places, round-precision=1]|
    S[table-format=+2.1, table-number-alignment=center,
    round-mode=places, round-precision=1]
    } \toprule

\multicolumn{2}{c|}{\Qsq} & \multicolumn{2}{c|}{$x$} & $\delta_{1}^{E_e}$ & $\delta_{1}^{y}$ & $\delta_{1}^{E-p_Z}$ & $\delta_{2}$ & $\delta_{3}$ & $\delta_{4}^\text{core}$ & $\delta_{4}^\text{tail}$ & $\delta_{5}$ & $\delta_{6}$ & $\delta_{7}$ \\
\multicolumn{2}{c|}{(\gev$^{2}$)} & \multicolumn{2}{c|}{} & {(\%)} & {(\%)} & {(\%)} & {(\%)} & {(\%)} & {(\%)} & {(\%)} & {(\%)} & {(\%)} & {(\%)} \\
\bottomrule
  \end{tabular}
  \caption{Systematic uncertainties for the double-differential cross sections of inclusive jet production in beauty events.
	 See Section~\protect\ref{sec:syst} for more details.}
    \label{tab:beauty_systematics_part1}
\end{sidewaystable}
\clearpage

\begin{sidewaystable}[!hp]
  \renewcommand{\arraystretch}{1.3}
  \centering
  \footnotesize
  \begin{tabular}{%
    S[table-format=3.0, table-number-alignment=right,
    round-mode=figures, round-precision=2]@{$\,:\,$}
    S[table-format=4.0, table-number-alignment=left,
    round-mode=figures, round-precision=2]|
    S[table-format=1.5, table-number-alignment=right, round-mode=off]@{$\,:\,$}
    S[table-format=1.4, table-number-alignment=left,  round-mode=off]|
    S[table-format=+2.1, table-number-alignment=center,
    round-mode=places, round-precision=1]|
    S[table-format=+2.1, table-number-alignment=center,
    round-mode=places, round-precision=1]|
    S[table-format=+2.1, table-number-alignment=center,
    round-mode=places, round-precision=1]|
    S[table-format=+2.1, table-number-alignment=center,
    round-mode=places, round-precision=1]|
    S[table-format=+2.1, table-number-alignment=center,
    round-mode=places, round-precision=1]|
    S[table-format=+2.1, table-number-alignment=center,
    round-mode=places, round-precision=1]|
    S[table-format=+2.1, table-number-alignment=center,
    round-mode=places, round-precision=1]|
    S[table-format=+2.1, table-number-alignment=center,
    round-mode=places, round-precision=1]|
    S[table-format=+2.1, table-number-alignment=center,
    round-mode=places, round-precision=1]|
    S[table-format=+2.1, table-number-alignment=center,
    round-mode=places, round-precision=1]|
    S[table-format=+2.1, table-number-alignment=center,
    round-mode=places, round-precision=1]
    }
  \toprule

\multicolumn{2}{c|}{\Qsq} & \multicolumn{2}{c|}{$x$} & $\delta_{8}^{Q^2, c}$ & $\delta_{8}^{Q^2, b}$ & $\delta_{8}^{\etajet, c}$ & $\delta_{8}^{\etajet, b}$ & $\delta_{8}^{\ETjet, c}$ & $\delta_{8}^{\ETjet, b}$ & $\delta_{9}$ & $\delta_{10}$ & $\delta_{11}$ & $\delta_{12}^\text{BR}$ & $\delta_{12}^\text{frag}$ \\
\multicolumn{2}{c|}{(\si{\GeV^{2}})} & \multicolumn{2}{c|}{} & {(\%)} & {(\%)} & {(\%)} & {(\%)} & {(\%)} & {(\%)} & {(\%)} & {(\%)} & {(\%)} & {(\%)} & {(\%)} \\
\bottomrule
  \end{tabular}
  \caption{Systematic uncertainties for the double-differential cross sections of inclusive jet production in beauty events (continued).}
  \label{tab:beauty_systematics_part2}
\end{sidewaystable}
\clearpage

\begin{sidewaystable}[!hp]
  \renewcommand{\arraystretch}{1.3}
  \centering
  \footnotesize
  \begin{tabular}{%
    S[table-format=3.0, table-number-alignment=right,
    round-mode=figures, round-precision=2]@{$\,:\,$}
    S[table-format=4.0, table-number-alignment=left,
    round-mode=figures, round-precision=2]|
    S[table-format=1.5, table-number-alignment=right, round-mode=off]@{$\,:\,$}
    S[table-format=1.4, table-number-alignment=left,  round-mode=off]|
	S[table-format=+2.1, table-number-alignment=center,
    round-mode=places, round-precision=1]|
    S[table-format=+2.1, table-number-alignment=center,
    round-mode=places, round-precision=1]|
    S[table-format=+2.1, table-number-alignment=center,
    round-mode=places, round-precision=1]|
    S[table-format=+2.1, table-number-alignment=center,
    round-mode=places, round-precision=1]|
    S[table-format=+2.1, table-number-alignment=center,
    round-mode=places, round-precision=1]|
    S[table-format=+2.1, table-number-alignment=center,
    round-mode=places, round-precision=1]|
	S[table-format=+2.1, table-number-alignment=center,
    round-mode=places, round-precision=1]|
    S[table-format=+2.1, table-number-alignment=center,
    round-mode=places, round-precision=1]|
    S[table-format=+2.1, table-number-alignment=center,
    round-mode=places, round-precision=1]|
    S[table-format=+2.1, table-number-alignment=center,
    round-mode=places, round-precision=1]
    }
  \toprule

\multicolumn{2}{c|}{\Qsq} & \multicolumn{2}{c|}{$x$} & $\delta_{1}^{E_e}$ & $\delta_{1}^{y}$ & $\delta_{1}^{E-p_Z}$ & $\delta_{2}$ & $\delta_{3}$ & $\delta_{4}^\text{core}$ & $\delta_{4}^\text{tail}$ & $\delta_{5}$ & $\delta_{6}$ & $\delta_{7}$ \\
\multicolumn{2}{c|}{(\gev$^{2}$)} & \multicolumn{2}{c|}{} & {(\%)} & {(\%)} & {(\%)} & {(\%)} & {(\%)} & {(\%)} & {(\%)} & {(\%)} & {(\%)} & {(\%)} \\
\bottomrule
  \end{tabular}
  \caption{Systematic uncertainties for the double-differential cross sections of inclusive jet production in charm events.
	 See Section~\protect\ref{sec:syst} for more details.}
    \label{tab:charm_systematics_part1}
\end{sidewaystable}
\clearpage

\begin{sidewaystable}[!hp]
  \renewcommand{\arraystretch}{1.3}
  \centering
  \footnotesize
  \begin{tabular}{%
    S[table-format=3.0, table-number-alignment=right,
    round-mode=figures, round-precision=2]@{$\,:\,$}
    S[table-format=4.0, table-number-alignment=left,
    round-mode=figures, round-precision=2]|
    S[table-format=1.5, table-number-alignment=right, round-mode=off]@{$\,:\,$}
    S[table-format=1.4, table-number-alignment=left, round-mode=off]|
	S[table-format=+2.1, table-number-alignment=center,
    round-mode=places, round-precision=1]|
    S[table-format=+2.1, table-number-alignment=center,
    round-mode=places, round-precision=1]|
    S[table-format=+2.1, table-number-alignment=center,
    round-mode=places, round-precision=1]|
    S[table-format=+2.1, table-number-alignment=center,
    round-mode=places, round-precision=1]|
    S[table-format=+2.1, table-number-alignment=center,
    round-mode=places, round-precision=1]|
    S[table-format=+2.1, table-number-alignment=center,
    round-mode=places, round-precision=1]|
    S[table-format=+2.1, table-number-alignment=center,
    round-mode=places, round-precision=1]|
	S[table-format=+2.1, table-number-alignment=center,
    round-mode=places, round-precision=1]|
    S[table-format=+2.1, table-number-alignment=center,
    round-mode=places, round-precision=1]|
    S[table-format=+2.1, table-number-alignment=center,
    round-mode=places, round-precision=1]|
    S[table-format=+2.1, table-number-alignment=center,
    round-mode=places, round-precision=1]
    }
  \toprule

\multicolumn{2}{c|}{\Qsq} & \multicolumn{2}{c|}{$x$} & $\delta_{8}^{Q^2, c}$ & $\delta_{8}^{Q^2, b}$ & $\delta_{8}^{\etajet, c}$ & $\delta_{8}^{\etajet, b}$ & $\delta_{8}^{\ETjet, c}$ & $\delta_{8}^{\ETjet, b}$ & $\delta_{9}$ & $\delta_{10}$ & $\delta_{11}$ & $\delta_{12}^\text{BR}$ & $\delta_{12}^\text{frag}$ \\
\multicolumn{2}{c|}{(\gev$^{2}$)} & \multicolumn{2}{c|}{} & {(\%)} & {(\%)} & {(\%)} & {(\%)} & {(\%)} & {(\%)} & {(\%)} & {(\%)} & {(\%)} & {(\%)} & {(\%)} \\
\bottomrule

  \end{tabular}
  \caption{Systematic uncertainties for the double-differential cross sections of inclusive jet production in charm events (continued).}
  \label{tab:charm_systematics_part2}
\end{sidewaystable}
\clearpage

\vspace*{\fill}
\begin{table}[!hp]
  \sisetup{explicit-sign=+}
  \centering
  \begin{tabular}{%
    S[table-format=3.1, table-number-alignment=center,
    round-mode=places, round-precision=1, explicit-sign]|
    S[table-format=1.5, table-number-alignment=center,
    round-mode=figures, round-precision=2, explicit-sign]|
    S[table-format=+1.1, table-number-alignment=center,
    round-mode=places, round-precision=1]|
    S[table-format=+1.1, table-number-alignment=center,
    round-mode=places, round-precision=1]|
    S[table-format=+1.1, table-number-alignment=center,
    round-mode=places, round-precision=1]|
    S[table-format=+1.1, table-number-alignment=center,
    round-mode=places, round-precision=1]|
    S[table-format=+1.1, table-number-alignment=center,
    round-mode=places, round-precision=1]|
    S[table-format=+1.1, table-number-alignment=center,
    round-mode=places, round-precision=1]}
  \toprule
      {\Qsq} & {$x$} &
	  {$\delta_{m_b}^{-}$} & {$\delta_{m_b}^{+}$} &
	  {$\delta_{\mu_R,\ \mu_F}^{-}$} & {$\delta_{\mu_R,\ \mu_F}^{+}$} &
	  {$\delta_{\alphaS}^{-}$} & {$\delta_{\alphaS}^{+}$} \\
      {(\si{\GeV^2})} & &
      {(\%)} & {(\%)} &
      {(\%)} & {(\%)} &
      {(\%)} & {(\%)} \\ \hline
      
        6.5     &     0.00015     &     7.43059     &     -5.94581     &     -3.26442     &     2.61251     &     0.715634     &     -0.302697  \\
        6.5     &     0.00028     &     8.11916     &     -6.83313     &     -1.53345     &     0.539915     &     0.107623     &     -0.414323 \\
        12     &     0.00043     &     6.88805     &     -5.24729     &     -1.40096     &     0.984167     &     0.283346     &     -0.246746  \\
        12     &     0.00065     &     4.2479     &     -2.70881     &     -0.773957     &     1.73063     &     0.349571     &     -0.0483051  \\ \hline

        25     &     0.00043     &     7.24744     &     -5.10525     &     -2.16036     &     2.2643     &     0.832604     &     -0.299157    \\
\bottomrule
  \end{tabular}
  \caption{Extrapolation uncertainties on the structure function \Ftwob
	 due to the variations of the beauty-quark mass, $m_b$, factorisation and renormalisation scales, $\mu_F$ and $\mu_R$, and the strong coupling constant, $\alphaS$.
	 The plus (minus) superscript indicates the upward (downward) variation of the corresponding parameter.
	 See Section~\protect\ref{sec:F2q} for more details.}
  \label{tab:f2b_extrapolation}
\end{table}
\vspace*{\fill}
\clearpage

\vspace*{\fill}
\begin{table}[!hp]
  \sisetup{explicit-sign=+}
  \centering
  \begin{tabular}{%
    S[table-format=3.1, table-number-alignment=center,
    round-mode=places, round-precision=1, explicit-sign]|
    S[table-format=1.5, table-number-alignment=center,
    round-mode=figures, round-precision=2, explicit-sign]|
    S[table-format=+1.1, table-number-alignment=center,
    round-mode=places, round-precision=1]|
    S[table-format=+1.1, table-number-alignment=center,
    round-mode=places, round-precision=1]|
    S[table-format=+2.1, table-number-alignment=center,
    round-mode=places, round-precision=1]|
    S[table-format=+2.1, table-number-alignment=center,
    round-mode=places, round-precision=1]|
    S[table-format=+1.1, table-number-alignment=center,
    round-mode=places, round-precision=1]|
    S[table-format=+1.1, table-number-alignment=center,
    round-mode=places, round-precision=1]}
   \toprule
      {\Qsq} & {$x$} &
	  {$\delta_{m_c}^{-}$} & {$\delta_{m_c}^{+}$} &
	  {$\delta_{\mu_R,\ \mu_F}^{-}$} & {$\delta_{\mu_R,\ \mu_F}^{+}$} &
	  {$\delta_{\alphaS}^{-}$} & {$\delta_{\alphaS}^{+}$} \\
      {(\si{\GeV^2})} & &
      {(\%)} & {(\%)} &
      {(\%)} & {(\%)} &
      {(\%)} & {(\%)} \\ \hline
      
            6.5     &     0.00015     &     9.87088     &     -6.76149     &     -19.6149     &     20.1694     &     4.81277     &     -1.00943    \\
            6.5     &     0.00028     &     9.16771     &     -9.62738     &     -17.7376     &     17.9594     &     4.31278     &     -3.01642    \\
            12     &     0.00043     &     7.39135     &     -6.86637     &     -17.5081     &     14.762     &     3.21823     &     -3.39084      \\
            12     &     0.00065     &     6.13402     &     -4.55921     &     -11.4383     &     10.8443     &     2.46933     &     -2.80966     \\ \hline
            
            25     &     0.00043     &     8.13524     &     -5.14307     &     -13.9441     &     14.8731     &     3.69538     &     -0.750982    \\
            25     &     0.0008     &     5.74692     &     -4.56089     &     -6.69074     &     4.07448     &     0.71064     &     -0.849047     \\
            30     &     0.0016     &     4.5311     &     -3.28866     &     -1.3191     &     -0.46732     &     -0.347832     &     0.295415     \\
            30     &     0.0025     &     2.40882     &     -3.36868     &     0.241768     &     -0.985451     &     -0.396695     &     -0.349398 \\
            30     &     0.0045     &     3.46164     &     0.207118     &     3.4829     &     0.232366     &     1.59029     &     1.51483        \\ \hline
            
            80     &     0.0016     &     3.56007     &     -1.77163     &     -1.73886     &     2.5976     &     0.629206     &     0.971529      \\
            80     &     0.0025     &     0.446259     &     -3.10217     &     -2.06975     &     -1.82627     &     -1.35577     &     -0.966476  \\
            80     &     0.0045     &     1.54712     &     -1.43923     &     -0.200399     &     -0.872649     &     -0.154778     &     -0.388347        \\
            80     &     0.008     &     -0.820048     &     -0.21686     &     -0.381542     &     -0.312406     &     -0.370169     &     -0.313662       \\ \hline
            
            160     &     0.0035     &     2.14812     &     -1.2226     &     -0.708124     &     0.635112     &     -0.306922     &     -0.660581 \\
            160     &     0.008     &     1.15603     &     -1.43803     &     0.709554     &     0.335673     &     -0.14048     &     -0.477166   \\
            160     &     0.02     &     -0.184419     &     0.145076     &     0.346151     &     0.835832     &     -0.0356663     &     1.14152  \\ \hline
            
            600     &     0.013     &     2.04565     &     -0.473381     &     -0.822974     &     0.477169     &     1.04498     &     0.577128   \\
            600     &     0.035     &     1.62343     &     0.328268     &     1.72403     &     -1.1632     &     2.16275     &     1.17683        \\
  \bottomrule
  \end{tabular}
  \caption{Extrapolation uncertainties on the structure function \Ftwoc
	 due to the variations of the charm-quark mass, $m_c$, factorisation and renormalisation scales, $\mu_F$ and $\mu_R$, and the strong coupling constant, $\alphaS$.
	 The plus (minus) superscript indicates the upward (downward) variation of the corresponding parameter.
	 See Section~\protect\ref{sec:F2q} for more details.}
  \label{tab:f2c_extrapolation}
\end{table}
\vspace*{\fill}
\clearpage

\vspace*{\fill}
\begin{table}[!hp]
  \sisetup{explicit-sign=+}
  \centering
  \begin{tabular}{%
    S[table-format=3.1, table-number-alignment=center,
    round-mode=places, round-precision=1, explicit-sign]|
    S[table-format=1.5, table-number-alignment=center,
    round-mode=figures, round-precision=2, explicit-sign]|
    S[table-format=+1.1, table-number-alignment=center,
    round-mode=places, round-precision=1]|
    S[table-format=+1.1, table-number-alignment=center,
    round-mode=places, round-precision=1]|
    S[table-format=+1.1, table-number-alignment=center,
    round-mode=places, round-precision=1]|
    S[table-format=+1.1, table-number-alignment=center,
    round-mode=places, round-precision=1]|
    S[table-format=+1.1, table-number-alignment=center,
    round-mode=places, round-precision=1]|
    S[table-format=+1.1, table-number-alignment=center,
    round-mode=places, round-precision=1]}
   \toprule
      {\Qsq} & {$x$} &
	  {$\delta_{m_b}^{-}$} & {$\delta_{m_b}^{+}$} &
	  {$\delta_{\mu_R,\ \mu_F}^{-}$} & {$\delta_{\mu_R,\ \mu_F}^{+}$} &
	  {$\delta_{\alphaS}^{-}$} & {$\delta_{\alphaS}^{+}$} \\
      {(\si{\GeV^2})} & &
      {(\%)} & {(\%)} &
      {(\%)} & {(\%)} &
      {(\%)} & {(\%)} \\
    \hline
      
            6.5     &     0.00015     &     7.33207     &     -5.96266     &     -3.17507     &     2.6283     &     0.721669     &     -0.319711   \\
            6.5     &     0.00028     &     8.08955     &     -6.84506     &     -1.26585     &     0.599867     &     0.321886     &     -0.328021 \\
            12     &     0.00043     &     6.79013     &     -5.20528     &     -1.38448     &     0.928851     &     0.198496     &     -0.255146  \\
            12     &     0.00065     &     4.74142     &     -2.2489     &     -0.385072     &     2.09405     &     0.762771     &     0.42062     \\ \hline
            
            25     &     0.00043     &     7.14862     &     -4.85686     &     -2.59669     &     2.37277     &     0.84558     &     -0.339474    \\
            25     &     0.0008     &     5.68601     &     -4.78752     &     -0.712448     &     0.519028     &     0.231946     &     -0.140304  \\
            30     &     0.0016     &     3.95593     &     -4.3942     &     0.32883     &     -1.65394     &     -0.718249     &     -0.550062    \\
            30     &     0.0025     &     3.2495     &     -2.97012     &     1.28609     &     -1.41078     &     0.0716334     &     -0.1773      \\
            30     &     0.0045     &     1.70603     &     -3.2043     &     0.504723     &     -4.21743     &     -2.38601     &     -2.21499     \\  \hline
            
            80     &     0.0016     &     3.08081     &     -3.0215     &     -1.34764     &     -0.0235217     &     0.0583863     &     -0.0784318        \\
            80     &     0.0025     &     2.52179     &     -2.37699     &     -0.118207     &     -0.77897     &     0.238571     &     0.152491   \\
            80     &     0.0045     &     2.19448     &     -1.96857     &     1.04569     &     -1.19848     &     0.0460239     &     0.152808    \\  \hline
            
            160     &     0.0035     &     2.16878     &     -1.35222     &     -0.187221     &     -0.140078     &     0.252404     &     -0.309872        \\
            160     &     0.008     &     1.89428     &     -1.7277     &     1.11817     &     -1.1589     &     -0.0256237     &     0.0991864    \\
            160     &     0.02     &     0.547174     &     -0.157237     &     2.8807     &     -1.75179     &     0.454428     &     -0.048942    \\ \hline
            
            600     &     0.013     &     0.748125     &     -1.28492     &     0.0856054     &     -0.383244     &     0.257881     &     -0.53666 \\
            600     &     0.035     &     1.04894     &     0.350599     &     2.67377     &     -1.35645     &     0.220264     &     0.758995     \\
    \bottomrule
  \end{tabular}
  \caption{Extrapolation uncertainties on the reduced beauty cross
      section, \sigredb, due to the variations of the beauty-quark mass, $m_b$,
      factorisation and renormalisation scales, $\mu_F$ and $\mu_R$,
      and the strong coupling constant, $\alphaS$.  The plus (minus)
      superscript indicates the upward (downward) variation of the
      corresponding parameter.
      See Section~\protect\ref{sec:F2q} for more details.}
  \label{tab:redb_extrapolation}
\end{table}
\vspace*{\fill}
\clearpage

\vspace*{\fill}
\begin{table}[!hp]
  \sisetup{explicit-sign=+}
  \centering
  \begin{tabular}{%
    S[table-format=3.1, table-number-alignment=center,
    round-mode=places, round-precision=1, explicit-sign]|
    S[table-format=1.5, table-number-alignment=center,
    round-mode=figures, round-precision=2, explicit-sign]|
    S[table-format=+1.1, table-number-alignment=center,
    round-mode=places, round-precision=1]|
    S[table-format=+1.1, table-number-alignment=center,
    round-mode=places, round-precision=1]|
    S[table-format=+2.1, table-number-alignment=center,
    round-mode=places, round-precision=1]|
    S[table-format=+2.1, table-number-alignment=center,
    round-mode=places, round-precision=1]|
    S[table-format=+1.1, table-number-alignment=center,
    round-mode=places, round-precision=1]|
    S[table-format=+1.1, table-number-alignment=center,
    round-mode=places, round-precision=1]}
  \toprule
      {\Qsq} & {$x$} &
	  {$\delta_{m_c}^{-}$} & {$\delta_{m_c}^{+}$} &
	  {$\delta_{\mu_R,\ \mu_F}^{-}$} & {$\delta_{\mu_R,\ \mu_F}^{+}$} &
	  {$\delta_{\alphaS}^{-}$} & {$\delta_{\alphaS}^{+}$} \\
      {(\si{\GeV^2})} & &
      {(\%)} & {(\%)} &
      {(\%)} & {(\%)} &
      {(\%)} & {(\%)} \\
    \hline
      
            6.5     &     0.00015     &     8.27811     &     -7.68984     &     -19.0508     &     18.2428     &     3.52725     &     -2.33602    \\
            6.5     &     0.00028     &     9.25155     &     -9.04904     &     -20.2907     &     18.6718     &     3.91992     &     -2.40452    \\
            12     &     0.00043     &     6.9896     &     -7.46971     &     -18.552     &     14.2468     &     2.05304     &     -3.28919       \\
            12     &     0.00065     &     5.67716     &     -4.77421     &     -14.1091     &     11.2984     &     0.942025     &     -2.7882     \\ \hline
            
            25     &     0.00043     &     7.75467     &     -4.26609     &     -13.8602     &     15.064     &     3.72001     &     -0.404724     \\
            25     &     0.0008     &     4.96219     &     -4.8387     &     -6.8742     &     3.96483     &     0.834978     &     -0.824302      \\
            30     &     0.0016     &     4.54938     &     -3.9389     &     -1.17377     &     -0.838045     &     -0.664073     &     0.320086   \\
            30     &     0.0025     &     2.35411     &     -3.39268     &     0.467269     &     -1.00482     &     0.129683     &     0.174879    \\
            30     &     0.0045     &     3.04599     &     0.445584     &     4.00225     &     0.552566     &     1.94388     &     1.87864       \\ \hline
            
            80     &     0.0016     &     3.71503     &     -1.78295     &     -1.98451     &     2.75014     &     0.624086     &     1.0471       \\
            80     &     0.0025     &     0.572026     &     -3.2291     &     -2.08445     &     -1.82983     &     -1.47994     &     -0.917636   \\
            80     &     0.0045     &     1.50888     &     -1.84298     &     -0.0647948     &     -0.888826     &     -0.315624     &     -0.423189       \\
            80     &     0.008     &     -1.01894     &     -0.157212     &     -0.727645     &     -0.706972     &     -0.401992     &     -0.215065       \\ \hline
            
            160     &     0.0035     &     2.08481     &     -1.2635     &     -0.634899     &     0.566289     &     -0.412742     &     -0.668766 \\
            160     &     0.008     &     1.3105     &     -2.17267     &     0.644225     &     -0.269256     &     -0.489317     &     -0.678415  \\
            160     &     0.02     &     -0.210289     &     0.266511     &     0.523725     &     -0.220228     &     0.136801     &     0.655033  \\ \hline
            
            600     &     0.013     &     2.25162     &     -0.544077     &     -0.703781     &     0.492326     &     1.07154     &     0.730328   \\
            600     &     0.035     &     1.59124     &     0.296342     &     1.7181     &     -1.18627     &     1.58573     &     1.38341        \\
  \bottomrule
  \end{tabular}
  \caption{Extrapolation uncertainties on the reduced charm cross
      section, \sigredc, due to the variations of the charm-quark mass, $m_c$,
      factorisation and renormalisation scales, $\mu_F$ and $\mu_R$,
      and the strong coupling constant, $\alphaS$.  The plus (minus)
      superscript indicates the upward (downward) variation of the
      corresponding parameter.
      See Section~\protect\ref{sec:F2q} for more details.}
  \label{tab:redc_extrapolation}
\end{table}
\vspace*{\fill}
\clearpage

\vspace*{\fill}
\begin{table}[!hp]
  \renewcommand{\arraystretch}{1.3}
  \centering
  \begin{tabular}{l|c|c}
  	\hline
  	Parameter               &         Variation          &                     Uncertainty \\
  	                        &                            & \multicolumn{1}{c}{(\si{\GeV})} \\ \hline
  	\multicolumn{3}{c}{Fit uncertainty}                      \\ \hline
  	Total                   &      $\Delta\chi^2=1$      &              $^{+0.14}_{-0.14}$ \\ \hline
  	\multicolumn{3}{c}{Model uncertainty}                \\ \hline
  	$f_s$                   & $0.31^{+0.07}_{-0.08}$     &              $^{+0.00}_{-0.00}$ \\
  	$Q^2_{\text{min}}$      & $\num{3.5} \to \SI{5.0}{\GeV^2}$ &        $^{+0.00}_{-0.00}$ \\
  	$Q^2_0$                 & $\num{1.4} \to \SI{1.9}{\GeV^2}$ &        $^{+0.01}_{-0.01}$ \\
  	$\delta m_{\text{ext}}$ &          see text          &           {\scriptsize $-0.07$} \\ \hline
  	Total                   &                            &              $^{+0.01}_{-0.07}$ \\ \hline
    \multicolumn{3}{c}{PDF parameterisation uncertainty}  \\ \hline
  	$D_{u_v}$               &        free in fit         &           {\scriptsize $+0.03$} \\
  	$D_{\bar D}$            &        free in fit         &           {\scriptsize $+0.03$} \\
  	$D_{\bar U}$            &        free in fit         &           {\scriptsize $+0.02$} \\ \hline
  	Total                   &                            &              $^{+0.05}_{-0.00}$ \\ \hline
  	\multicolumn{3}{c}{Theory uncertainty}               \\ \hline
  	$m_c(m_c)$              & $\SI{1.26\pm 0.06}{\GeV}$  &              $^{+0.02}_{-0.02}$ \\
  	$\alphaS$              &      $0.105\pm 0.002$      &              $^{+0.02}_{-0.02}$ \\
  	$\mu$                   &  $\times 2,\ \times 1/2$   &              $^{+0.07}_{-0.04}$ \\ \hline
  	Total                   &                            &              $^{+0.08}_{-0.05}$ \\ \hline
  \end{tabular}
  \caption{List of uncertainties for the beauty-quark mass determination. A
    description of the uncertainties not explicitly mentioned in the
    text is given elsewhere~\protect\cite{epj.c73.2311}.}
  \label{tab:result}
\end{table}
\vspace*{\fill}
\clearpage


%% file: zeusF2bc-fig.tex
\setlength{\unitlength}{1.0mm}
\vspace*{\fill}
\begin{figure}[!hp]
  \centering
  \includegraphics[width=\textwidth]{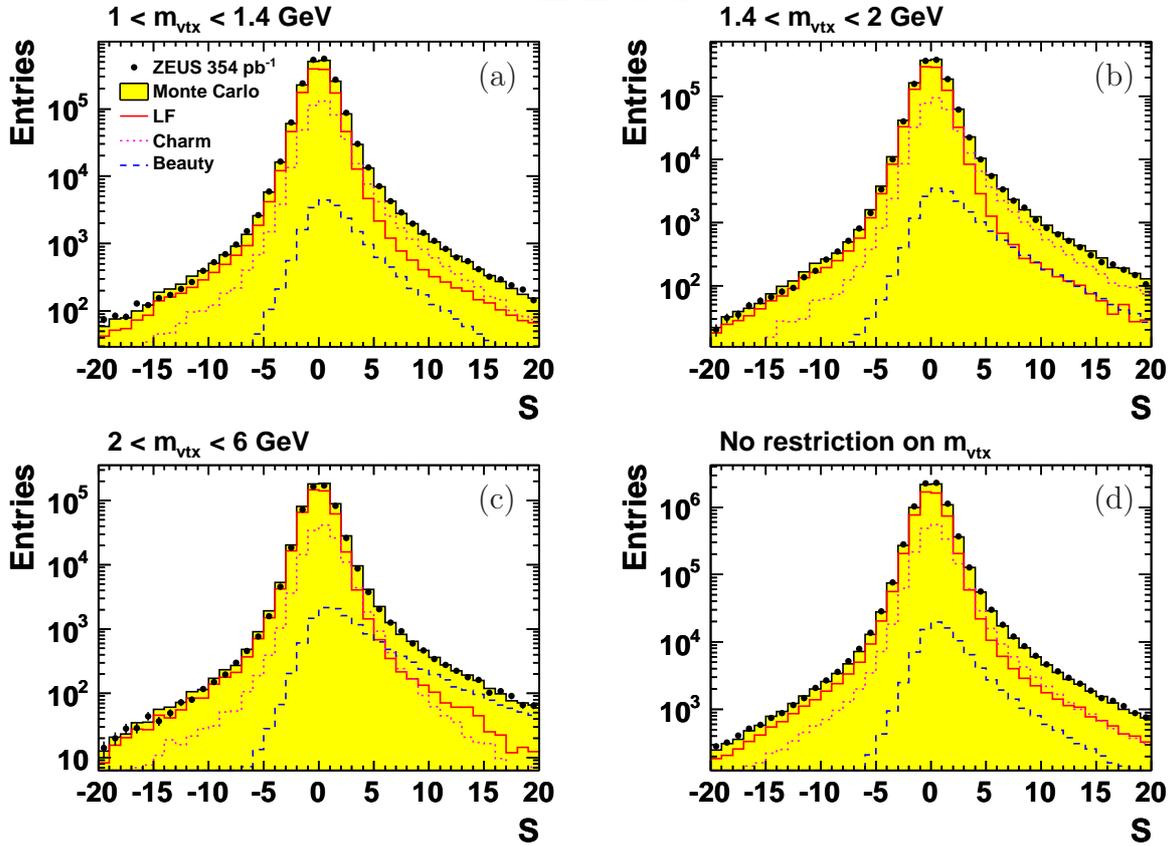}
  \put( -98,102){(a)}
  \put( -17,102){(b)}
  \put( -98, 46){(c)}
  \put( -17, 46){(d)}
  \caption{Distributions of the decay-length significance, $S$, for
    (a) $1 < \mvtx < \SI{1.4}{\GeV}$, (b) $1.4 < \mvtx < \SI{2}{\GeV}$,
    (c) $2 < \mvtx < \SI{6}{\GeV}$ and (d) no restriction on \mvtx. The data are
    compared to the sum of all MC distributions as well as the
    individual contributions from the beauty, charm and light-flavour (LF)
    MC subsamples. All samples were normalised according to the
    scaling factors obtained from the fit (see text).}
  \label{fig:significance}
\end{figure}
\vspace*{\fill}

\vspace*{\fill}
\begin{figure}[!hp]
  \centering
  \includegraphics[width=\textwidth]{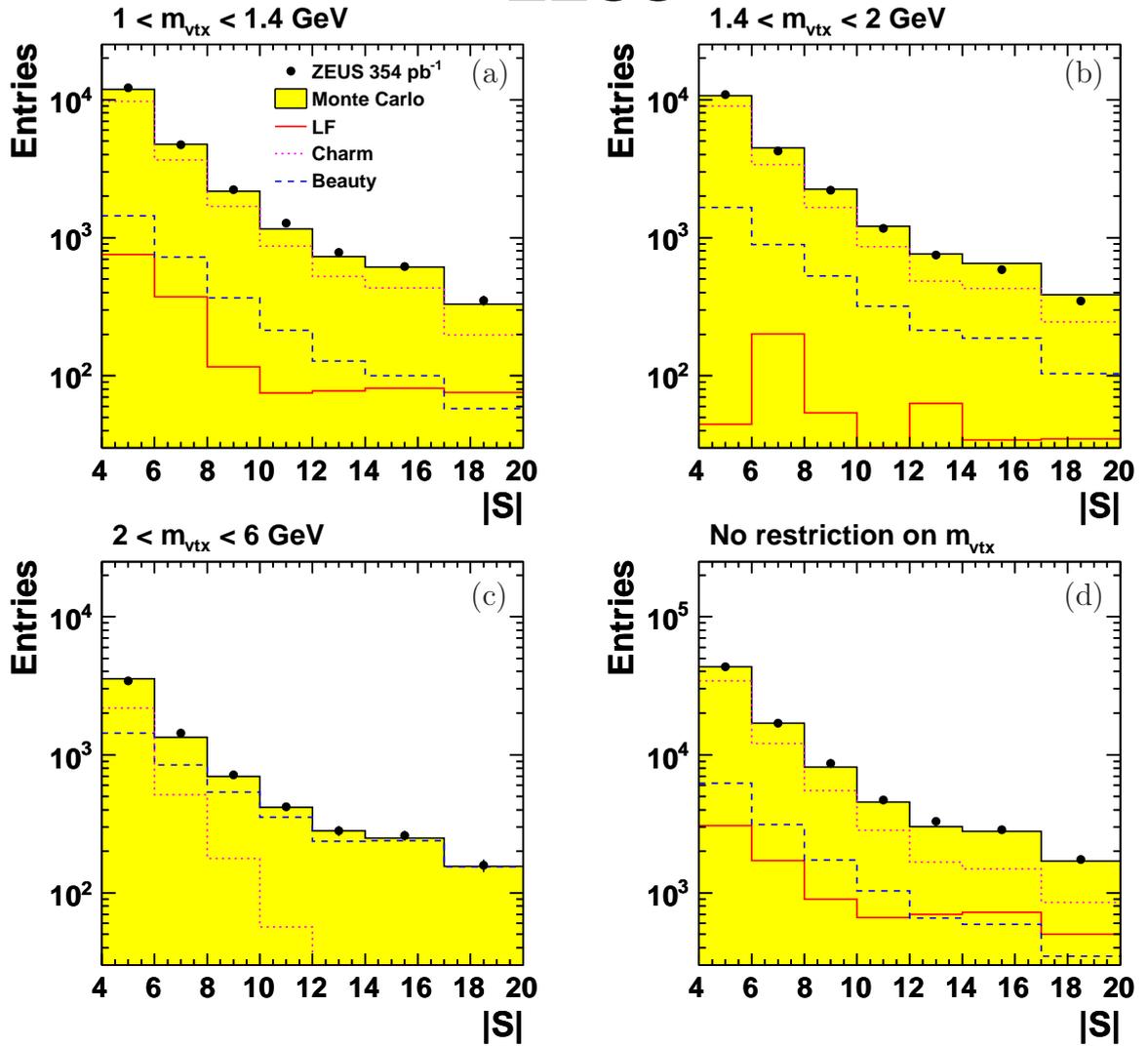}
  \put( -92,134){(a)}
  \put( -12,134){(b)}
  \put( -92, 63){(c)}
  \put( -12, 63){(d)}
  \caption{Distribution of the subtracted decay-length significance in
    four ranges of \mvtx.  
    For more details, see the caption of Fig.~\protect\ref{fig:significance}.}
  \label{fig:subtracted_sig}
\end{figure}
\vspace*{\fill}
\clearpage

\vspace*{\fill}
\begin{figure}[!hp]
  \centering
  \includegraphics[width=\textwidth]{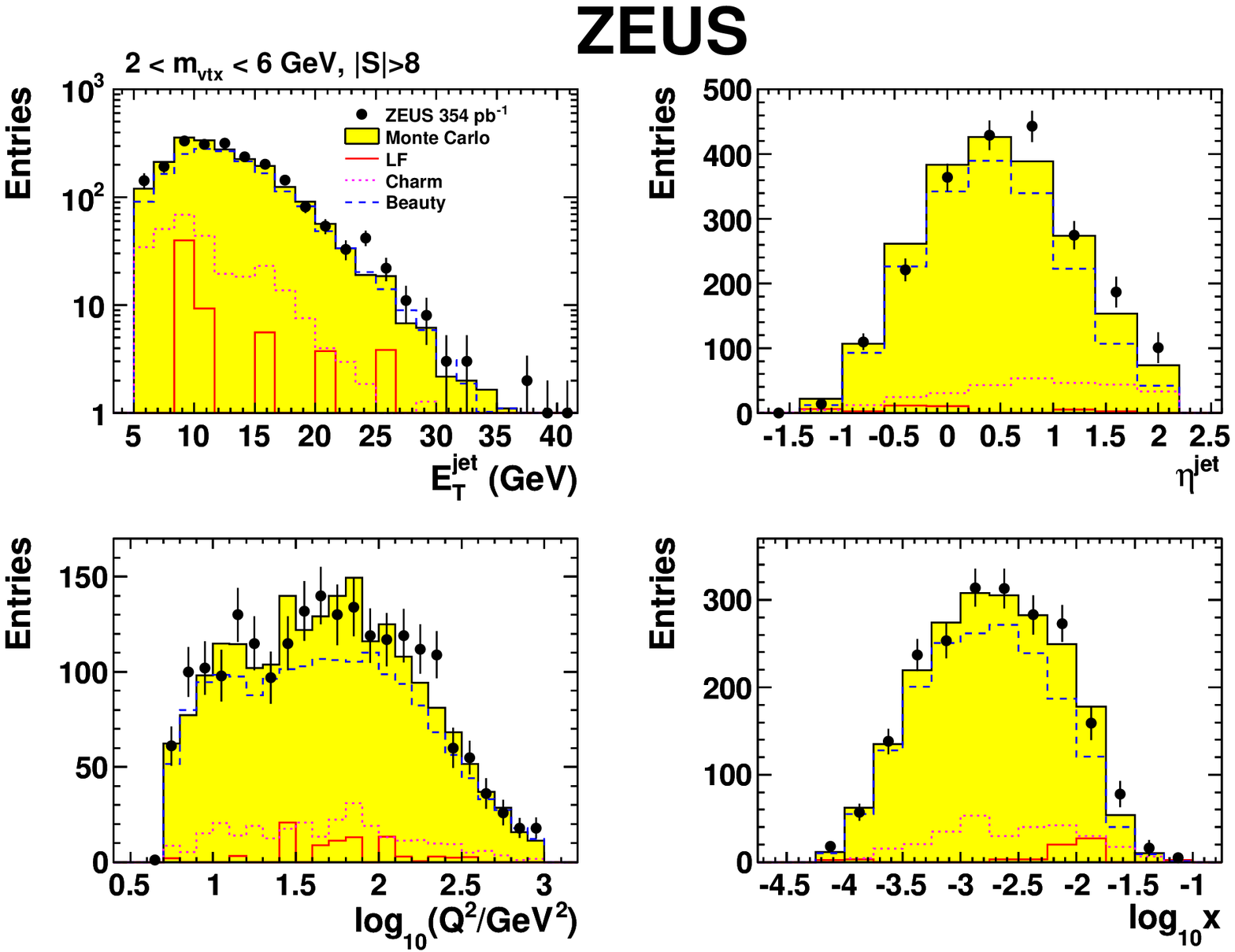}
  \put(-93,103){(a)}
  \put(-13,103){(b)}
  \put(-93, 46){(c)}
  \put(-13, 46){(d)}
  \caption{Distributions of (a) \ETjet, (b) \etajet, (c) $\log_{10}\Qsq$ and
    (d) $\log_{10}x$ of the selected secondary vertices for a
    beauty-enriched subsample with $2 < \mvtx < \SI{6}{\GeV}$ and
    $|S| > 8$. 
    For more details, see the caption of Fig.~\protect\ref{fig:significance}.}
  \label{fig:beauty_enriched}
\end{figure}
\vspace*{\fill}
\clearpage

\vspace*{\fill}
\begin{figure}[!hp]
  \centering
  \includegraphics[width=\textwidth]{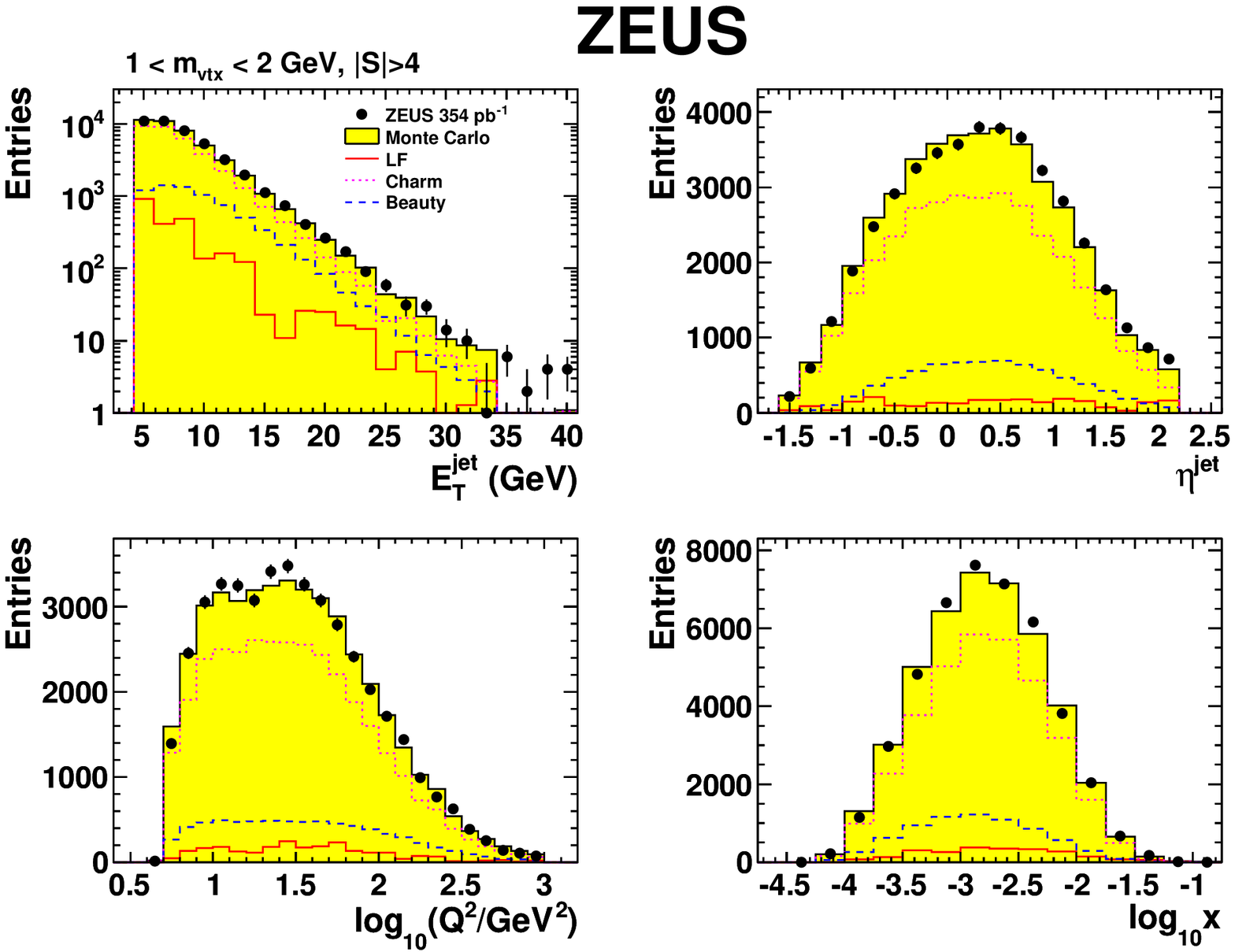}
  \put( -93,103){(a)}
  \put( -13,103){(b)}
  \put( -93, 46){(c)}
  \put( -13, 46){(d)}
  \caption{Distributions of (a) \ETjet, (b) \etajet, (c) $\log_{10}\Qsq$ and
    (d) $\log_{10}x$ of the selected secondary vertices for a
    charm-enriched subsample with $1 < \mvtx < \SI{2}{\GeV}$ and
    $|S| > 4$. 
    For more details, see the caption of Fig.~\protect\ref{fig:significance}.}
  \label{fig:charm_enriched}
\end{figure}
\vspace*{\fill}
\clearpage

\vspace*{\fill}
\begin{figure}[!hp]
  \centering
  \includegraphics[width=0.45\textwidth]{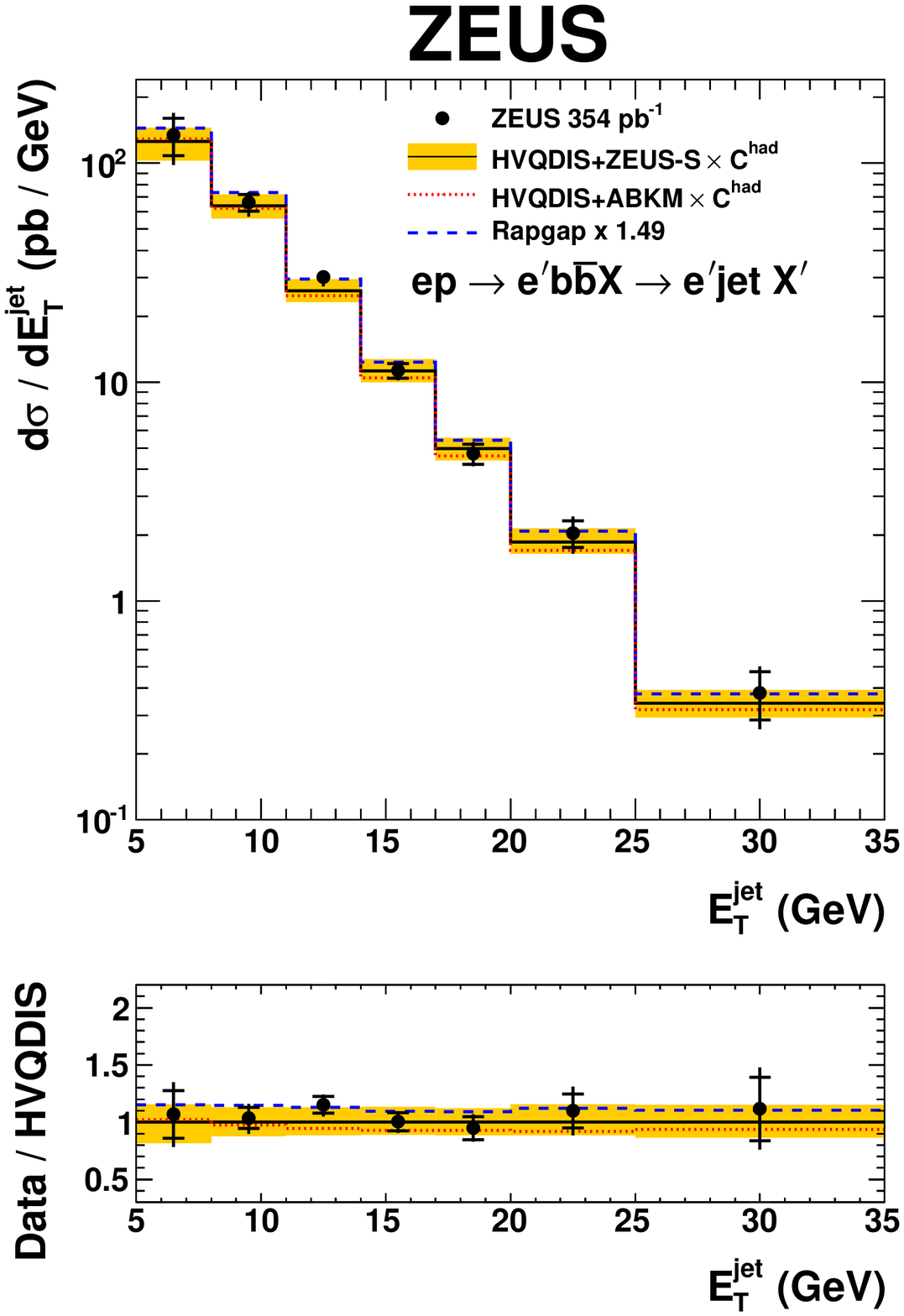}
  \hspace*{1cm}\includegraphics[width=0.45\textwidth]{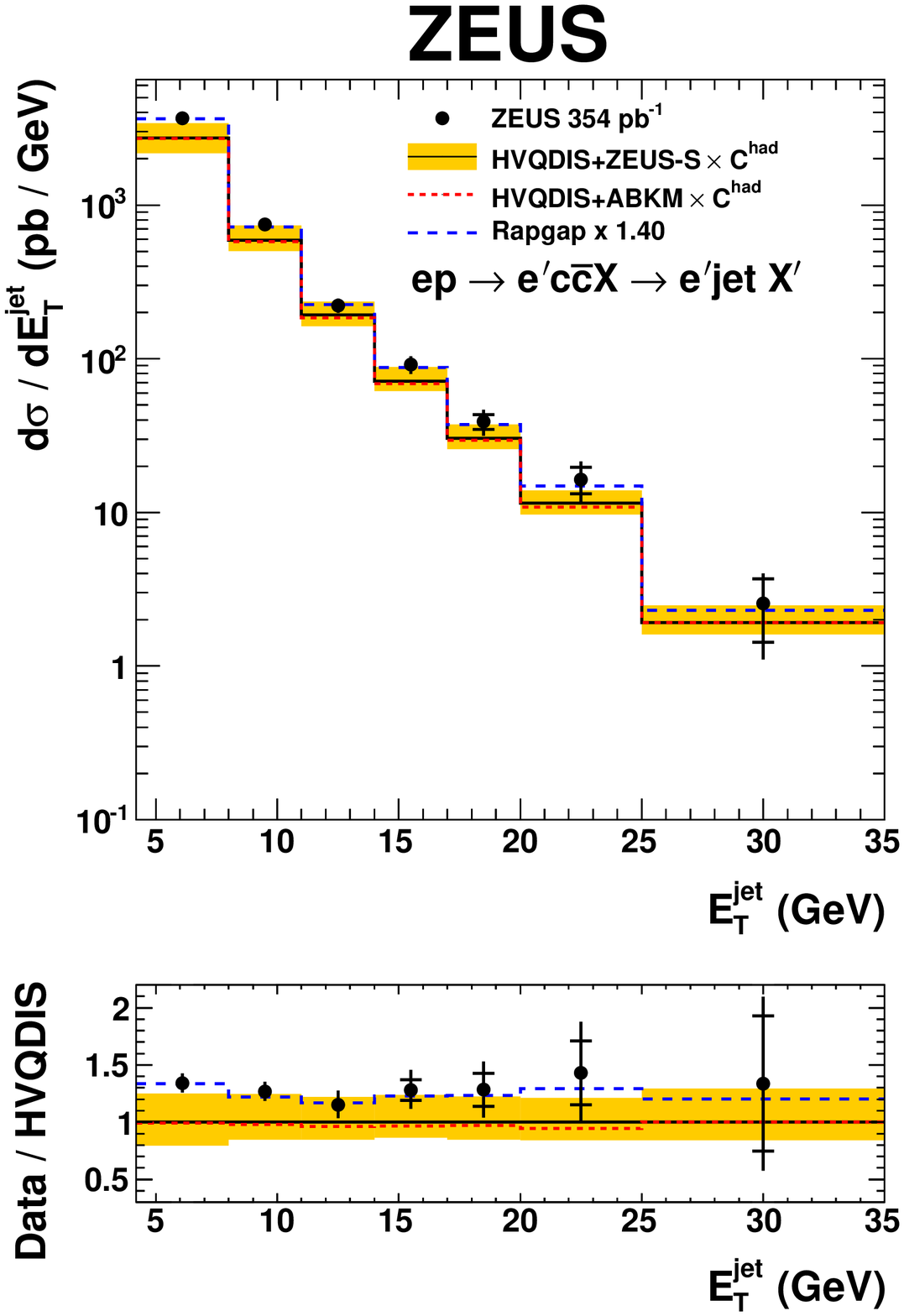}
  \put( -92, 95){(a)}
  \put(  -8, 95){(b)}
  \caption{Differential cross sections for inclusive jet production in 
      (a) beauty events and (b) charm events as a function of \ETjet.
    The cross sections are given for $5 < \Qsq < \SI{1000}{\GeV^{2}}$,
    $0.02 < y < 0.7$, $\ETjet > \SI[parse-numbers=false]{5 (4.2)}{\GeV}$ and $-1.6 < \etajet < 2.2$. 
    The data are shown as points. The inner error bars are the statistical
    uncertainties, while the outer error bars show the statistical and
    systematic uncertainties (not including the error on the integrated luminosity) added in quadrature. The solid line shows
    the \HVQDIS prediction with the ZEUS-S PDF, corrected for
    hadronisation effects, with the uncertainties
    indicated by the band; the dotted line shows the same prediction
    using the ABKM PDF; the dashed line shows the prediction
    from \RAPGAP scaled to match the measured integrated cross sections.}
  \label{fig:diff_et}
\end{figure}
\vspace*{\fill}
\clearpage

\vspace*{\fill}
\begin{figure}[!hp]
  \centering
  \includegraphics[width=0.45\textwidth]{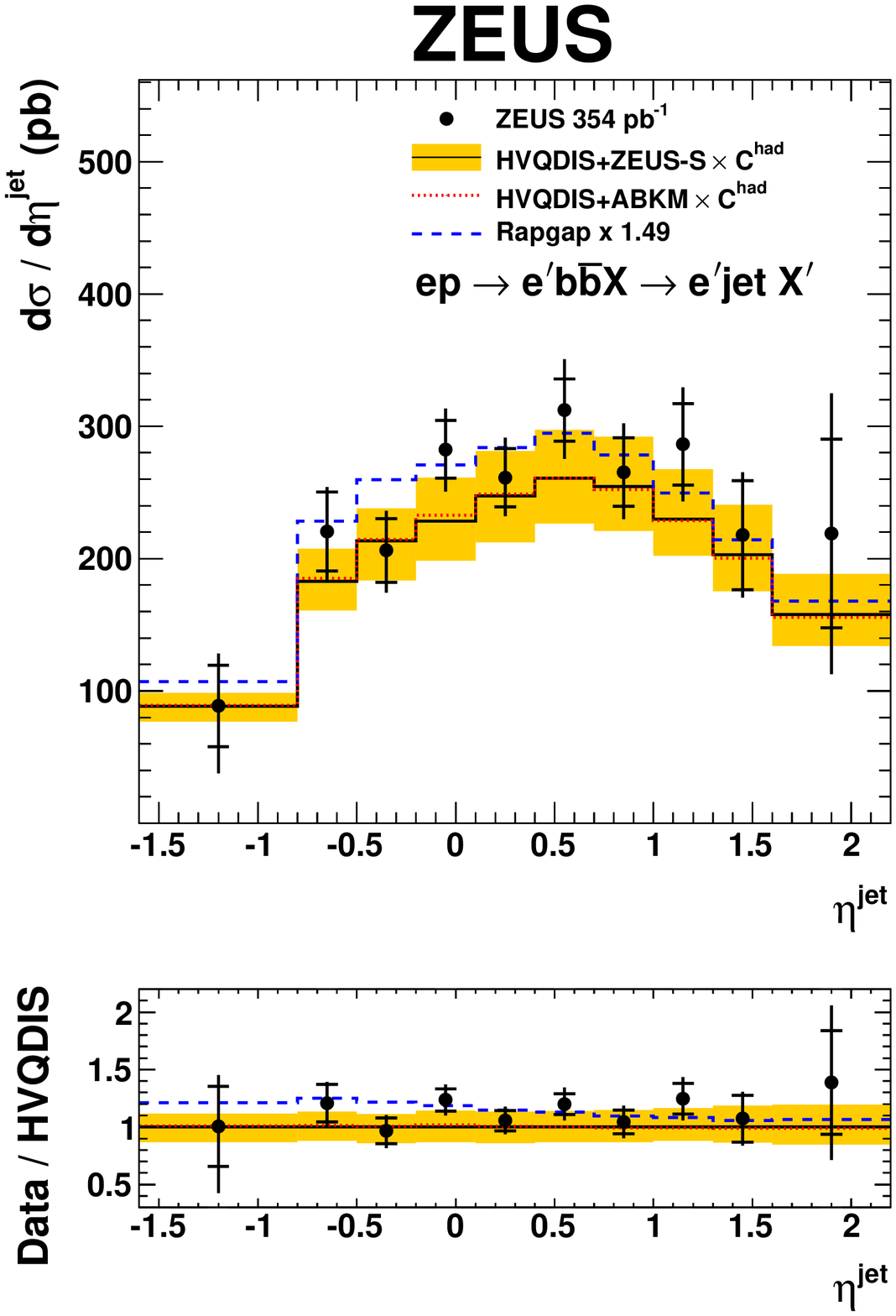}
  \hspace*{1cm}\includegraphics[width=0.45\textwidth]{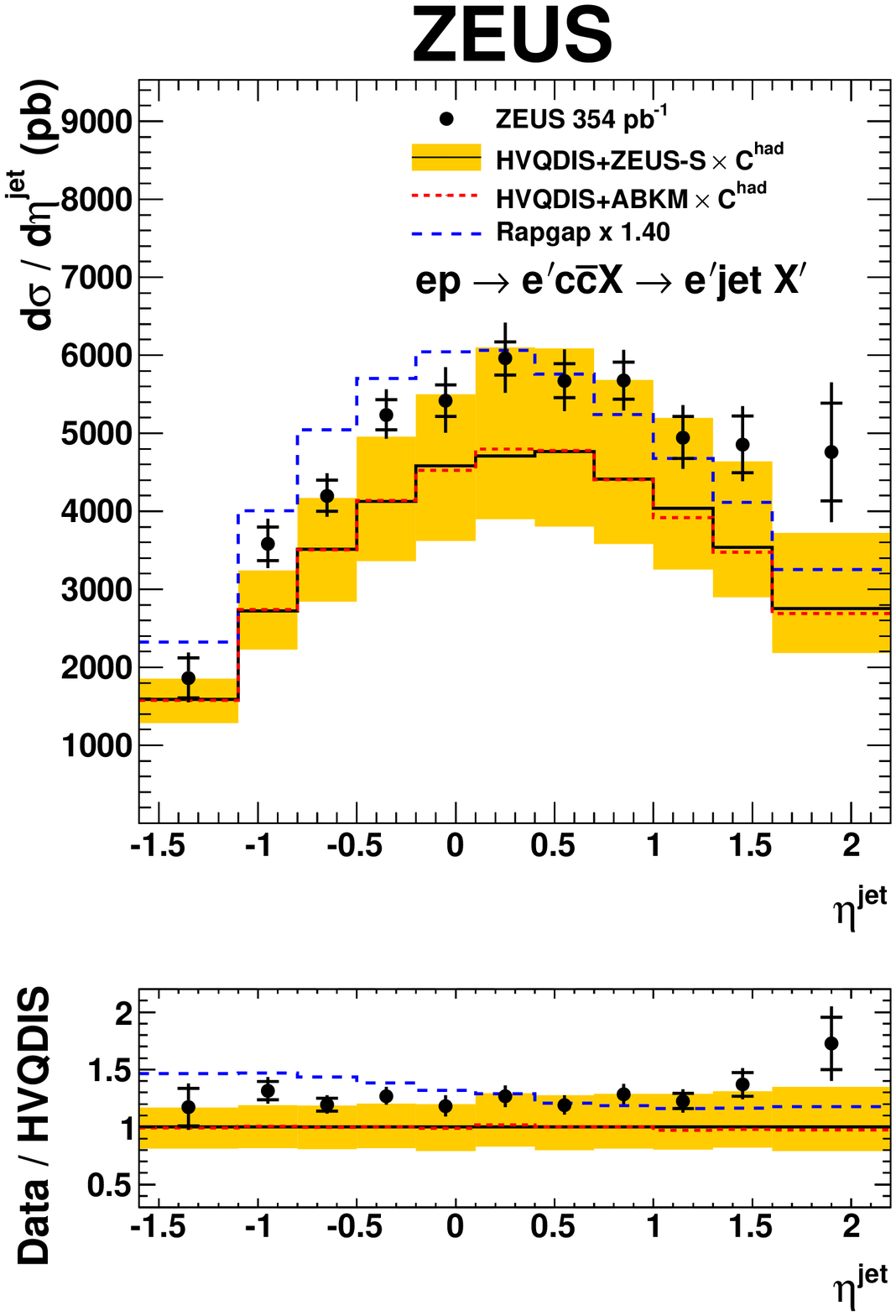}
  \put( -91, 95){(a)}
  \put(  -8, 95){(b)}
  \caption{Differential cross sections for inclusive jet production in 
    (a) beauty events and (b) charm events as a function of \etajet.
    For more details, see the caption of Fig.~\ref{fig:diff_et}.}
  \label{fig:diff_eta}
\end{figure}
\vspace*{\fill}
\clearpage

\vspace*{\fill}
\begin{figure}[!hp]
  \centering
  \includegraphics[width=0.45\textwidth]{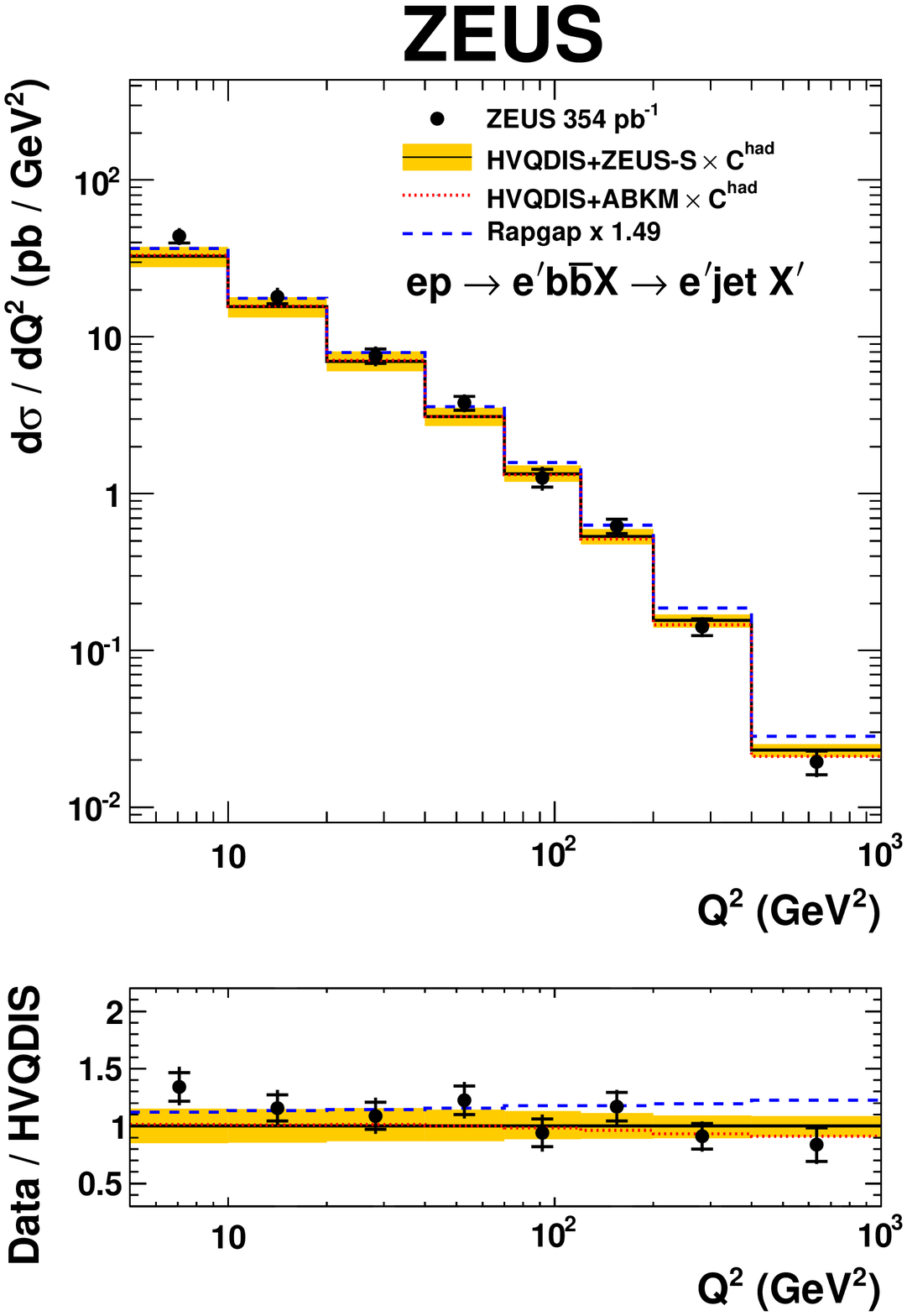}
  \hspace*{1cm}\includegraphics[width=0.45\textwidth]{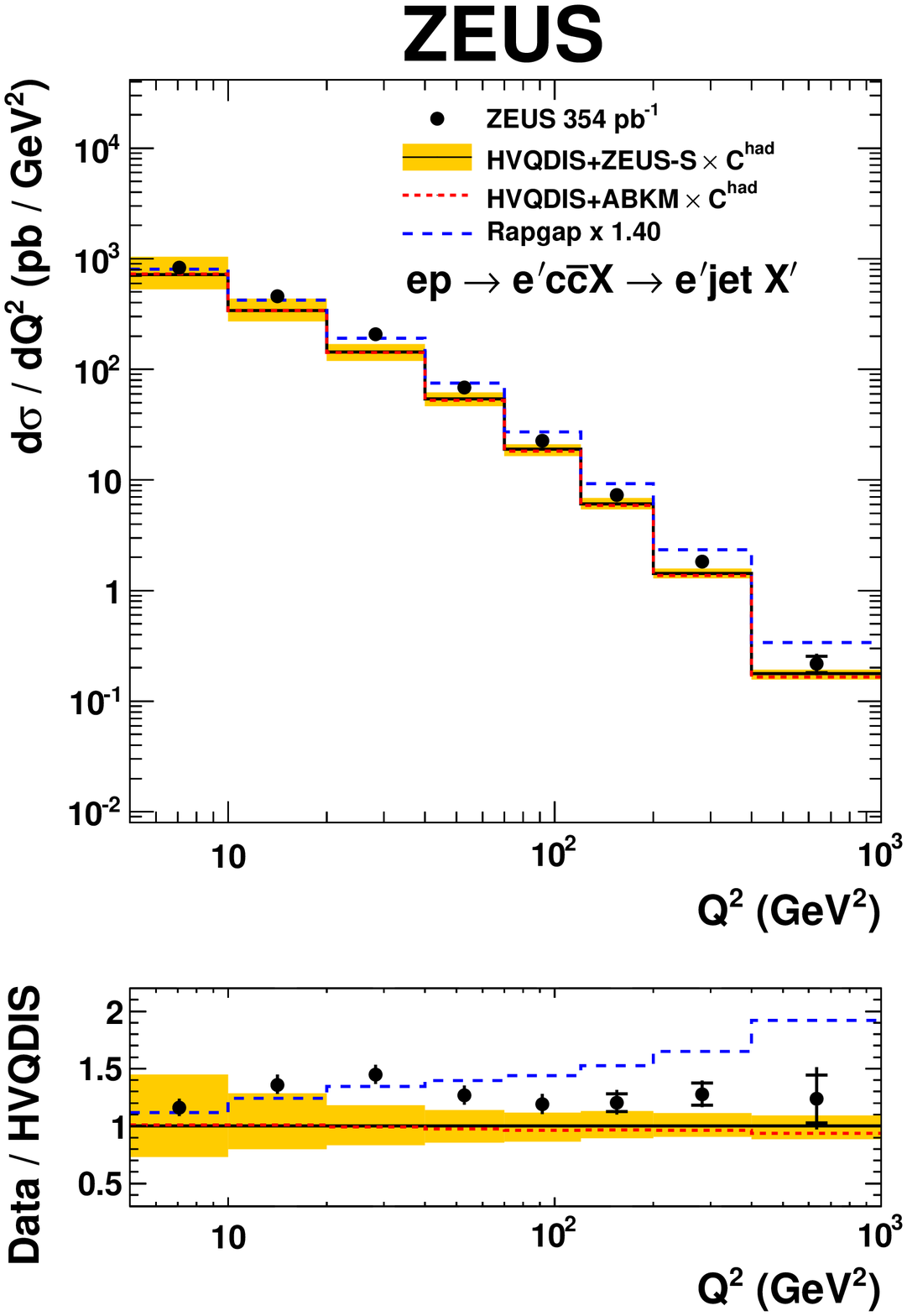}
  \put( -93, 94){(a)}
  \put( -10, 94){(b)}
  \caption{Differential cross sections for inclusive jet production in 
    (a) beauty events and (b) charm events as a function of \Qsq.
    For more details, see the caption of Fig.~\ref{fig:diff_et}.}
  \label{fig:diff_q2}
\end{figure}
\vspace*{\fill}
\clearpage

\vspace*{\fill}
\begin{figure}[!hp]
  \centering
  \includegraphics[width=0.45\textwidth]{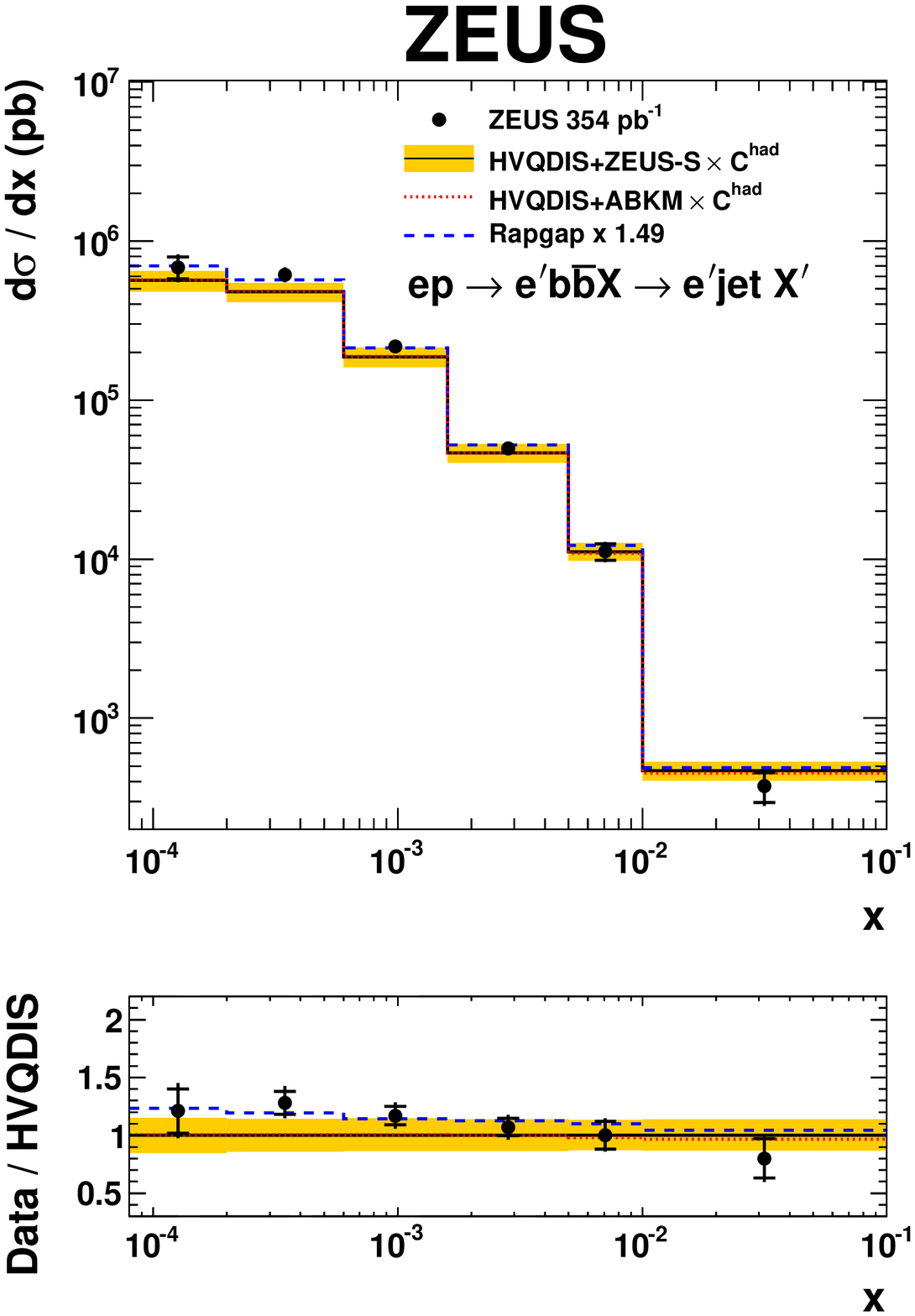}
  \hspace*{1cm}\includegraphics[width=0.45\textwidth]{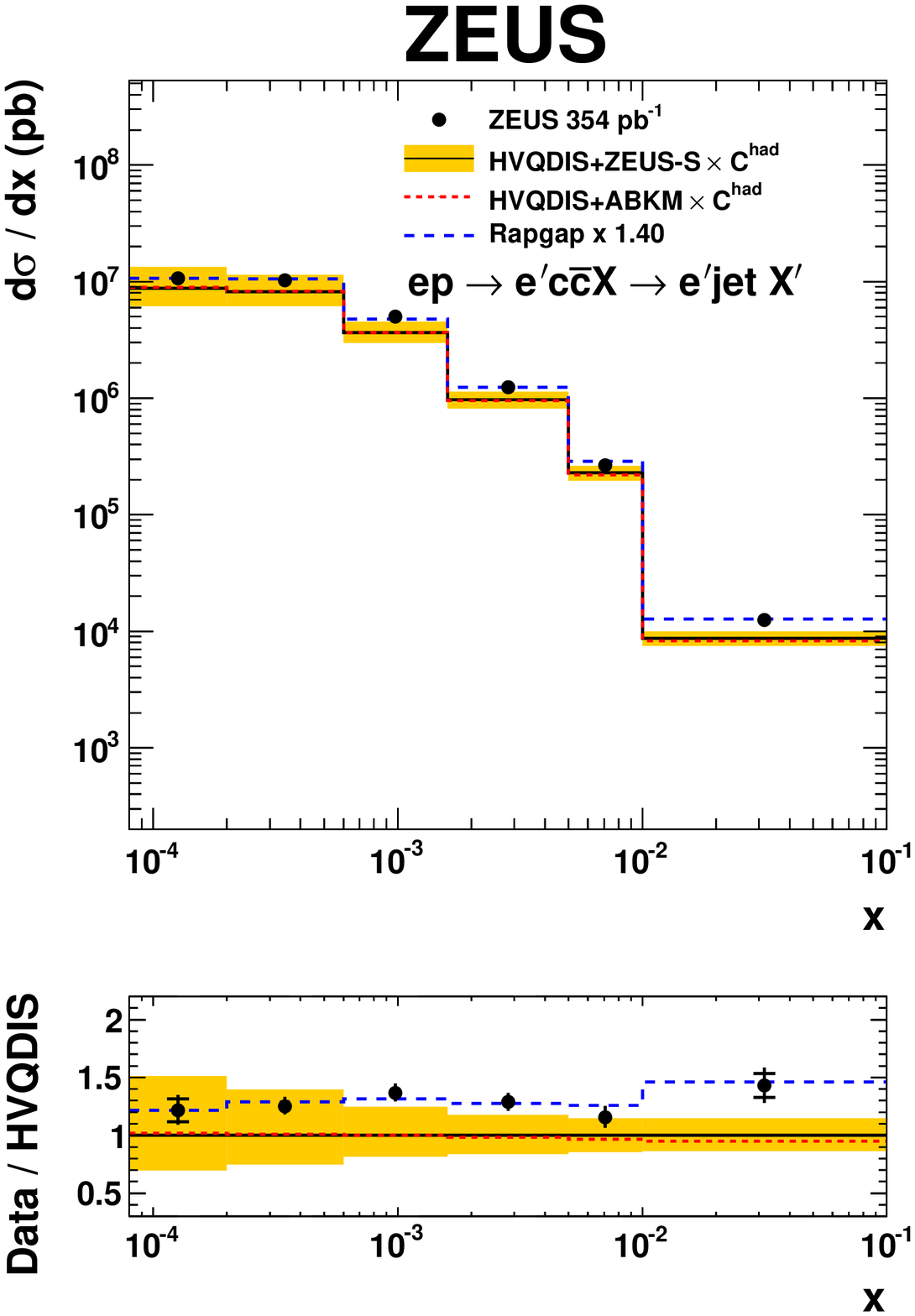}
  \put( -93, 93){(a)}
  \put( -10, 93){(b)}
  \caption{Differential cross sections for inclusive jet production in 
    (a) beauty events and (b) charm events as a function of $x$.
    For more details, see the caption of Fig.~\ref{fig:diff_et}.}
  \label{fig:diff_x}
\end{figure}
\vspace*{\fill}
\clearpage

\vspace*{\fill}
\begin{figure}[!hp]
  \centering
  \includegraphics[width=0.9\textwidth]{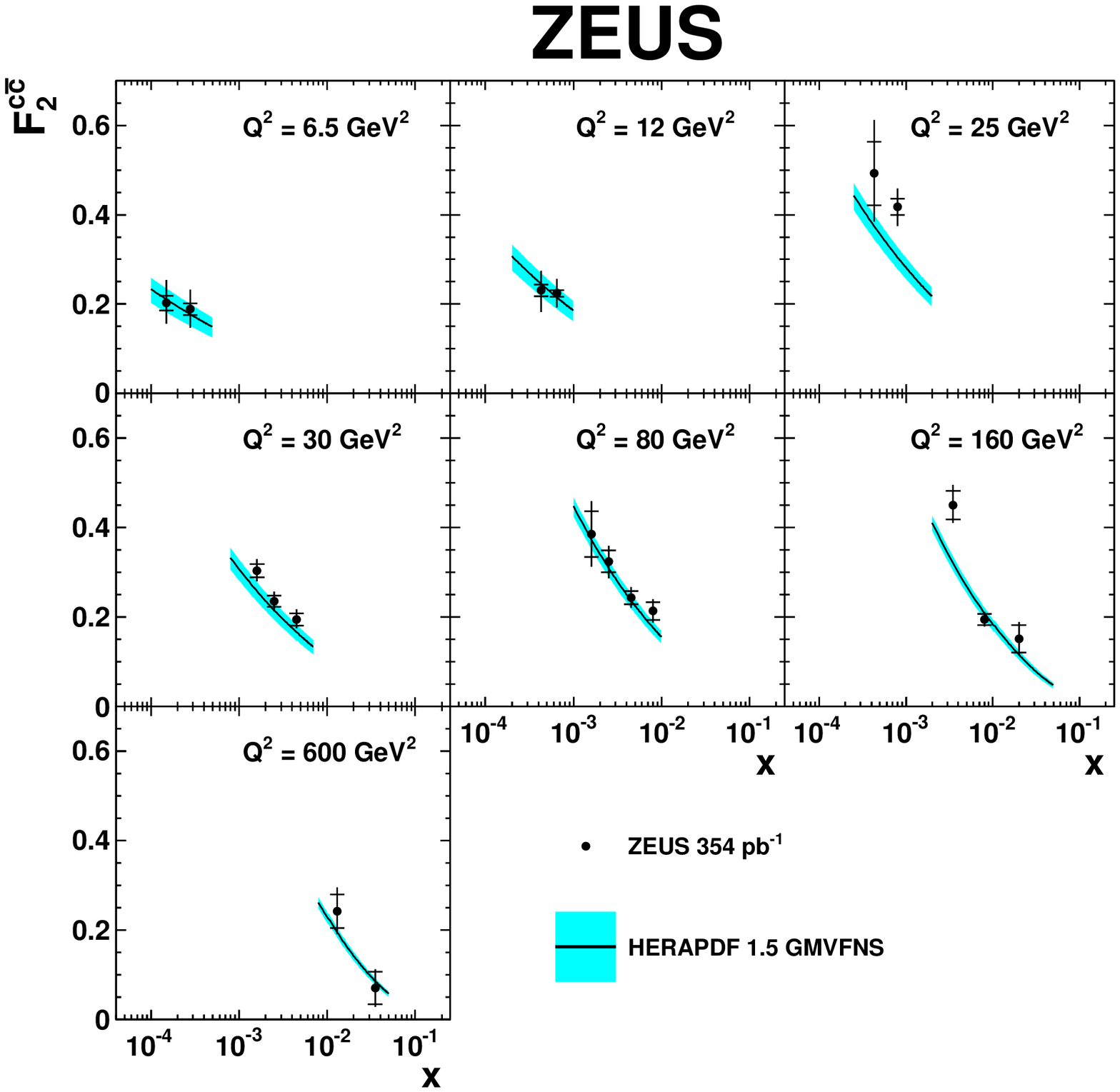}
  \caption{The structure function \Ftwoc (filled symbols) as a
    function of $x$ for seven different values of \Qsq. The inner
    error bars are the statistical uncertainty while the outer error
    bars represent the statistical, systematic (not including the error on the integrated luminosity) and extrapolation
    uncertainties added in quadrature. Also shown are the NLO QCD HERAPDF~1.5
    predictions based on the
    general-mass variable-flavour-number scheme
    (solid line and shaded area for the
    uncertainties).}
  \label{fig:f2c_x}
\end{figure}
\vspace*{\fill}
\clearpage

\vspace*{\fill}
\begin{figure}[!hp]
  \centering
  \includegraphics[width=1\textwidth]{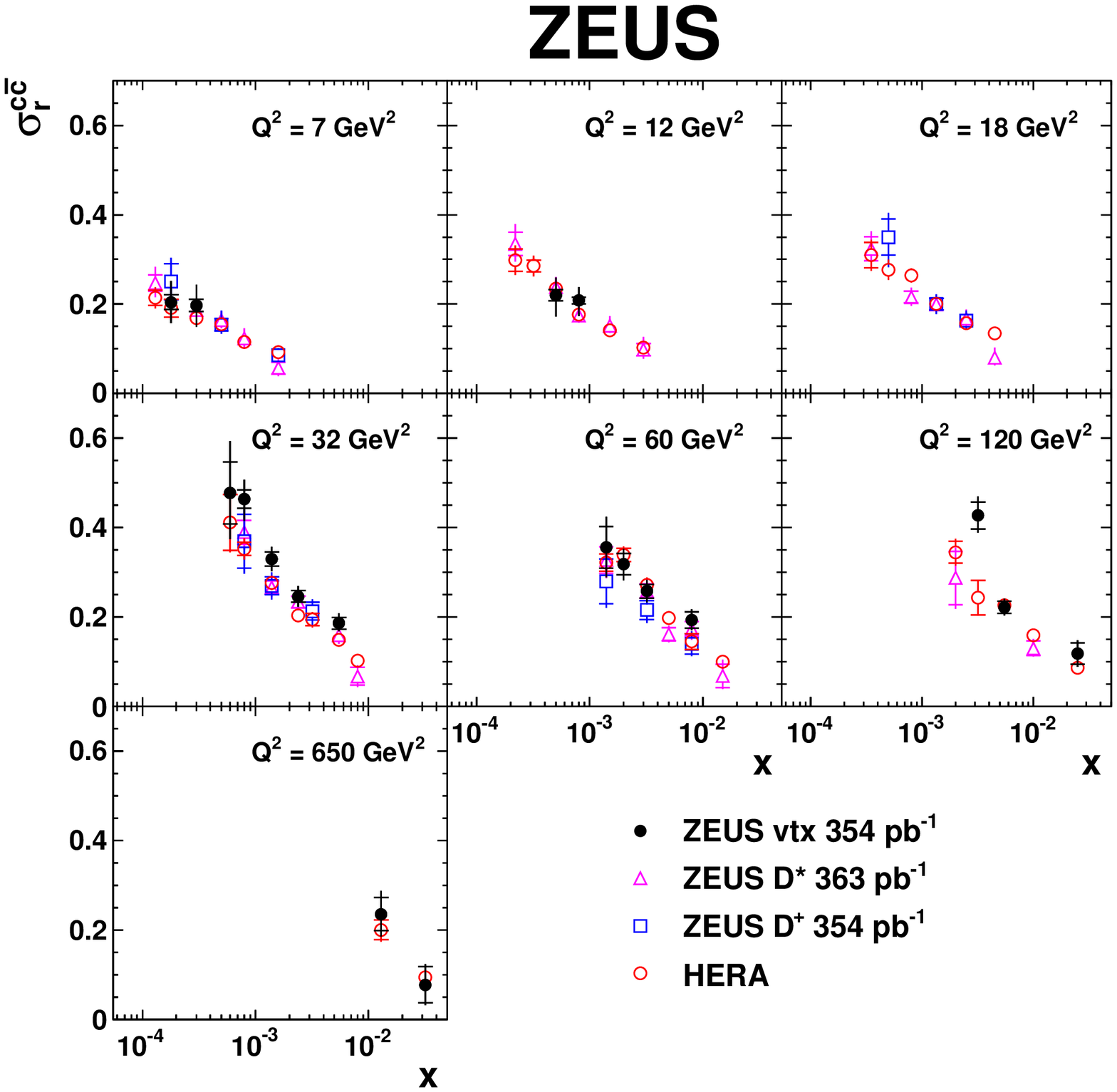}
  \caption{Reduced charm cross section, \sigredc, as a function of $x$
    for fixed values of \Qsq.  Results from the current analysis (filled
    circles) are compared to the ZEUS $D^{*\pm}$
    data~\protect\cite{jhep05.2013.097} (empty triangles), the ZEUS
    $D^+$ measurement~\protect\cite{jhep05.2013.023} (empty squares) and
    the combination of previous HERA
    results~\protect\cite{epj.c73.2311} (empty circles).  The inner
    error bars in the ZEUS measurements show the statistical
    uncertainties.  The inner error bars of the combined HERA data
    represent the uncorrelated part of the uncertainty.  The outer
    error bars include statistical, systematic (not including the error on the integrated luminosity) and theoretical
    uncertainties added in quadrature. }
  \label{fig:sigma_red_c}
\end{figure}
\vspace*{\fill}
\clearpage

\vspace*{\fill}
\begin{figure}[!hp]
  \centering
  \includegraphics[width=0.9\textwidth]{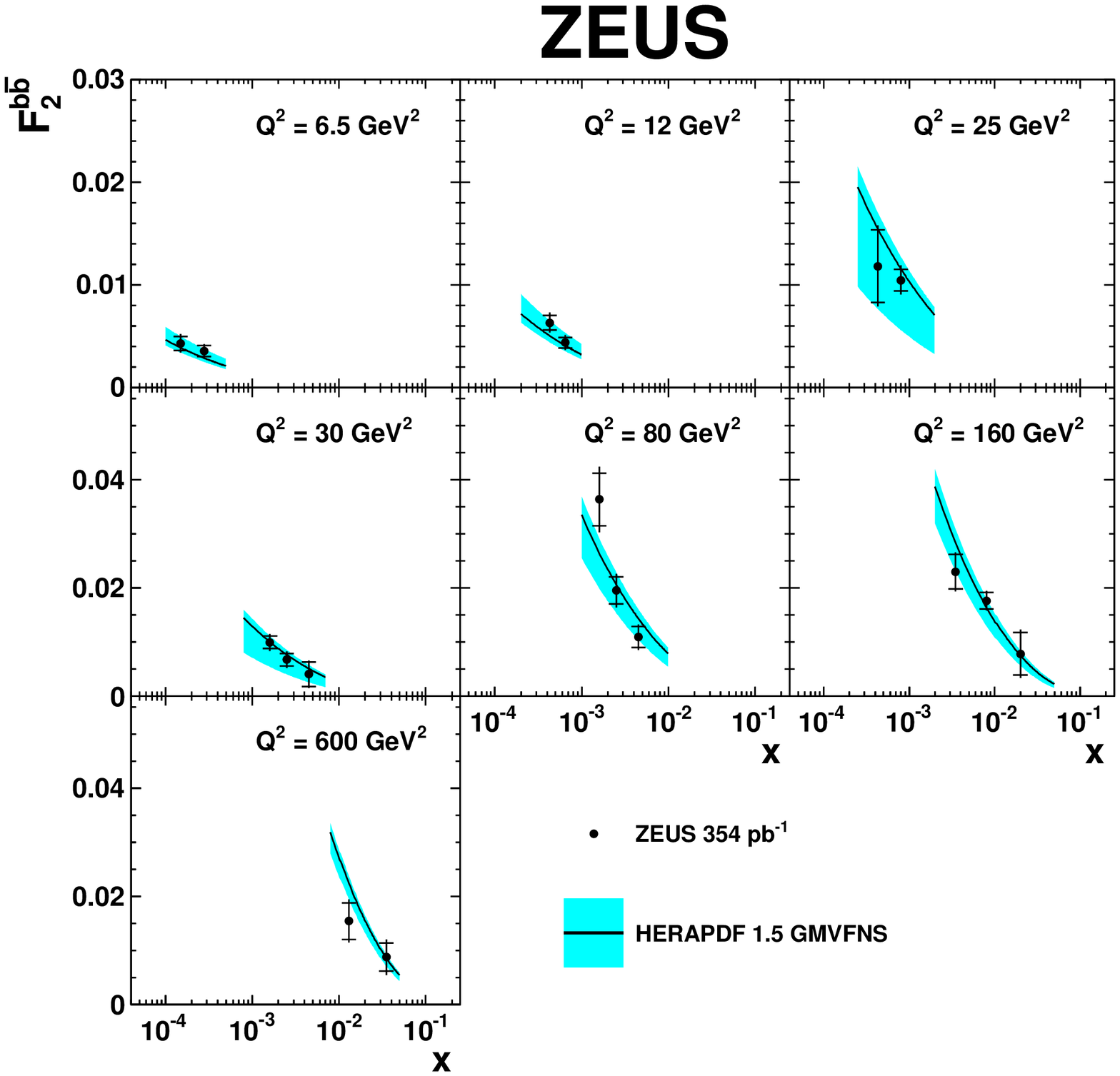}
  \caption{The structure function \Ftwob (filled symbols) as a
    function of $x$ for seven different values of \Qsq. The inner
    error bars are the statistical uncertainty while the outer error
    bars represent the statistical, systematic (not including the error on the integrated luminosity) and extrapolation
    uncertainties added in quadrature. Also shown are the NLO QCD HERAPDF~1.5
    predictions based on the
    general-mass variable-flavour-number scheme
    (solid line and shaded area for the
    uncertainties).}
  \label{fig:f2b_x}
\end{figure}
\vspace*{\fill}
\clearpage

\vspace*{\fill}
\begin{figure}[!hp]
  \centering
  \includegraphics[width=0.8\textwidth]{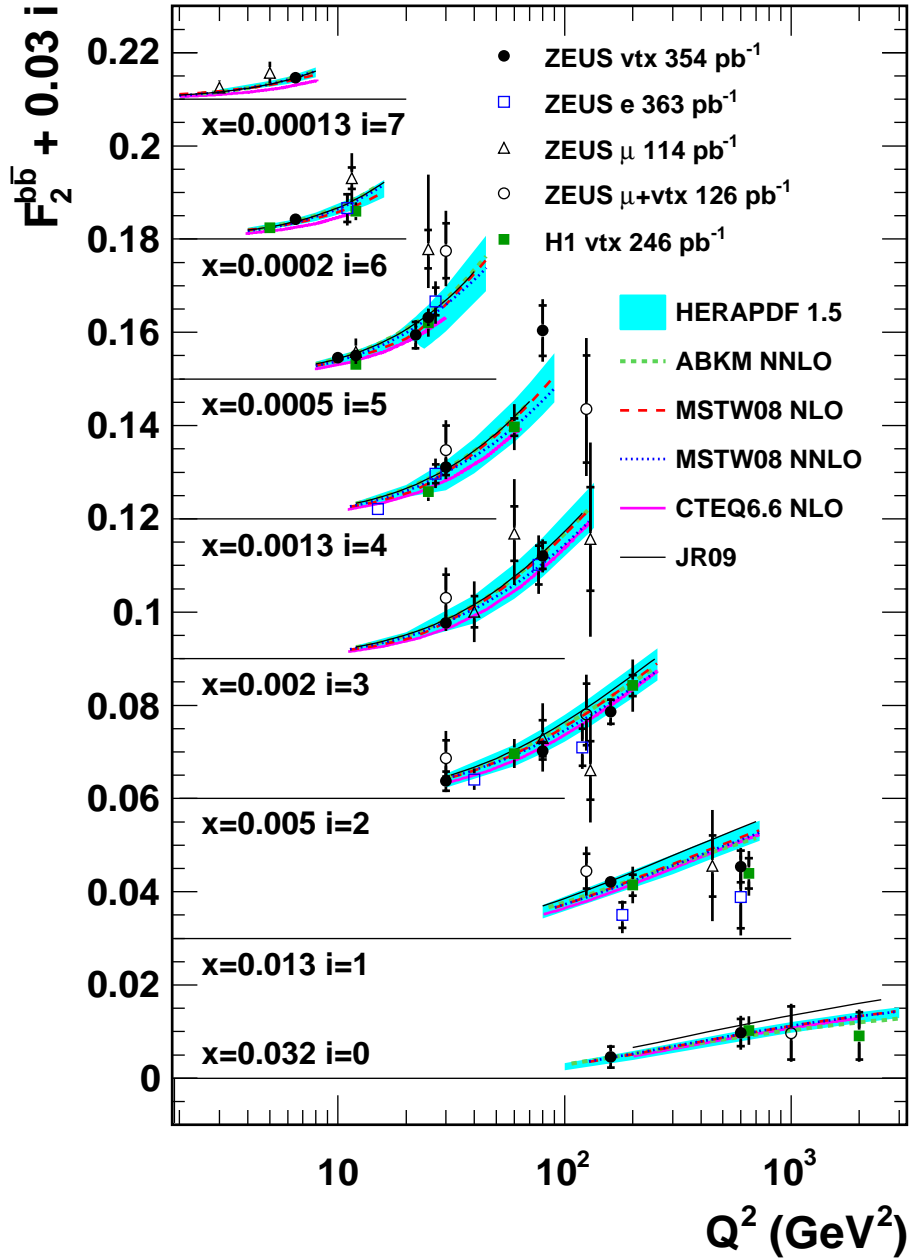}
  \caption{The structure function \Ftwob (filled circles) as
    a function of \Qsq for fixed values of $x$ compared to previous
    results (open squares~\cite{Abramowicz:2011rs}, open triangles~\cite{Abramowicz:2010zq}, open circles~\cite{Chekanov:2009kj} and filled squares~\cite{epj:c40:349, epj:c45:23, Aaron:2009ut}). The inner error bars are the statistical uncertainty
    while the outer error bars represent the statistical, systematic (not including the error on the integrated luminosity)
    and extrapolation uncertainties added in quadrature.  The data
    have been corrected to the same reference $x$ as the previous
    analysis~\protect\cite{Abramowicz:2010zq}.  The measurements are
    compared to several NLO and NNLO QCD predictions~\cite{herapdf.web,PhysRevD.78.013004,
      springerlink:10.1140/epjc/s10052-009-1072-5,
      Thorne:2008xf, jimenez:2009,
      Alekhin2009166, Alekhin:2009ni, Alekhin:2010iu}.}
  \label{fig:f2b_q2}
\end{figure}
\vspace*{\fill}
\clearpage

\vspace*{\fill}
\begin{figure}[!hp]
  \centering
  \includegraphics[width=0.9\textwidth]{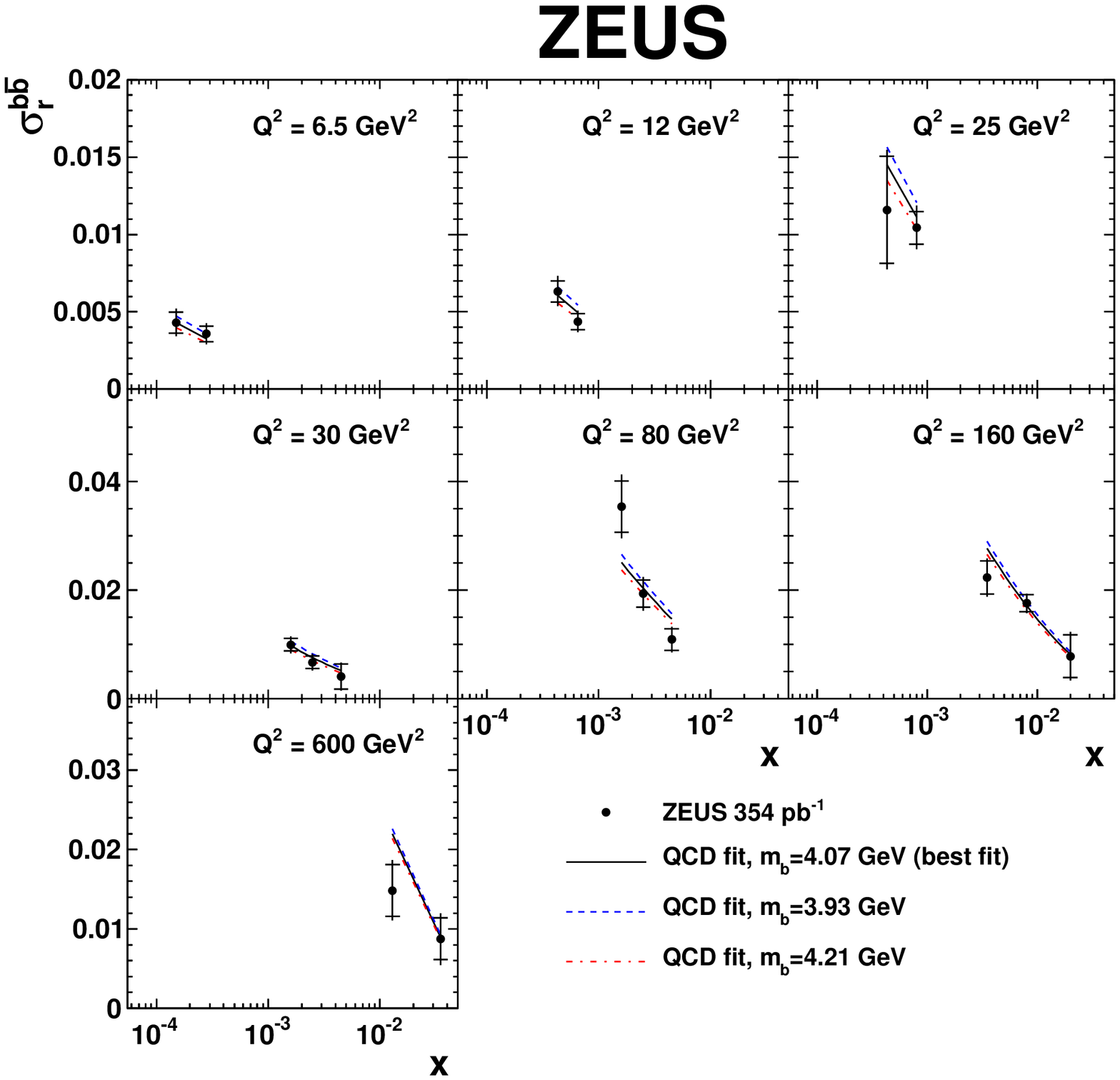}
  \caption{Reduced beauty cross section, \sigredb, (filled symbols)
    as a function of $x$ for seven different values of \Qsq. The
    inner error bars are the statistical uncertainty while the outer
    error bars represent the statistical, systematic (not including the error on the integrated luminosity) and extrapolation
    uncertainties added in quadrature.
    Also shown are the results of the QCD fit described in Section~\ref{sec:mb}.
    The central line indicates the best fit, the lower and upper line give the fit for a higher and lower beauty mass, respectively.}
  \label{fig:sigma_red_b}
\end{figure}
\vspace*{\fill}
\clearpage

\vspace*{\fill}
\begin{figure}[!hp]
  \centering
  \includegraphics[width=1\textwidth]{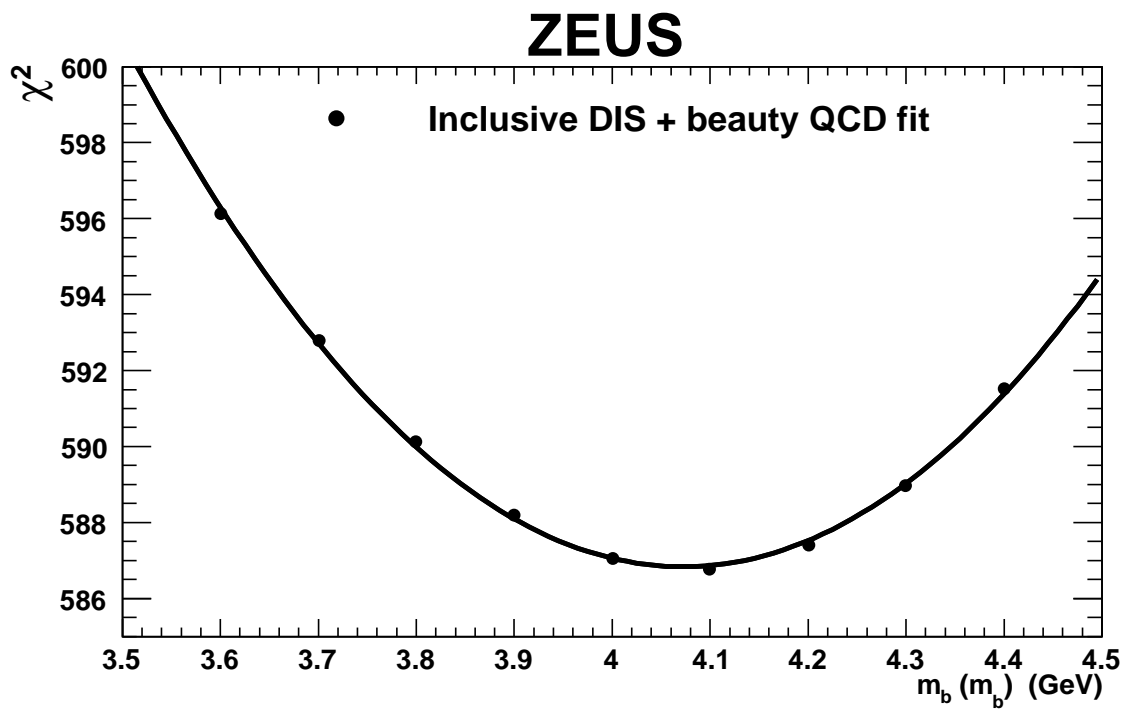}
  \caption{The values of $\chi^2$ for the PDF fit to the combined HERA
    DIS data including the beauty measurements, as a function of the
    running beauty quark mass $m_b(m_b)$. The FFNS ABM scheme is used,
    where the beauty quark mass is defined in the
    \MSbar scheme. The solid line is a second order polynomial parameterisation
    of the points.}
  \label{fig:chi2}
\end{figure}
\vspace*{\fill}
\clearpage
